\documentclass[10pt,preprint]{aastex}
\usepackage{subfigure}
\def\etal{{et~al.}}
\begin{document}
\setcounter{figure}{0}
\title{Proper Motions of Dwarf Spheroidal
Galaxies from \textit{Hubble Space Telescope} Imaging.  V:
Final Measurement for Fornax.\footnote{Based on observations with NASA/ESA
\textit{Hubble Space Telescope}, obtained at the Space Telescope
Science Institute, which is operated by the Association of
Universities for Research in Astronomy, Inc., under NASA contract NAS
5-26555.}}

\author{Slawomir Piatek} \affil{Dept. of Physics, New Jersey Institute
of Technology,
Newark, NJ 07102 \\ E-mail address: piatek@physics.rutgers.edu}

\author{Carlton Pryor}
\affil{Dept. of Physics and Astronomy, Rutgers, the State University
of New Jersey, 136~Frelinghuysen Rd., Piscataway, NJ 08854--8019 \\
E-mail address: pryor@physics.rutgers.edu}

\author{Paul Bristow}
\affil{Instrument Division, European Southern Observatory, 
Karl-Schwarzschild-Str. 2, D-85748, Garching bei Munchen, Germany \\
E-mail address: Paul.Bristow@eso.org }

\author{Edward W.\ Olszewski}
\affil{Steward Observatory, University of Arizona,
    Tucson, AZ 85721 \\ E-mail address: eolszewski@as.arizona.edu}

\author{Hugh C.\ Harris}
\affil{US Naval Observatory, Flagstaff Station, P. O. Box 1149,
Flagstaff, AZ 86002-1149 \\ E-mail address: hch@nofs.navy.mil}

\author{Mario Mateo} \affil{Dept. of Astronomy, University of
Michigan, 830 Denninson Building, Ann Arbor, MI 48109-1090 \\
E-mail address: mmateo@umich.edu}

\author{Dante Minniti}
\affil{Universidad Catolica de Chile, Department of Astronomy and
Astrophysics, Casilla 306, Santiago 22, Chile \\
E-mail address: dante@astro.puc.cl}

\author{Christopher G.\ Tinney}
\affil{Anglo-Australian Observatory, PO Box 296, Epping, 1710,
Australia \\ E-mail address: cgt@aao.gov.au}

\begin{abstract}

	The measured proper motion of Fornax, expressed in the
equatorial coordinate system, is $(\mu_{\alpha},\mu_{\delta})=(47.6\pm
4.6,-36.0\pm 4.1)$~mas~century$^{-1}$.  This proper motion is a
weighted mean of four independent measurements for three distinct
fields.  Each measurement uses a quasi-stellar object as a reference
point.  Removing the contribution of the motion of the Sun and of the
Local Standard of Rest to the measured proper motion produces a
Galactic rest-frame proper motion of
$(\mu_{\alpha}^{\mbox{\tiny{Grf}}}, \mu_{\delta}^{\mbox{\tiny{Grf}}})
= (24.4\pm 4.6,-14.3\pm 4.1)$~~mas~century$^{-1}$.  The implied space
velocity with respect to the Galactic center has a radial component of
$V_{r}=-31.8 \pm 1.7$~km~s$^{-1}$ and a tangential component of
$V_{t}=196 \pm 29$~km~s$^{-1}$.  Integrating the motion of Fornax in a
realistic potential for the Milky Way produces orbital elements. The
perigalacticon and apogalacticon are 118 (66, 137)~kpc and 152 (144,
242)~kpc, respectively, where the values in the parentheses represent
the 95\% confidence intervals derived from Monte Carlo
experiments. The eccentricity of the orbit is 0.13 (0.11, 0.38), and
the orbital period is 3.2 (2.5, 4.6)~Gyr.  The orbit is retrograde and
inclined by $101^{\circ}$ ($94^{\circ}$, $107^{\circ}$) to the
Galactic plane.  Fornax could be a member of a proposed ``stream'' of
galaxies and globular clusters, however the membership of another
proposed galaxy in the stream, Sculptor, has been previously ruled
out.  Fornax is in the Kroupa-Theis-Boily plane that contains eleven
of the Galactic satellite galaxies, but its orbit will take it out of
that plane.

\end{abstract}


\section{Introduction}
\label{intro}

	The Local Group is a dynamic environment.  Galaxies swarm in
the gravitational well of the cluster and move about each other.  The
Earth-bound observer had only a one-dimensional view of this plethora
of motions until a few decades ago: measuring the radial components of
velocities was ``easy,'' measuring the tangential components was hard.
Most of the difficulty lies in measuring the proper motion of a galaxy
--- which must be done with respect to cosmic standards of rest, such
as high-redshift galaxies and quasi-stellar objects (QSOs).  Since the
size of a proper motion decreases with increasing distance, all else
being equal, measuring motions of even the nearest galaxies required
time baselines of several tens of years using images taken with the
ground-based telescopes or just years when using those taken with the
Hubble Space Telescope (HST).

	The analysis of the astrometric data is complex.  The
available ground-based images were acquired with a variety of
telescopes and imagers and under different atmospheric conditions and
air masses and, thus, they contain non-uniform distortions.
Understanding and characterizing these distortions is essential
because the expected measured proper motion is very small --- on the
order of a few tens of mas~century$^{-1}$.  Although space-based data
do not contain distortions caused by the atmosphere, the analysis of
such data is still complicated because of several factors.  1. Useful
reference points are scarce.  Even bright and compact background
galaxies have larger positional uncertainties than star-like objects.
The field of view of \textit{HST} is small, thus, a typical image
contains too few such galaxies to serve as useful standards of rest.
Instead, a QSO with its star-like point-spread function (PSF) is the
standard of rest of choice.  Therefore, the availability of QSOs
within the appropriate brightness range behind the target
galaxy determines the number of independent images --- fields.
2. Stars are scarce, ranging in number from a few tens to a few
hundreds per field.  This scarcity limits the ability to measure the
PSF, to transform stellar coordinates between epochs, and to determine
the average proper motion of the galaxy.  3. The images still contain
distortions that are inherent to the detectors and optics of
\textit{HST}.  The expected motion of stars with respect to the QSO is
on the order of a few thousandth of a pixel in several years,
therefore, a model for these distortions should be accurate to a
sub-pixel scale.  None of the existing models has such accuracy.
Despite these obstacles, clever observations and data analysis have
allowed the list of nearby galaxies with a measured proper motion to
grow steadily.

	In alphabetical order, the list now contains:  Draco
\citep{si93}, Canis Major \citep{di05b}, Carina \citep{p03}, Fornax
\citep{p02,di04}, LMC \citep{jo94,kr94,kb97,an00,dr01,pe02,ka06}, M33
\citep{bler05}, Sagittarius \citep{di05a}, Sculptor \citep{sc95,p06},
SMC \citep{ir96,ka06b}, and Ursa Minor \citep{si93,sc97,p05}.  This
sample is becoming large enough to reveal the three-dimensional nature
of motions in the Local Group.  Notable is the measurement for M33, the
Triangulum Galaxy, the third most luminous galaxy in the Local Group at
a distance from the Milky Way comparable to that for M31.
\citet{bler05} use radio observations with the Very Long Baseline Array
of two groups of water masers located on opposite sides of the galaxy
to derive a proper motion on the order of $50~\mu$as~yr$^{-1}$.  This
is the smallest proper motion of a galaxy ever measured.  The
uncertainties in the model of the rotation of M33 and not measurement
errors dominate the uncertainty of this proper motion.

	Just as the motions of planets around the Sun contain
information about the mass distribution in and the formation of the
Solar System, space velocities of galaxies contain such information
for the Local Group.  In addition, the space velocities give galactic
orbits, and these can demonstrate if the galaxies evolved in seclusion
or if they influenced each other through close encounters, collisions,
or mergers.

	Dark matter continues to evade direct detection.  Its
gravitational interaction with the luminous matter and light is the key
observational clue for its existence.  On a galactic scale, the
rotation curve of a spiral galaxy or the velocity dispersion profile of
an elliptical galaxy are such key observations, whereas, on the scale
of galaxy groups, it is the velocity dispersion of the member
galaxies.  Knowing all three components of the space velocities of the
galaxies provides a check on the assumption of velocity isotropy in the
mass estimates based on the velocity dispersion.  It also offers
additional ways of constraining the amount and distribution of dark
matter.  For example, for the Milky Way and its satellite dwarf
galaxies one may:  1. Derive a lower limit for the mass of the Milky
Way from the assumption that a satellite galaxy is bound or derive an
estimate of the mass of the Milky Way out to the radius of the
satellites from their velocity dispersion.  2. Derive a limit on the
amount of dark matter in a satellite galaxy that approaches the Milky
Way closely enough that the Galactic tidal force is significant
compared to the self-gravity of the galaxy.

The space velocities of galaxies can also test more general features of
models for a cold dark matter cosmology by:  1.  Comparing the
distributions of the observed orbital elements with those predicted by
numerical simulations of the formation of the Local Group
\citep{gill04,mo06}.  2.  Investigating possible close encounters and
collisions among the satellite galaxies \citep{zh98,di04}.

	Two interesting ideas about the formation of the satellite
galaxies of the Milky Way challenge the picture that these galaxies are
surviving substructure, predicted by cold dark matter cosmology.
\citet{lb95} propose that some Galactic satellite galaxies and globular
clusters move along similar orbits --- they are members of a
``stream''  formed when the tidal force of the Milky Way tore a
progenitor satellite galaxy into fragments.  The fragments continue to
move together on the orbit of the progenitor in the absence of
collisions, significant orbital precession, or dynamical friction.
This is a testable idea:  \citet{lb95} predict proper motions for the
members of proposed streams.  The other testable idea is based on the
``planar'' distribution of the Galactic dwarf galaxies.  \citet{kr05}
note that the 11 galaxies nearest to the Milky Way lie within a disk
whose thickness-to-radius ratio is $\leq$~0.15.  \citet{kr05} argue
that such a distribution of satellites is very unlikely for the
substructure predicted by cold dark matter cosmologies, though this is
disputed by \citet{li05} and \citet{ze05}.  If the orbits of the
satellites are indeed within the disk, the geometry of the plane
predicts the coordinates of their poles.  Note that the galaxies do not
have to share the same orbit --- be in a stream --- for their orbital
poles to be the same, all that is required is that their orbits must be
in the same plane.

	This article is one of a series that is attempting to
better-define the kinematics of the Local Group and reports a
measurement of the proper motion for the Fornax dSph galaxy.  The
measurement derives from a larger set of data than and, thus, replaces
the preliminary measurement reported in \citet{p02}.
Section~\ref{sec:fornax} presents Fornax, emphasizing those of its
physical and structural properties that are relevant to our analysis.
Section~\ref{sec:data} describes the observations and data and
Section~\ref{sec:mpm} explains the process of deriving the proper
motion making use of several key performance diagnostics.  The next
three sections, \S \ref{sec:pm}, \S \ref{sec:orbit}, and \S
\ref{sec:disc}, present the measured proper motion and space velocity,
calculate the implied Galactic orbit, and discuss the implications of
the motion, respectively.  The final section, \S \ref{sec:summary}, is
a summary.

\section{Physical and Structural Properties of Fornax}
\label{sec:fornax}

	Among the Galactic dSphs, Fornax is one of the ``giants.''
The luminosity of Fornax, given in Table~1, is comparable to that of
Sagittarius, at least three times greater than that of the
next most luminous dSph, Leo I, and about two orders of magnitude
greater than that of the least luminous --- Draco.  Fornax and Sgr
are the only dSphs with confirmed globular clusters.  Reviews of
research on dwarf galaxies in the Local Group by \citet{ma98} and
\citet{vdb00}, and a study of structure of the Galactic dSphs by
\citet{ih95} are convenient sources for a side-by-side
comparison of the dSphs.

	Discovered by Shapley \citep{sh38a,sh38b}, Fornax is at the
celestial location $(\alpha, \delta)=(02^{\mbox{h}}39^{\mbox{m}}53\fs
1, -34^{\circ} 30^{\prime}16.0^{\prime\prime})$ \citep{vdb99} in the
J2000.00 equatorial coordinate system.  This location corresponds to
$(\ell,b)=(237.245^{\circ}, -65.663^{\circ})$ in the Galactic
coordinate system.  The galaxy is only about 21$^{\circ}$ on the sky
from the Sculptor dSph.

	\citet{sa00} derived a distance modulus of $(m-M)_{0}=20.70 \pm
0.12$ for Fornax using the magnitude of the tip of the red giant
branch.  The data consist of ground-based observations in the $B$, $V$,
and $I$ bands of four overlapping fields, each with an effective area
of $10.7^{\prime}\times 10.7^{\prime}$.  The implied heliocentric
distance is $138 \pm 8$~kpc.  Our study adopts this distance when
calculating distance-dependent quantities.  An earlier measurement of
the distance modulus by \citet{bu99}, using archival \textit{HST}
photometry in the F555W and F814W bands of cluster 4 in Fornax and its
surroundings, produced a similar value.

	Several studies have investigated the structure of Fornax using
wide-field imaging.  A ``benchmark'' investigation is that by
\citet{ih95}, who used star counts from a photographic plate taken with
UK Schmidt telescope to derive the key structural parameters.  Their
field of view is about 36~deg$^{2}$.  In a similar manner, the study
also derives structural parameters for seven other dSphs.  Table~1
lists the values for the ellipticity, position angle, and core and
tidal radii from \citet{ih95}.  This contribution adopts these values
where they are needed.

	An analysis of deeper images, albeit with smaller areal
coverage, by \citet{w03} produces a similar set of structural
parameters as that in \citet{ih95}.  The images cover an area of
8.5~deg$^{2}$ and reach a limiting magnitude of 23.5 in the $V$ band.
The resulting isopleth map shows a steeper decline in the projected
surface density on the south and east sides of the galaxy
\textit{versus} the north and west sides --- a behavior noted previously
by several studies, for example, \citet{d94} or \citet{shs98}.

	More recently, \citet{co05} surveyed Fornax producing images
with a total areal coverage of 10~deg$^{2}$ in the $V$ and $I$ bands
and reaching a limiting magnitude $V=20$.  The surface density of stars
selected to lie along the giant branch shows that: 1) stars beyond the
\citet{ih95} tidal radius are found to the north-west and south-east of
the galaxy, which is the orientation of the minor axis and 2) there is
an additional overdensity of stars in a shell about 1.3$^\circ$ to the
north-west of the galaxy.  \citet{co04} argued that the asymmetry in
the surface density profile noted in the previous paragraph is caused
by another shell about $17^{\prime}$ to the south-east and that this
feature is dominated by a stellar population with an age of about
2~Gyrs.  \citet{co05} propose that these two shells and the other
extra-tidal stars are the signature of the infall of a gas-rich galaxy
with $M_V \simeq -8$.  Deeper imaging by \citet{ol06} confirms that the
inner shell contains an excess of stars formed in a burst about 1.4~Gyr
ago.  The location of these stars in the color-magnitude diagram
indicates that they have [Fe/H]$ = -0.7$, which is higher than that
expected for a low-luminosity dwarf galaxy and implies that these stars
formed from enriched gas originating from Fornax.  The conflict between
the evidence for infall --- shells and extratidal stars --- and the
high metallicity of the inner shell may perhaps be resolved by star
formation stimulated by a minor merger.

	\citet{ma91} measured radial velocities for 44 stars in two
fields located in the direction of Fornax.  One field coincides with
the center of the galaxy and the other is about two core radii away
from the center in the south-west direction.  After excluding 12 likely
non-member stars from the sample, the measured radial velocities imply
a heliocentric systemic velocity of $53 \pm 1.8$~km~s$^{-1}$.  The
sample shows no evidence for a significant rotation around the minor
axis.  The velocity dispersion is $9.9 \pm 1.7$~km~s$^{-1}$ in the
central field and $12.0 \pm 2.8$~km~s$^{-1}$ in the outer --- thus, the
velocity dispersion does not vary significantly with the projected
radius.  The study derives a mass-to-light ratio ($M/L_V$) of $12.3 \pm
4.5$ in solar units for the center of Fornax, but notes that $M/L_V$s in
the range between 5.3 and 26 are also acceptable given the
uncertainties in the fitted structural parameters.  More recently,
\citet{w06} measured radial velocities for an additional 176 stars
scattered throughout the galaxy and combined these measurements with
those in \citet{ma91} in their analysis.  The combined sample with the
most stringent exclusion of non-members implies $53.3 \pm
0.8$~km~s$^{-1}$ for the heliocentric radial velocity, which our
contribution adopts when calculating quantities that depend on this
parameter.  The global velocity dispersion is $11.6 \pm
0.6$~km~s$^{-1}$; the velocity dispersion calculated in radial bins
does not vary with projected radius.

	\citet{p02} and \citet{di04} independently measured the proper
motion of Fornax.  The measurement by \citet{p02} of $(\mu_{\alpha},
\mu_{\delta}) = (49 \pm 13, -59 \pm 13)$~mas~century$^{-1}$ derives
from \textit{HST} images of three distinct fields using STIS or WFPC2.
For two of these fields, the data are at two epochs separated by 1 - 2
years.  For the third field, the data are at three epochs with a
total time baseline of 2 years.  The measurement of \citet{p02} is
preliminary because it was based on partial data while the
observing program was still ongoing.  The measurement quoted
in this contribution, which derives from the complete set of data,
replaces the older preliminary measurement.  The measurement by
\citet{di04} of $(\mu_{\alpha}, \mu_{\delta}) = (59 \pm 16, -15 \pm
16)$~mas~century$^{-1}$ derives from a combination of ground-based and
\textit{HST} imaging.  The time baseline between epochs varies between
20 and 50 years.  Some of the space-based data are the same as those in
\citet{p02}, but they are such a tiny fraction of the whole dataset that
the measurement in \citet{di04} can be considered independent from that
in \citet{p02} or this article.

\section{Observations and Data}
\label{sec:data}

\begin{figure}[ht!]
\centering
\includegraphics[angle=-90,scale=0.8]{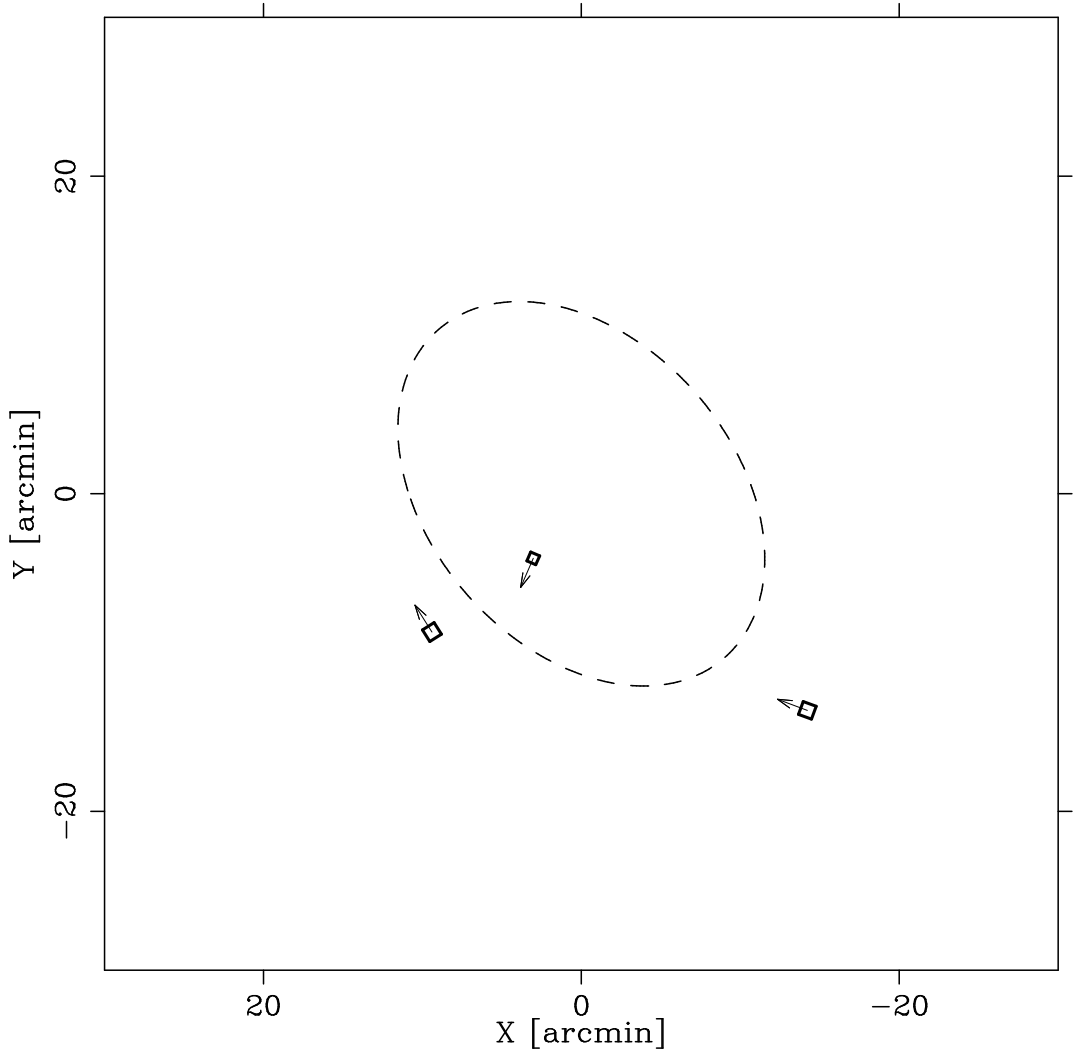}
\caption{A $60^{\prime} \times 60^{\prime}$ section of the sky in the
direction of Fornax.  North is up and East is to the left.  The dashed
ellipse represents the measured core.  The three squares depict the
fields that this article studies.  From North to South, they correspond
to the fields FOR~J$0238-3443$, FOR~J$0240-3434$, and
FOR~J$0240-3438$.  An arrow indicates the direction of the positive $Y$
axis of the CCD.}
\label{fig:fields}
\end{figure}

	The data consist of images taken with \textit{HST} in three
distinct fields in the direction of Fornax.  Each field contains at
least one known QSO.  Figure~\ref{fig:fields} shows the configuration
of the fields --- each represented by a square --- on a $60^{\prime}
\times 60^{\prime}$ section of the sky.  The center of Fornax is at
$(X,Y)=(0^{\prime},0^{\prime})$.  North is up and East is to the left.
The dashed ellipse in the figure represents the core.  Its semi-major
axis, ellipticity, and position angle are those in Table~1.
One of the three fields --- the smallest in size --- is within the
core and close to the minor axis.  An arrow emanating from a field
points in the direction of the positive $Y$ axis in the CCD.  The name
of each field and the equatorial coordinates of its center are in the
first three columns of Table~2.

\begin{figure}[h]
\centering
\includegraphics[angle=-90,scale=0.8]{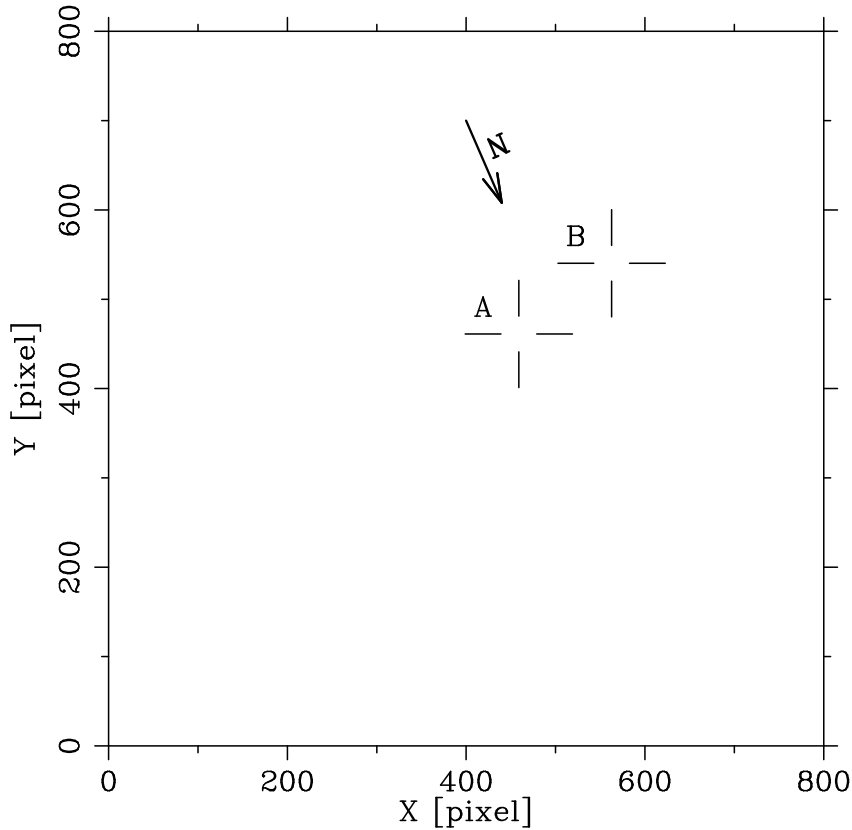}
\caption{A sample image from the epoch 1999 data for the
FOR~J$0240-3434$ field.  It is an average of two images at one dither
position with cosmic rays removed.  Each image of the lensed QSO lies
within a cross-hair.  North is in the direction indicated by an
arrow and east is 90$^\circ$ counter-clockwise from north.}
\label{fig:f2}
\end{figure}

	The most northern field is FOR~J$0240-3434$.  It is
approximately centered on image A of a lensed QSO discovered by
\citet{ti95} and located at $(\alpha, \delta) =
(02^{\mbox{h}}40^{\mbox{m}}07\fs 73, -34^{\circ}
34^{\prime}19.8^{\prime\prime})$ \citep{ye92}.  Image B of the QSO is
$6.1^{\prime \prime}$ away from image A.  \citet{ti95} measured the
$B$-band magnitude and $B-H$ color of image A to be $19.00\pm0.05$ and
$1.42\pm0.08$, respectively, and of image B to be $19.77\pm0.05$ and
$1.92\pm0.08$, respectively.  The redshift of the QSO measured from
both images is 1.4.

	HST observed the FOR~J$0240-3434$ field on March 10, 1999;
March 8, 2001; and March 8, 2003 using the Planetary Camera (PC2) of the
Wide Field and Planetary Camera 2 (WFPC2) and the F606W filter
(columns 4 --- 6 of Table~2).  The observations produced 18, 16, and
16 images, respectively, each with an exposure time of 160~s (column
7 of Table~2).  The position angle of 156.6$^\circ$ for the Y axis of
the CCD is the same to within 0.01$^\circ$ for all of the exposures.
Figure~\ref{fig:f2} is a sample image from the 1999 epoch.

	The other two fields shown in Figure~\ref{fig:fields} are, in
order of decreasing declination FOR~J$0240-3438$ and FOR~J$0238-3443$.
Again, each field contains and is approximately centered on a known
QSO.  \citet{ti97} provide information on these QSOs.  The QSO in the
FOR~J$0240-3438$ field is at $(\alpha,
\delta)=(02^{\mbox{h}}40^{\mbox{m}}38\fs 7, -34^{\circ}
38^{\prime}58^{\prime\prime})$ (J2000.0), has a $B$-band magnitude of
20.2, and a redshift of 0.38.  The QSO in the FOR~J$0238-3443$ field is
at $(\alpha, \delta)=(02^{\mbox{h}}38^{\mbox{m}}43\fs 8,-34^{\circ}
43^{\prime}53^{\prime\prime})$ (J2000.0), has a $B$-band magnitude of
20.2, and a redshift of 2.0.

\begin{figure}[h]
\centering
\includegraphics[angle=-90,scale=0.8]{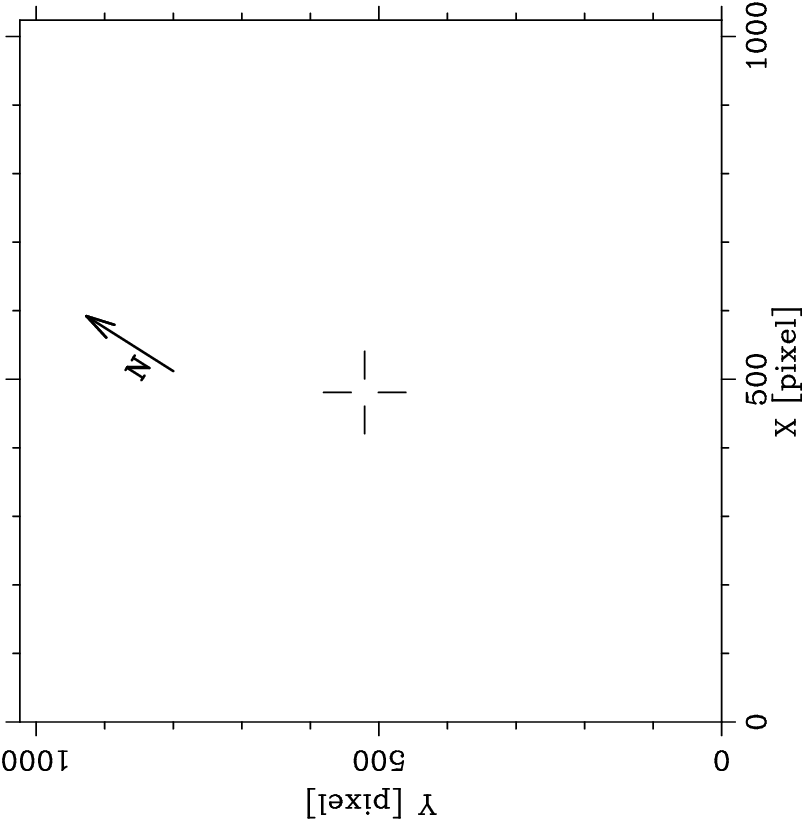}
\caption{A sample image from the epoch 2000 data for the
FOR~J$0240-3438$ field.  It is a sum of three images at one dither
position with cosmic rays removed.  The cross-hair indicates the QSO.}
\label{fig:f3}
\end{figure}
\begin{figure}[h]
\centering
\includegraphics[angle=-90,scale=0.8]{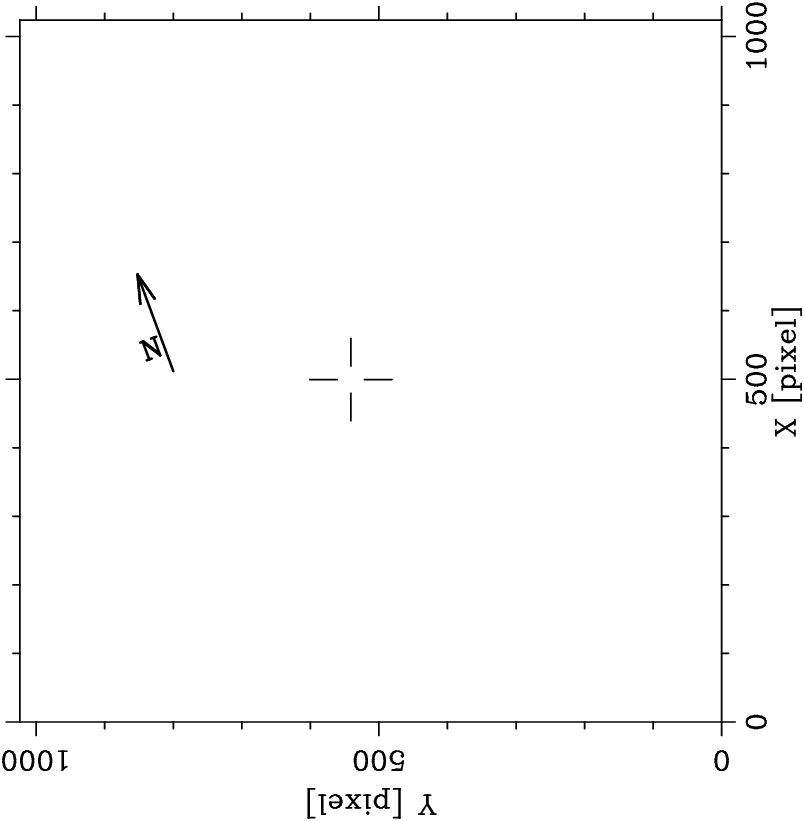}
\caption{A sample image from the epoch 2000 data for the
FOR~J$0238-3443$ field.  It is a sum of three images at one dither
position with cosmic rays removed.  The cross-hair indicates the QSO.}
\label{fig:f4}
\end{figure}
HST observed the FOR~J$0240-3438$ field on January 31, 2000; January
25, 2001; and January 29, 2002.  It observed the FOR~J$0238-3443$ field
on March 8, 2000; March 8, 2001; and March 9, 2003.  For all of these
observations, HST used the Space Telescope Imaging Spectrograph (STIS)
with no filter (50CCD).  There are 24 images per epoch of the
FOR~J$0240-3438$ field, each with an exposure time of 192~s, for a
total of 72 images.  The position angle of the Y axis for all of the
images is 32.4$^\circ$ to within 0.05$^\circ$.  There are also 24
images per epoch in the FOR~J$0238-3443$ field for a total of 72
images.  The average exposure times at the three epochs are 193~s,
192~s, and 189~s, respectively, and the position angle of the Y axis is
69.7$^{\circ}$ to within 0.06$^\circ$.  Table~2 summarizes the above
information.  Figures~\ref{fig:f3} and \ref{fig:f4} are sample images
from the earliest observations in the two fields.

\subsection{Charge Transfer Efficiency}

	During the readout of a CCD, the electrical charge present in a
pixel at the end of an exposure transfers from pixel to pixel towards
the transfer register.  For STIS, this motion is in the positive $Y$
direction, or ``up,'' whereas for PC2, it is in the negative, or
``down,'' direction.  Once in the transfer register, the charge moves
along the $X$ direction to the circuit that measures its amount.  The
transfer of the charge is imperfect:  CCD pixels contain ``charge
traps'' which both capture and release charge.  Thus, the quantity of
charge passing through a pixel can either increase, if more charge is
captured than released, or decrease, if the reverse is true.  Because
the charge traps have a finite capacity, they have a smaller fractional
effect on bright sources than on faint sources.  The origin of charge
traps is not completely understood, but it is known that the efficiency
of charge transfer in STIS and WFPC2 decreases with time
\citep{wh99,go06} --- most likely because of defects in the crystal
lattice of the CCD created by cosmic radiation \citep{ja91}.
Therefore, the effect of charge traps on the photometry and astrometry
increases with time.

	\citet{br05} describe the various effects caused by charge
traps and why these effects arise, emphasizing the impact on
time-dependent astrometry such as the measurement of a proper motion.
\begin{figure}
\centering
\includegraphics[angle=-90,scale=1.0]{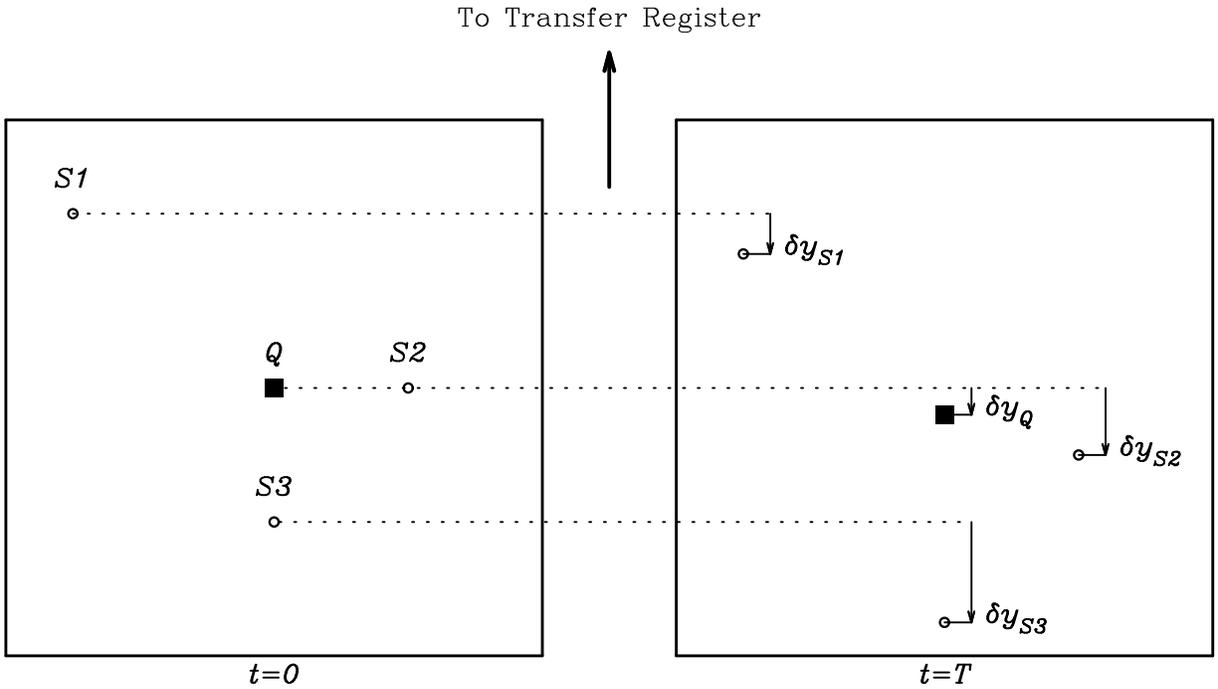}
\caption{Hypothetical images taken at $t=0$ (left) and at $t=T$ (right)
with a CCD containing charge traps whose number increases with time.
There are four sources in an image: three stars and a QSO.  The open
circles, labeled $S1$, $S2$, and $S3$, represent the three stars while
the solid square, labeled $Q$, represents the QSO.  The location
depicted for a source in an image corresponds to its measured
centroid.  The signal in the QSO is much larger than the signal in each
star, the latter assumed to be about the same.  Because the number of
charge traps has increased between $t=0$ and $t=T$, a measured centroid
in the ``later'' image has a smaller value of $Y$, that is, the
centroid has shifted in the direction of decreasing $Y$ values.  Arrows
in the figure represent these shifts.  The magnitude of the shift
depends on the location of a source in the image and on its signal.}
\label{fig:f5}
\end{figure}
Figure~\ref{fig:f5} illustrates how charge traps can create a spurious
proper motion.  Consider a field that contains three stars with similar
signal values and a QSO with a much larger signal.  In the figure, the
stars and the QSO are represented by open circles and a solid square,
respectively, and are labeled as $S1$, $S2$, $S3$, and $Q$.  The QSO is
the reference point.  Assume that the stars, the QSO, and the observer
are at rest with respect to each other, which means that the observer
should detect no motion for any star with respect to the QSO.
Figure~\ref{fig:f5} depicts two hypothetical images of this field taken
at $t=0$ and at $t=T$ with exactly the same pointings of the telescope
and with a CCD that contains charge traps whose number has increased
with time.  Let the measured centroid of a source be depicted by the
location of the source in an image.  Because in the time interval $T$
separating the two images the number of charge traps has increased, a
centroid in the later image has a smaller value of $Y$ --- it has
shifted away from the transfer register.  Here we ignore a possible
shift in the $X$ direction resulting from inefficient charge transfer
in the serial register since this inefficiency is negligible for STIS
\citep{bwn02} and one-third as large as in the parallel direction 
for WFPC2 \citep{he04}.  The arrows $\delta y_{Q}, \delta y_{S1},
\delta y_{S2}$, and $\delta y_{S3}$ represent these shifts for the QSO
and the three stars, respectively.  The magnitude of the shift is
larger the smaller the signal of the object and the greater its
distance from the serial register.  Thus, for example, $S3$ has a
spurious motion in the coordinate system of the CCD of $(\delta y_{S3}
- \delta y_{Q})/T$, which implies that the observer would measure a
proper motion for $S3$ with respect to the QSO.

	Both WFPC2 and STIS contain charge traps whose number
increases with time as a consequence of the ongoing production of
lattice defects by the cosmic radiation \citep{ja91}.  The transfer
register is at the bottom of the image for PC2 and at the top for
STIS.  Therefore, the centroids shift towards larger values of $Y$, or
``up,'' for PC2 and towards smaller values of $Y$, or ``down,'' for
STIS.  Accurate astrometry requires either that the shifts are
accounted for in the analysis or that a CCD image be ``restored'' so
that its pixels have the same values as immediately before the
readout.  \citet{ba02} developed a software package that restores an
image taken with STIS, which was used to correct all of our images
taken with STIS prior to our analysis.  No corresponding software
exists for restoring an image taken with PC2.  In this case, our
analysis measures and removes the shifts --- the details are in
Section~\ref{sec:pmpc}.

	To estimate how greatly the decreasing charge transfer
efficiency can affect the measured proper motion, this study performs a
parallel analysis ignoring the effects of charge traps and comments on
the difference.

\section{Measuring Proper Motion}
\label{sec:mpm}

	\citet{p06} and references therein describe our method of
deriving a proper motion from undersampled images taken with
\textit{HST}.  This article outlines briefly the four major components
of the method.

1.  Determine the initial estimates for the centroids of objects ---
stars and the QSOs.  This task uses stand-alone astronomy software
packages such as DAOPHOT, ALLSTAR, and DAOMASTER
\citep{st87,st92,st94}.  For each field and epoch, this step produces a
list of objects at that epoch and the translational offsets between the
chronologically first image of an epoch --- the fiducial image --- and
those that follow.  The centroid of an object on the list is in the
coordinate system of the fiducial image; however, its centroid in any
other image derives from the known offsets.

2.  Construct an effective point-spread function (ePSF; \citet{ak00})
for each field and epoch.  Using a select set of stars and the QSOs,
our method builds a single, position-independent or constant ePSF that
is used for all of the images of a field at an epoch.  An object
contributes to the construction of the ePSF if its signal-to-noise
ratio (S/N) is greater than some limit --- 15 for these data --- and if
its inner $3\times3$ array of pixels (see \citet{p02} for details about
the structure of the data) have data quality flags equal to 0.  If
there are $N$ images for a given field and epoch, an object may
contribute up to $N$ times to the construction of the ePSF --- fewer if
the object is flagged out in some images.

3.  Derive accurate centroids for objects by fitting an ePSF to the
image of each object using least squares.  The fitting procedure for a
set of images at an epoch involves an interplay between the centroids,
the ePSF, and the transformation between the coordinate system of an
image and the fiducial image.  An improvement in the centroids leads to
more accurate transformations and a better estimate of the ePSF and
this is iterated until a stable solution ensues.  Averaging the
centroids from each image transformed to the coordinate system of the
fiducial image produces the coordinate for an object at one epoch and
its uncertainty.  The fiducial coordinate systems for different epochs
are generally offset from each other because of small changes in the
telescope pointing and rotation and are also not congruent because of
changes in image scale and telescope motion (the aberration of starlight).

4.  Determine the motion of the QSO in a ``standard'' coordinate
system.  In this coordinate system, the stars of the dSph are at rest
while all other objects move.  A proper motion of the dSph derives from
the reflex motion of the QSO in this coordinate system.  The standard
coordinate system is defined as the fiducial coordinate system of the
first epoch.  The $X$ and $Y$ coordinates of a star of the dSph in this
system are the average of its coordinates at the three epochs
transformed into the standard coordinate system.  A six-parameter
linear transformation converts coordinates to the standard coordinate
system.  The transformation also includes a fitted linear motion for
the QSO and, iteratively, for objects whose large scatter in their
transformed coordinates indicate that they are not members of the
dSph.  For data taken with PC2, an additional term in the
transformation that is linear with $y$ and depends on the flux of
an object corrects the effects caused by charge traps.

	Both STIS and WFPC2 exhibit geometric distortions which are
several pixels at the edge of the field.  These are large compared to
the approximately 0.001~pixel precision in measuring the positions of
objects required to determine the proper motion of a dSph.  Even the
accuracy with which the distortions are known is larger than the
required precision, so our observations must minimize the impact of the
distortions.  Thus, the observations attempted to place objects on the
same pixels of the CCD at each epoch.  If the execution of this plan
were perfect, no correction for the distortions would have been
necessary.  Although the roll angles are similar to a high degree,
there are linear offsets between epochs on the order of a few tens of
pixels.  Therefore, our method corrects the distortions using the
prescriptions from the STIS Data Handbook \citep{bwn02} for the data
taken with STIS and from \citet{ak03} for the data taken with WFPC2.

	An error in the manufacture of the WFPC2 CCDs makes every 34th
row in the PC2 images narrower than the other rows \citep{sh95}.
Dubbed the ``34th-row defect,'' it can produce a spurious contribution
to the measurement of the proper motion.  Our method corrects the
centroids for the effect of this defect using the prescription from
\citet{ak99}.

\subsection{Flux Residuals}
\label{sec:rf}

	\citet{p02} developed diagnostics for the performance of their
algorithm that determines the centroids of objects.  The remainder of
this section describes the ``flux residual'' and ``position residual''
diagnostics and how the algorithm performed in the three fields.

	The flux residual diagnostic, ${\cal RF}$, defined by
Equation~22 in \citet{p02} is a measure of how well the constructed
ePSF matches an image of an object.  If the match is perfect, ${\cal
RF} = 0$.  However, if the image of an object is wider than the ePSF,
${\cal RF} > 0$; otherwise, ${\cal RF} < 0$.  If ${\cal RF}$ does
depart from zero, the departure increases with increasing S/N
for an object.

	Plotting ${\cal RF}$ as a function of position in a CCD can
reveal potential flaws in the data and analysis.  In the presence of
only random noise, a plot of ${\cal RF}$ \textit{versus} $X$ or $Y$
would show that the points scatter around ${\cal RF} = 0$.  Any other
distribution is undesirable and can mean, for example, that the true
PSF varies with position in a CCD, that the constructed ePSF has an
incorrect shape, or that the PSF of an object is distinctly different
from the ePSF, as it would be in the case of a galaxy with a resolved
core.

\subsubsection{Flux Residual Diagnostic for the FOR~J$0240-3434$ Field}
\label{sec:rf-pc}

\begin{figure}
\centering
\includegraphics[angle=-90,scale=0.55]{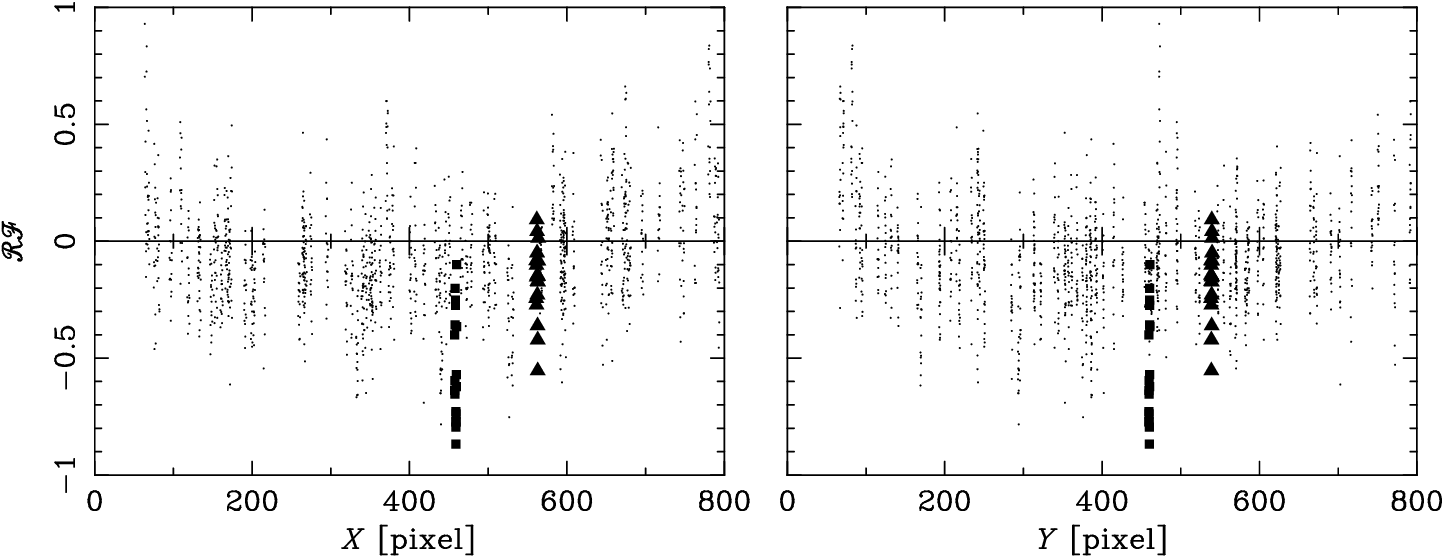}

\vspace*{\baselineskip}
\includegraphics[angle=-90,scale=0.55]{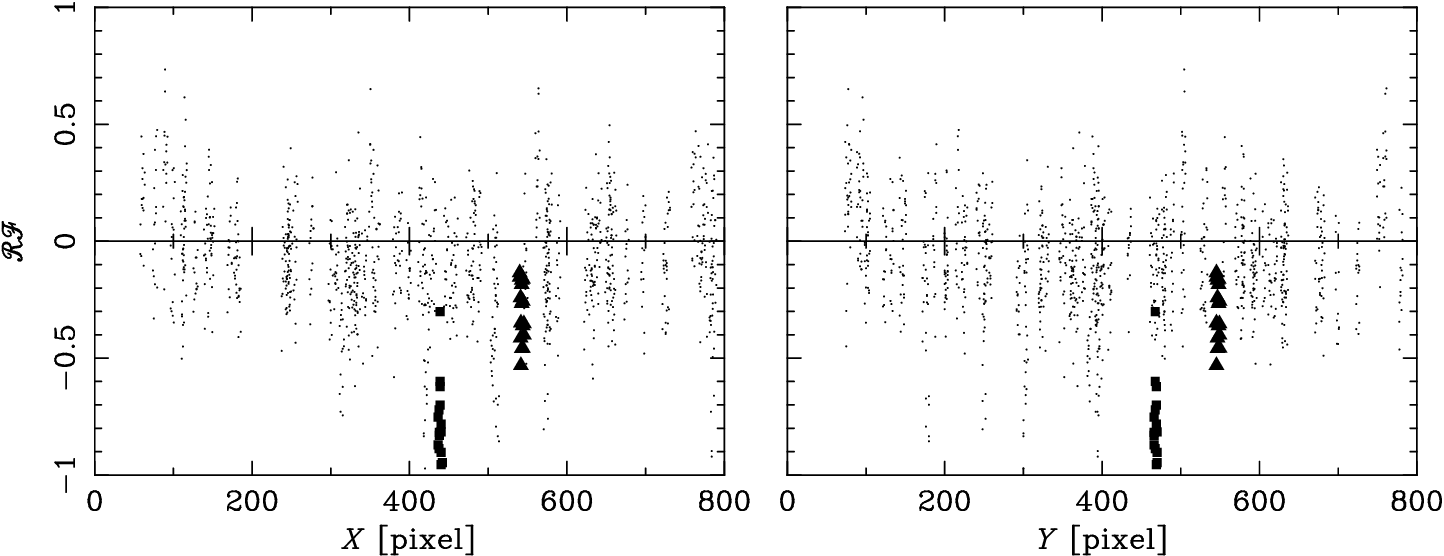}

\vspace*{\baselineskip}
\includegraphics[angle=-90,scale=0.55]{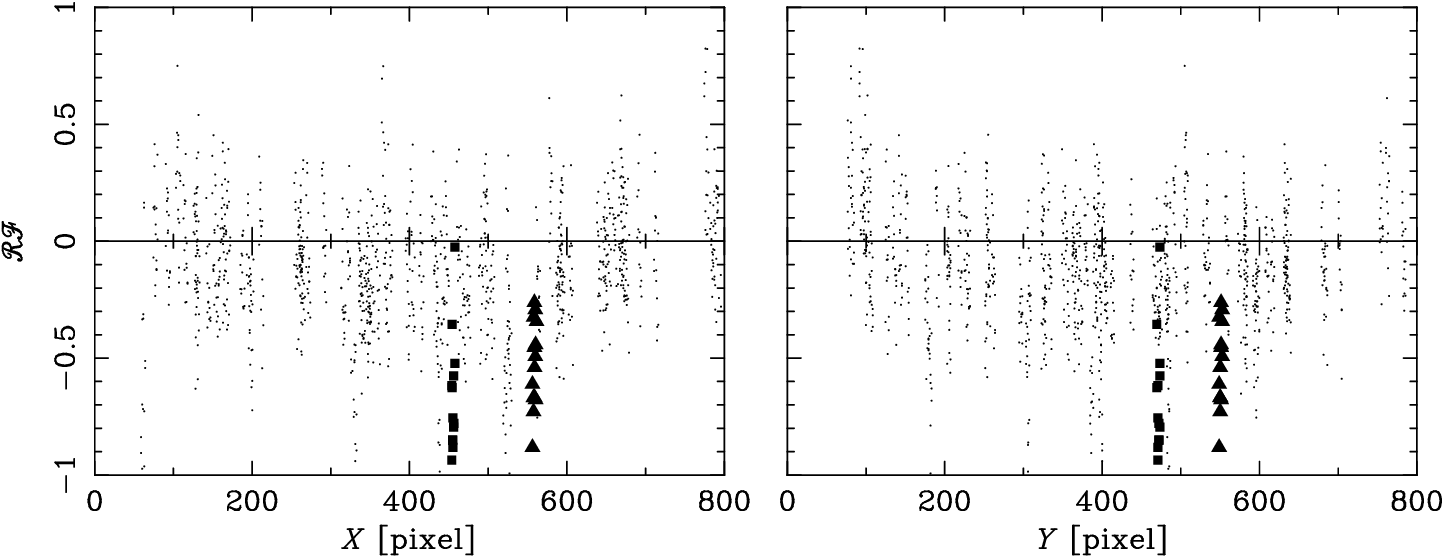}
\caption{Left column: Flux residual ${\cal RF}$ as a function of $X$.
Right column: Flux residual ${\cal RF}$ as a function of $Y$.  The
plots are for the FOR~J$0240-3434$ field that was imaged with PC2.
From top to bottom, the rows of plots are for 18 images at epoch 1999,
16 at 2001, and 16 at 2003, respectively.  The figure shows
points only for those objects that have S/N~$> 15$.  The solid squares
mark points corresponding to image A of the QSO, while the solid
triangles do the same for image B.}
\label{rf-pc}
\end{figure}

	Figure~\ref{rf-pc} plots ${\cal RF}$ \textit{versus} $X$ in the
left column and ${\cal RF}$ \textit{versus} $Y$ in the right column for
the FOR~J$0240-3434$ field imaged with PC2.  From top to bottom, the
plots are for the epochs 1999, 2001, and 2003.  Solid squares mark the
points corresponding to image A of the QSO, whereas the solid triangles
do the same for image B.
All of the plots show a modest ``bow-shaped'' dependence of ${\cal RF}$
on $X$ and $Y$ indicating that the true PSF varies with location.  The
constructed ePSF is narrower than the true PSF around the edges of the
CCD and wider near the center.  The values of ${\cal RF}$ for both
images of the QSO tend to be negative; those for image A are more
negative than those for image B, most likely because of the variation
of the PSF across the image.  The ${\cal RF}$ for the QSO is more
negative than those of most nearby stars, which could be due to the
higher S/N of the QSO or, perhaps, to a narrower PSF caused by the
bluer color of the QSO.

\subsubsection{Flux Residual Diagnostic for the FOR~J$0240-3438$ Field}
\label{sec:rf-stis1}

\begin{figure}
\centering
\includegraphics[angle=-90,scale=0.55]{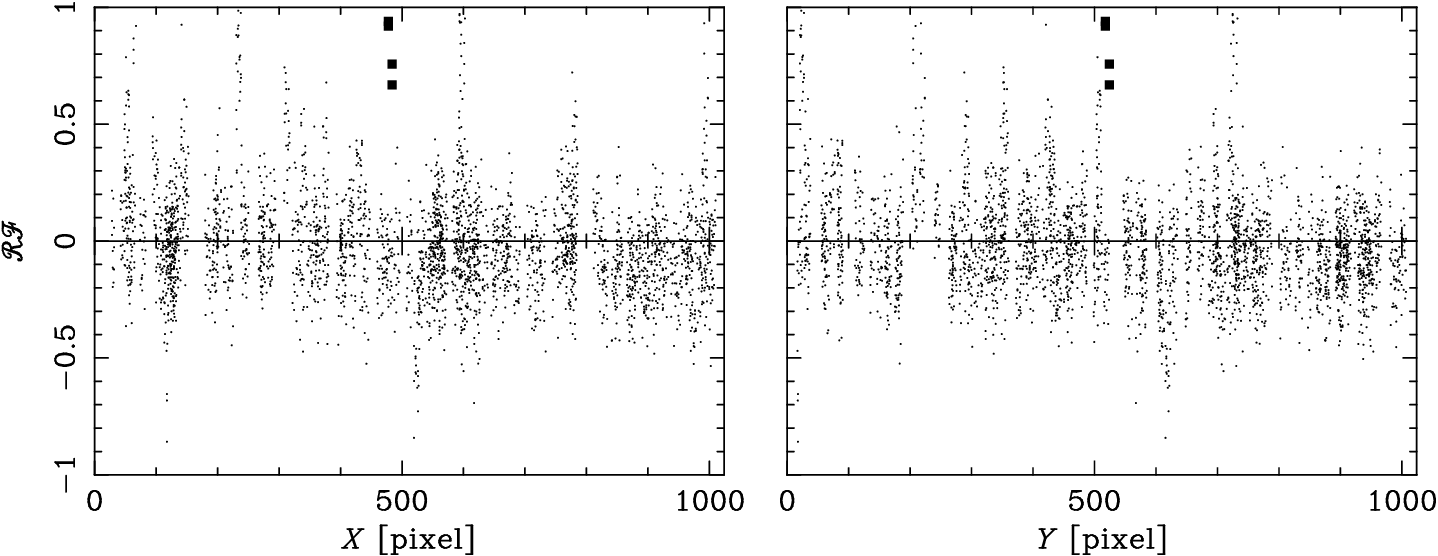}

\vspace*{\baselineskip}
\includegraphics[angle=-90,scale=0.55]{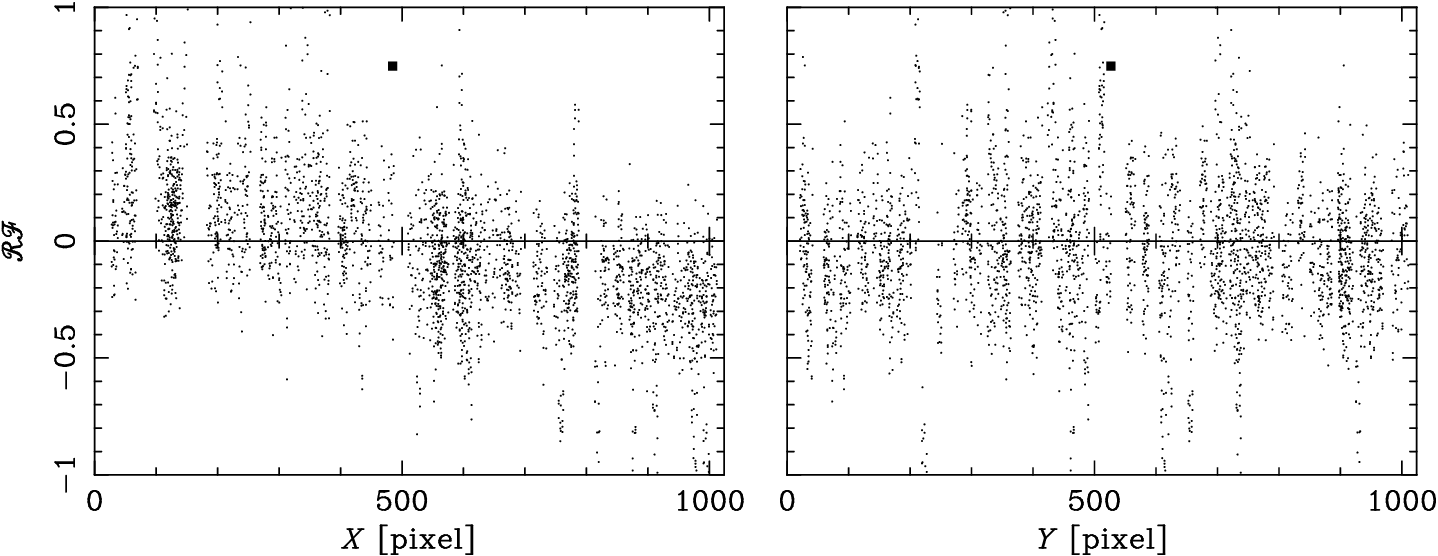}

\vspace*{\baselineskip}
\includegraphics[angle=-90,scale=0.55]{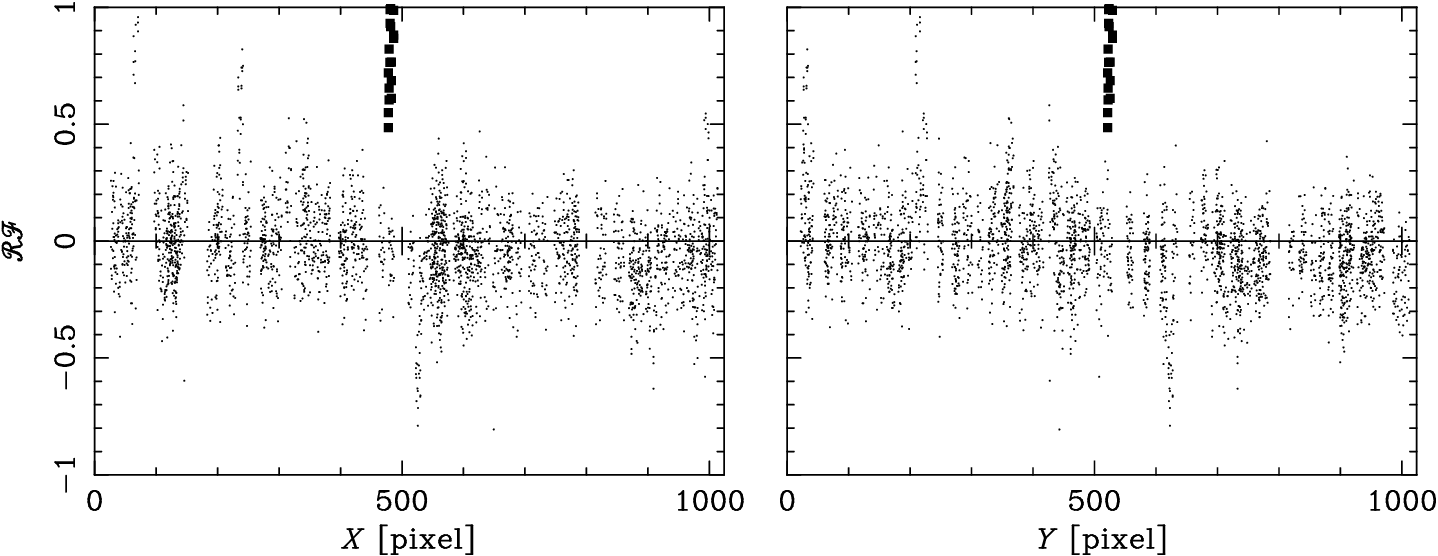}
\caption{Left column: Flux residual ${\cal RF}$ as a function of $X$.
Right column: Flux residual ${\cal RF}$ as a function of $Y$.  The
plots are for the FOR~J$0240-3438$ field, which was imaged with STIS.
From top to bottom, the rows of plots are for 24 images at each of
epoch 2000, 2001, and 2002, respectively.  The figure shows points only
for those objects that have $S/N > 15$.  The solid squares mark points
corresponding to the QSO.}
\label{rf-stis1}
\end{figure}

Figure \ref{rf-stis1} plots ${\cal RF}$
\textit{versus} $X$ (left column) and  ${\cal RF}$
\textit{versus} $Y$ (right column) for the FOR~J$0240-3438$ field imaged with STIS.
From top to bottom, the rows of plots are for the epochs 2000, 2001,
and 2002.  The values of ${\cal RF}$ for the QSO are positive for all
epochs and most of these points are outside of the top boundary in the
upper two rows of plots.  The PSF of the QSO is thus wider than
the ePSF, most likely because, with a redshift of 0.38, this QSO is
near enough for its core to be resolved.

	The values of ${\cal RF}$ for stars in the top and bottom rows
of plots are well behaved.  They show no clear trends with either $X$
or $Y$ and scatter around ${\cal RF}=0$.  A few bright
stars have ${\cal RF}$s comparable to those for the QSO.  However, the
plots in the middle row are not well behaved.  The plot of ${\cal RF}$
\textit{versus} $X$ shows a clear trend: ${\cal RF}$s tend to be
positive for $X\lesssim 500$~pixel and negative otherwise.  A line with
negative slope crossing ${\cal RF}=0$ around $X\approx500$~pixel could
fit the trend.  Although the plot of ${\cal RF}$ \textit{versus} $Y$
does not show trends, the scatter around ${\cal RF}=0$ is larger than
that in the corresponding plots at the other two epochs.  The likely
cause for this behavior is a dependence of the PSF with $X$.  Remembering
that the ePSF is a global average of individual PSFs, the PSF is wider
than average for $X\lesssim 500$~pixel and narrower than average for
greater values of $X$.  A variation of the PSF with $X$ in STIS seems
isolated to this field and epoch.  We are unable to trace the source of
this dependence.  Position residual diagnostics, ${\cal RX}$ and ${\cal
RY}$, discussed in Section~\ref{sec:rx-ry}, will show if the variable
PSF had a significant impact on the measured centroids.

\subsubsection{Flux Residual Diagnostic for the FOR~J$0238-3443$ Field}
\label{sec:rf-stis2}

\begin{figure}
\centering
\includegraphics[angle=-90,scale=0.55]{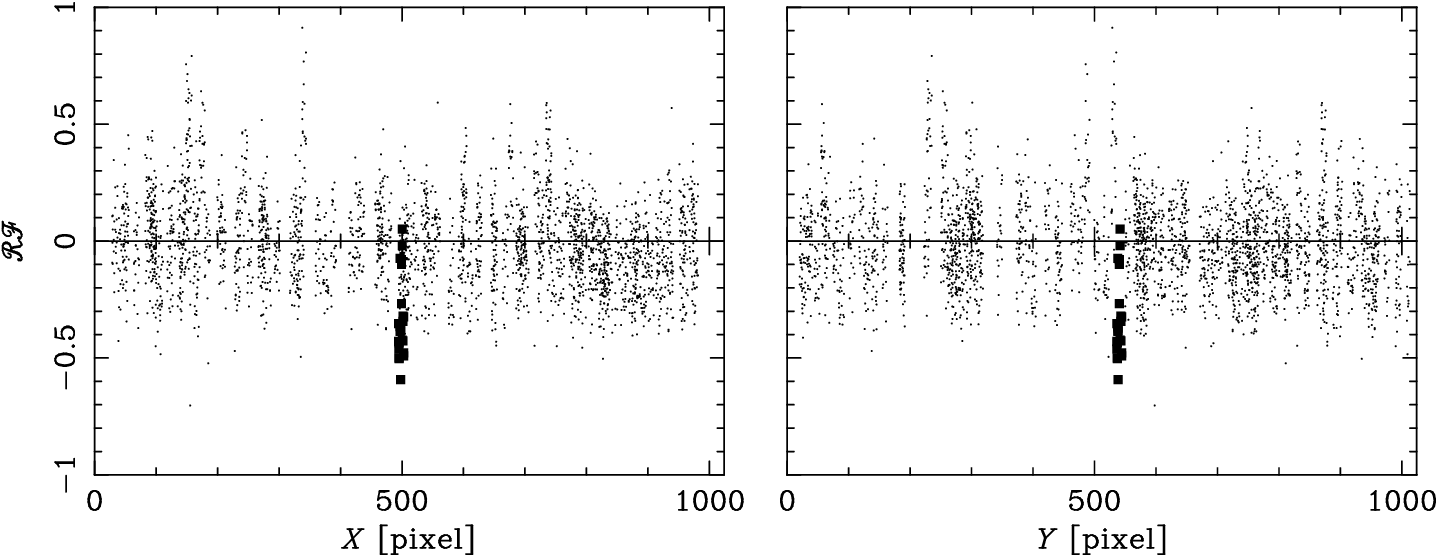}

\vspace*{\baselineskip}
\includegraphics[angle=-90,scale=0.55]{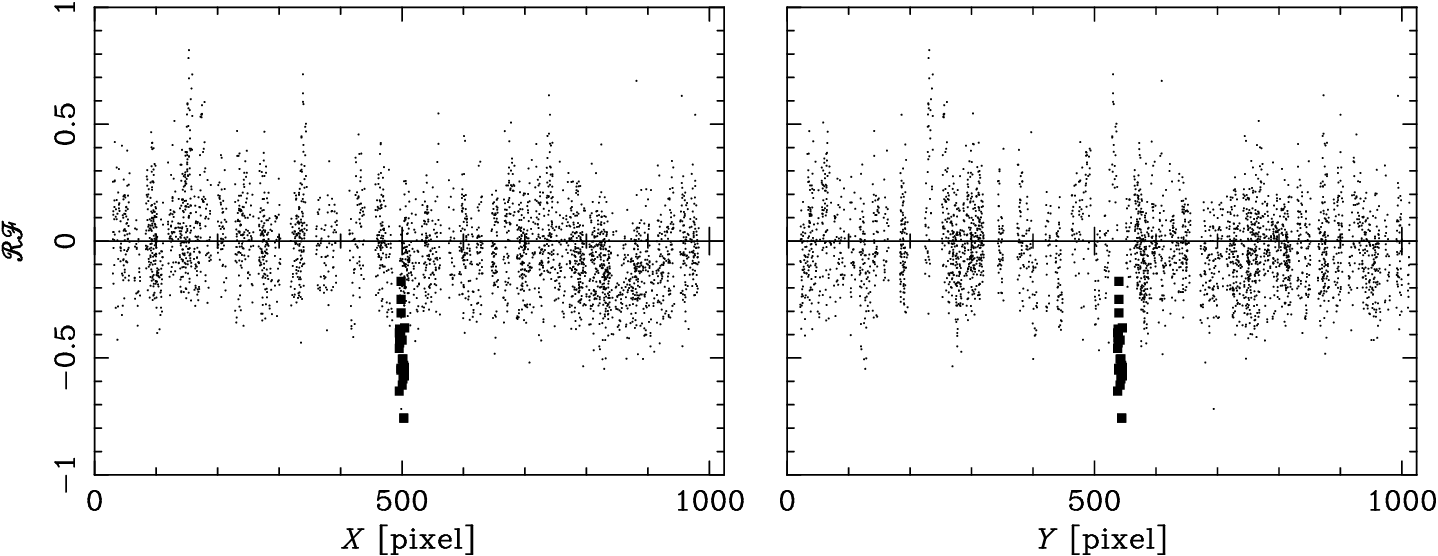}

\vspace*{\baselineskip}
\includegraphics[angle=-90,scale=0.55]{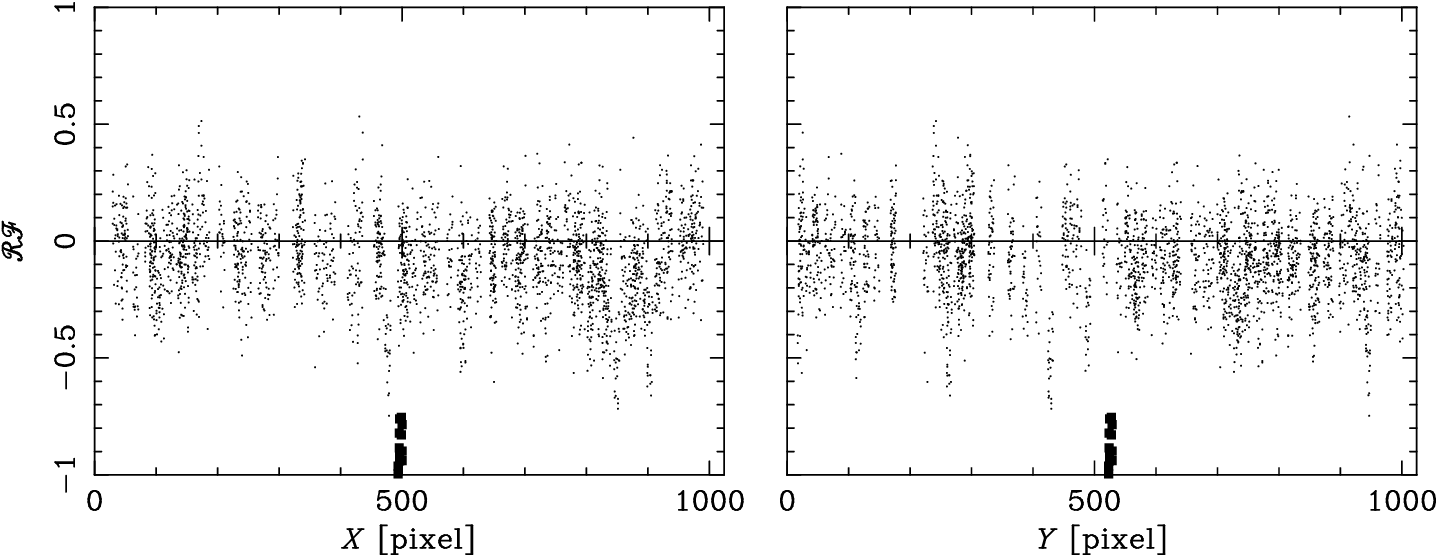}
\caption{Left column: Flux residual ${\cal RF}$ as a function of $X$.
Right column: Flux residual ${\cal RF}$ as a function of $Y$.  The
plots are for the FOR~J$0238-3443$ field, which was imaged with STIS.
From top to bottom, the rows of plots are for 24 images at each of the
epochs 2000, 2001, and 2003, respectively.  The figure shows points only
for those objects that have S/N~$> 15$.  The solid squares mark points
corresponding to the QSO.}
\label{rf-stis2}
\end{figure}

	Figure~\ref{rf-stis2} for the FOR~J$0238-3443$ field is
analogous to Figure~\ref{rf-stis1}.  From top to bottom, the rows of
plots are for the epochs 2000, 2001, and 2003.

	None of the plots shows a trend between ${\cal RF}$ and either
$X$ or $Y$.  The values of ${\cal RF}$ for the QSO tend to be negative
--- most distinctly in the bottom row of plots --- indicating that the
PSF of the QSO is narrower than the ePSF.

\subsection{Position Residuals}
\label{sec:rx-ry}

	The position residual diagnostic consists of plots of the
position residuals ${\cal RX}\equiv \langle X_{0}\rangle-X_{0}$ and
${\cal RY}\equiv \langle Y_{0}\rangle-Y_{0}$ \textit{versus} the pixel
phase $\Phi_{x}\equiv X_{0}-\mbox{Int}(X_{0})$ and $\Phi_{y}\equiv
Y_{0}-\mbox{Int}(Y_{0})$.  Here $\langle X_{0}\rangle$ and $\langle
Y_{0}\rangle$ are the components of the mean centroid in the fiducial
coordinate system --- the system of the first image in time at a given
epoch.  The function Int$(x)$ returns the integer part of a variable
$x$.  The coordinate transformation to a fiducial system involves
translation, rotation, and scale change.  The coefficients of the
transformation result from a least-squares fit using objects common to
the fiducial image and the one transformed to it.

	In the presence of only random errors, position residuals
would scatter around ${\cal RX}=0$ or ${\cal RY}=0$ with constant
$rms$ for all values of $\Phi_{x}$ or $\Phi_{y}$.  Trends and
structures in the plots are symptomatic of systematic errors.
	
\subsubsection{Position Residual Diagnostic for the FOR~J$0240-3434$ Field}
\label{sec:rxry-pc}

\begin{figure}
\centering
Fig.~9a\hspace{0.1truein}
\includegraphics[angle=-180,scale=0.4]{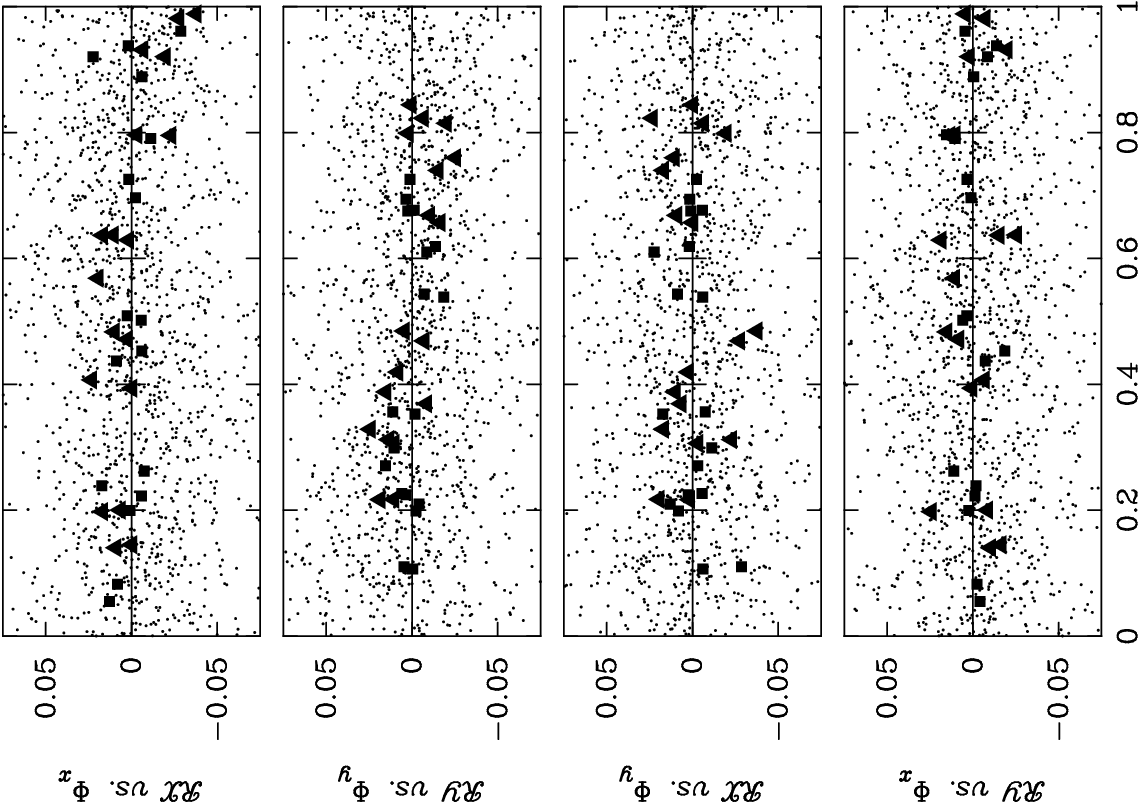}

\vspace*{\baselineskip}
Fig.~9b\hspace{0.1truein}
\includegraphics[angle=-180,scale=0.4]{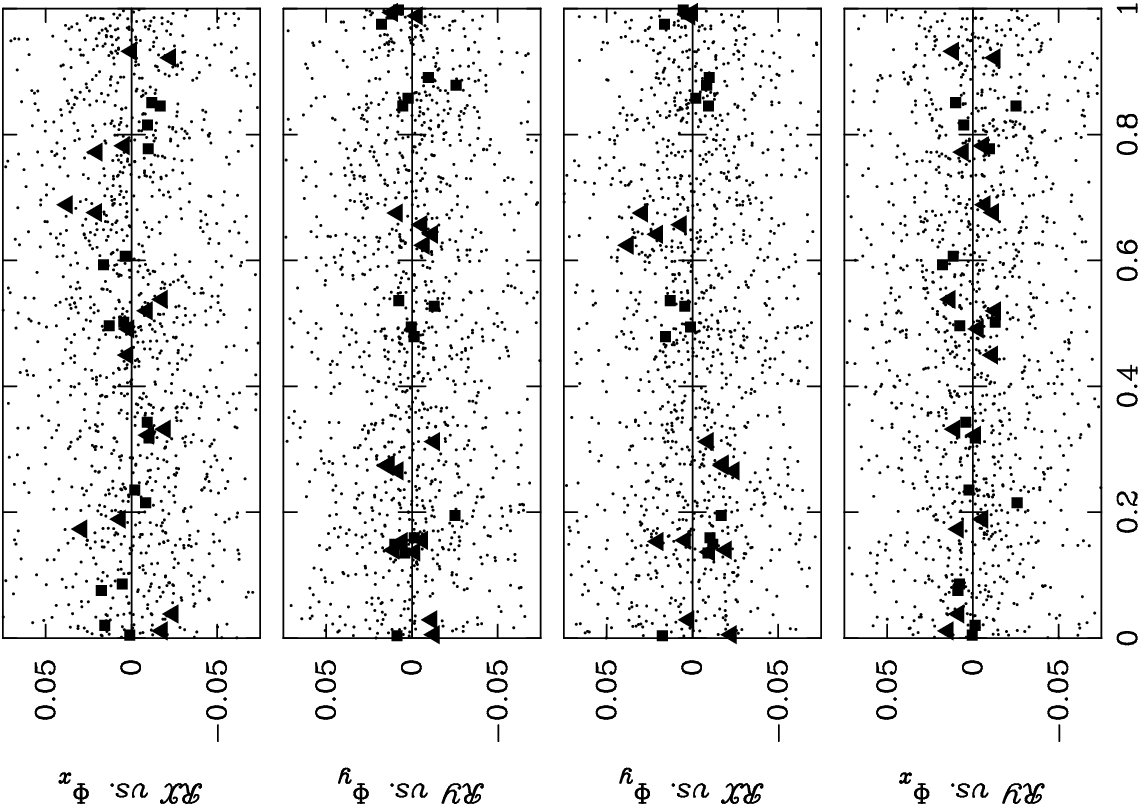}

\vspace*{\baselineskip}
Fig.~9c\hspace{0.1truein}
\includegraphics[angle=-180,scale=0.4]{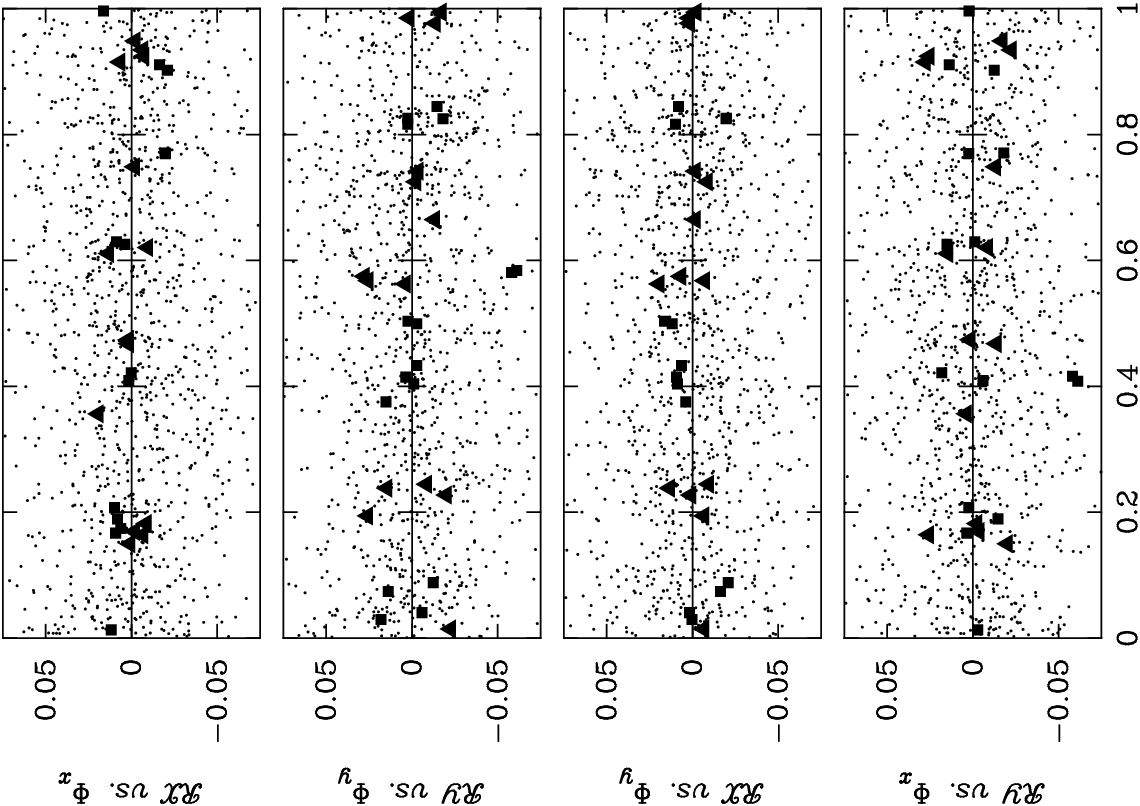}
\caption{Plots for the FOR~J$0240-3434$ field of the position
residuals, ${\cal RX}$ and ${\cal RY}$, as a function of the pixel
phase, $\Phi_{x}$ and $\Phi_{y}$.  Panels $a$, $b$, and $c$ are
for the epochs 1999, 2001, and 2002, respectively.  Solid squares and
triangles correspond to images A and B of the QSO, respectively.  The
plots show points only for those objects with S/N~$> 15$.}
\label{rxry-pc}
\end{figure}

	Figure~\ref{rxry-pc} shows plots of the position residuals
\textit{versus} pixel phase for the FOR~J$0240-3434$ field.  Panels
$a$, $b$, and $c$ are for epochs
1999, 2001, and 2002, respectively.  Only two plots, ${\cal RX}$
\textit{versus} $\Phi_{x}$ and ${\cal RY}$ \textit{versus} $\Phi_{y}$,
in Figure~\ref{rxry-pc}$a$ show evidence for trends among the points
for images A and B of the QSO.  In the first of these two plots, the
values of ${\cal RX}$ are mostly positive up to $\Phi_{x}\approx 0.8$
and then negative for larger values of $\Phi_{x}$.  In the second, the
${\cal RY}$s show a sinusoidal dependence on $\Phi_{y}$.

	No plot for any epoch shows trends or structures for points
corresponding to stars, though we caution that the small number of
points and their large scatter due to the modest S/N of the majority of
stars would make such trends and structures hard to see.

	Even though Figure~\ref{rf-pc} implies that the PSF varies with
position in the image at all three epochs for this field,
Figure~\ref{rxry-pc} shows that the QSO has pixel phase errors only at
the first epoch.  Thus, the origin of the pixel phase errors for the
QSO are not clearly caused by the variation of the PSF with position in
the field.  Neither are they clearly caused by the variation of the PSF
between images at one epoch because, as Figure~\ref{rf-pc} shows, that
variation is similar at all three epochs.

\subsubsection{Position Residual Diagnostic for the FOR~J$0240-3438$ Field}
\label{sec:rxry-stis1}

\begin{figure}
\centering
Fig.~10a\hspace{0.1truein}
\includegraphics[angle=-180,scale=0.4]{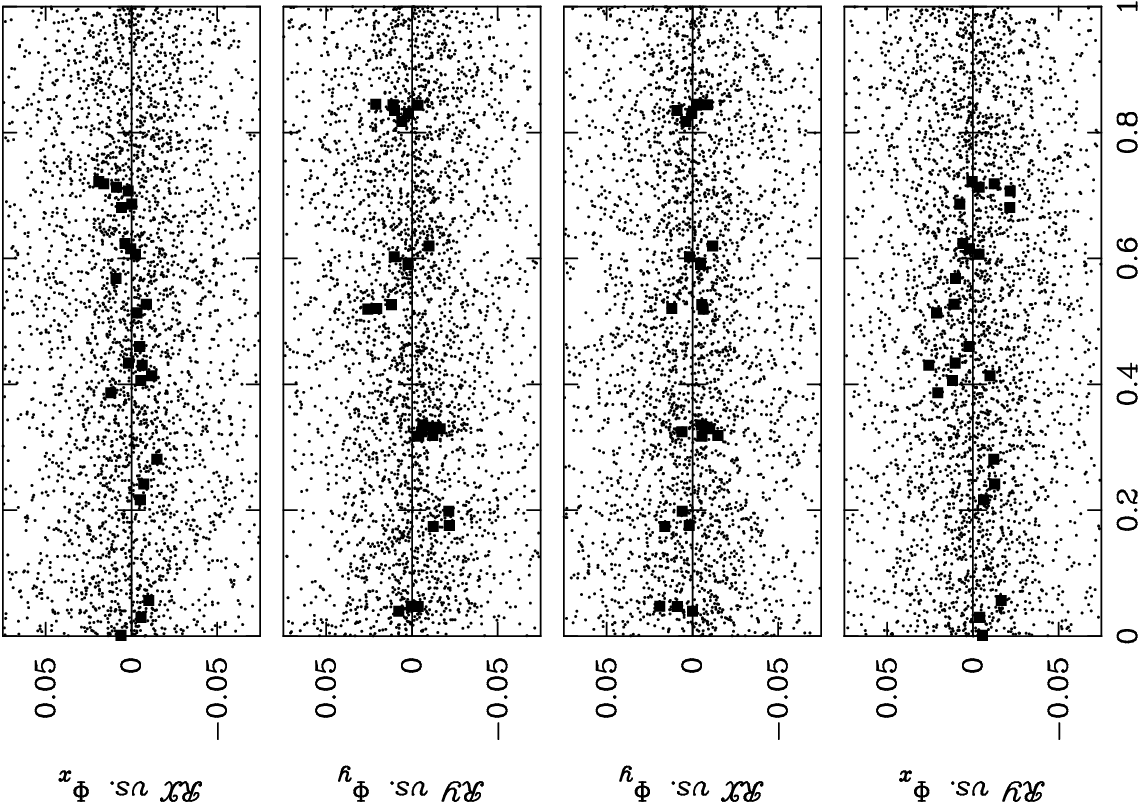}

\vspace*{\baselineskip}
Fig.~10b\hspace{0.1truein}
\includegraphics[angle=-180,scale=0.4]{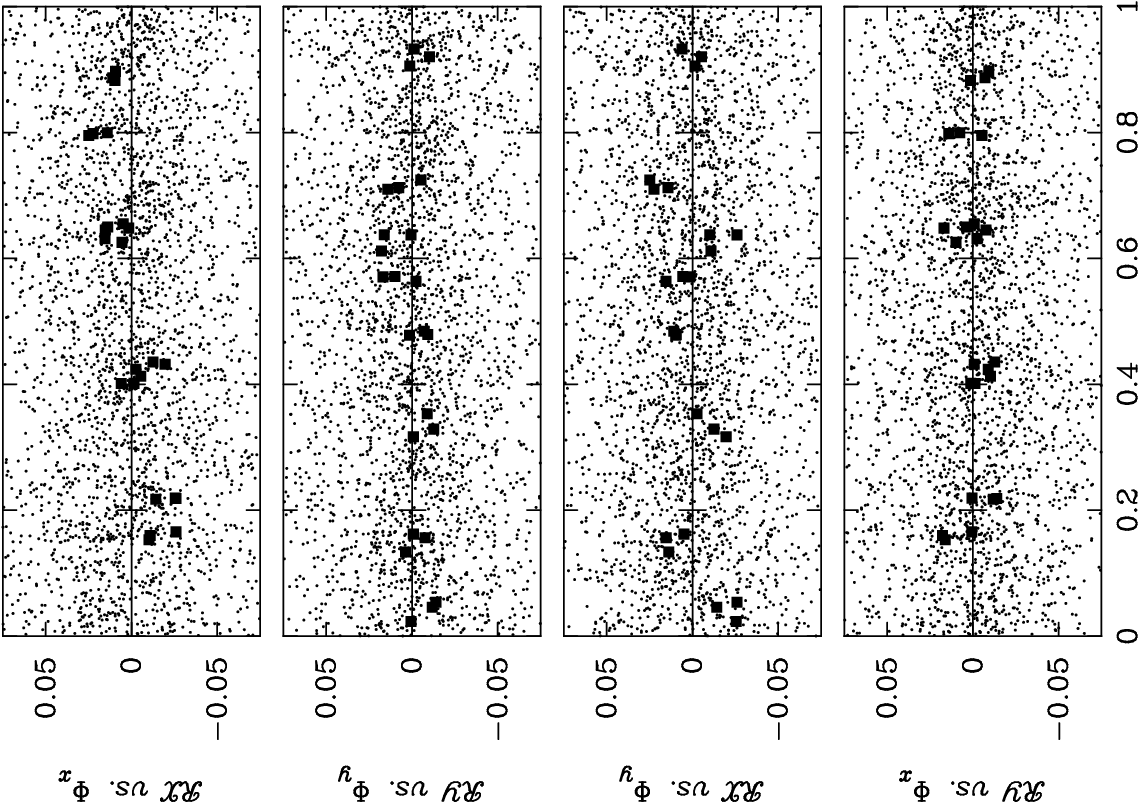}

\vspace*{\baselineskip}
Fig.~10c\hspace{0.1truein}
\includegraphics[angle=-180,scale=0.4]{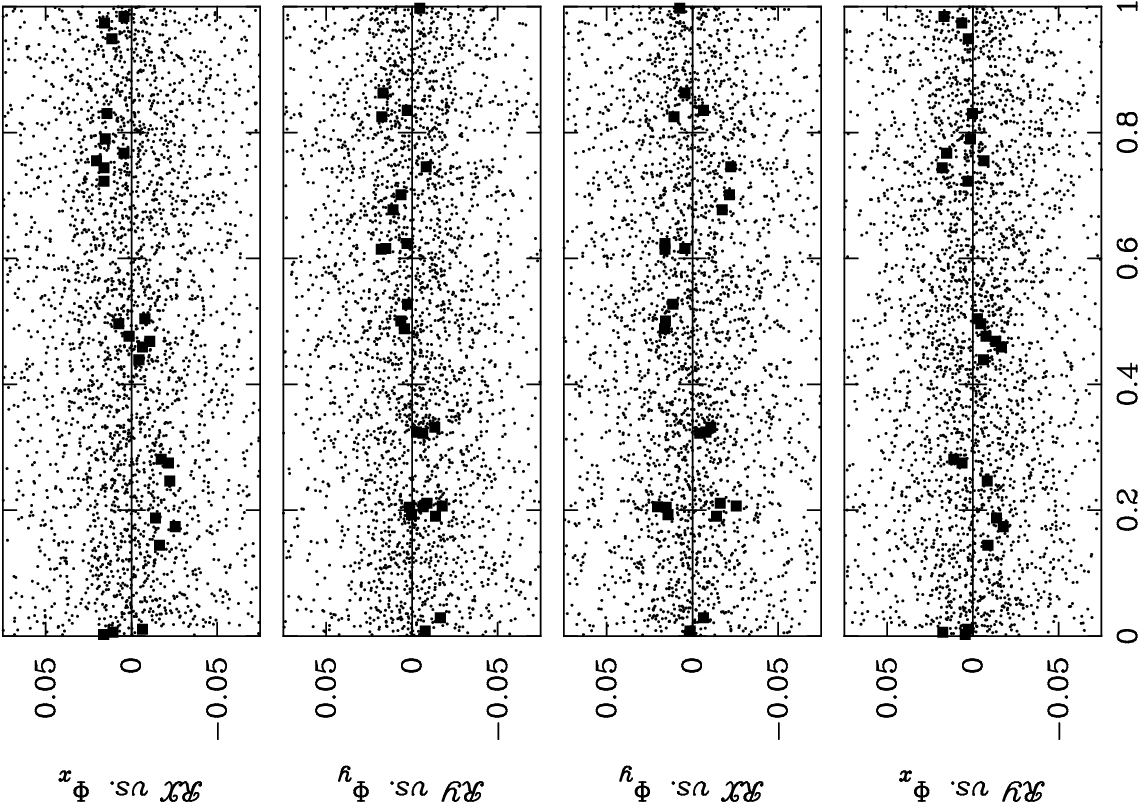}
\caption{Plots for the FOR~J$0240-3438$ field of the position
residuals, ${\cal RX}$ and ${\cal RY}$, as a function of the pixel
phase, $\Phi_{x}$ and $\Phi_{y}$.  The panels $a$, $b$, and $c$ are for
the epochs 2000, 2001, and 2002, respectively.  The solid squares mark
points for the QSO.  The plots show points only for those objects with
S/N~$> 15$.}
\label{rxry-stis1}
\end{figure}

	Figure \ref{rxry-stis1} plots ${\cal RX}$ and ${\cal RY}$
\textit{versus} $\Phi_{x}$ and $\Phi_{y}$ for the FOR~J$0240-3438$
field.  Panels $a$, $b$, and $c$ are for epochs 2000, 2001, and 2002,
respectively.

	None of the plots in Figure~\ref{rxry-stis1}$a$ shows
convincing evidence for systematic errors.  However, the plots of ${\cal RX}$
\textit{versus} $\Phi_{x}$ in Figures~\ref{rxry-stis1}$b$ and
\ref{rxry-stis1}$c$ do.  In the first plot, ${\cal RX}$ has a linear trend
with $\Phi_{x}$ with positive slope.  In the second plot, the trend is
more sinusoidal.  In both plots, ${\cal RX}$s corresponding to
stars show greater scatter than ${\cal RY}$s for the same objects.  To
a lesser degree, the plot of ${\cal RY}$ \textit{versus} $\Phi_{y}$ in
Figure~\ref{rxry-stis1}$c$ shows a sinusoidal trend too.

As is the case for the previous field, there is not a clear correlation
between a variation of the PSF with position and the presence of pixel
phase errors for the QSO.  For the 2001 epoch, both a variation of the
PSF with $X$ and pixel phase errors in $X$ are present.  In contrast,
for the 2002 epoch, pixel phase errors are present even though the PSF
does not vary with position.  The 2002 epoch has better agreement than
the 2000 and 2001 epochs between the PSF of the QSO and that of the
stars and it also shows less variation of the PSF
between the images of an epoch.

\subsubsection{Position Residual Diagnostic for the FOR~J$0238-3443$ Field}
\label{sec:rxry-stis2}

\begin{figure}
\centering
Fig.~11a\hspace{0.1truein}
\includegraphics[angle=-180,scale=0.4]{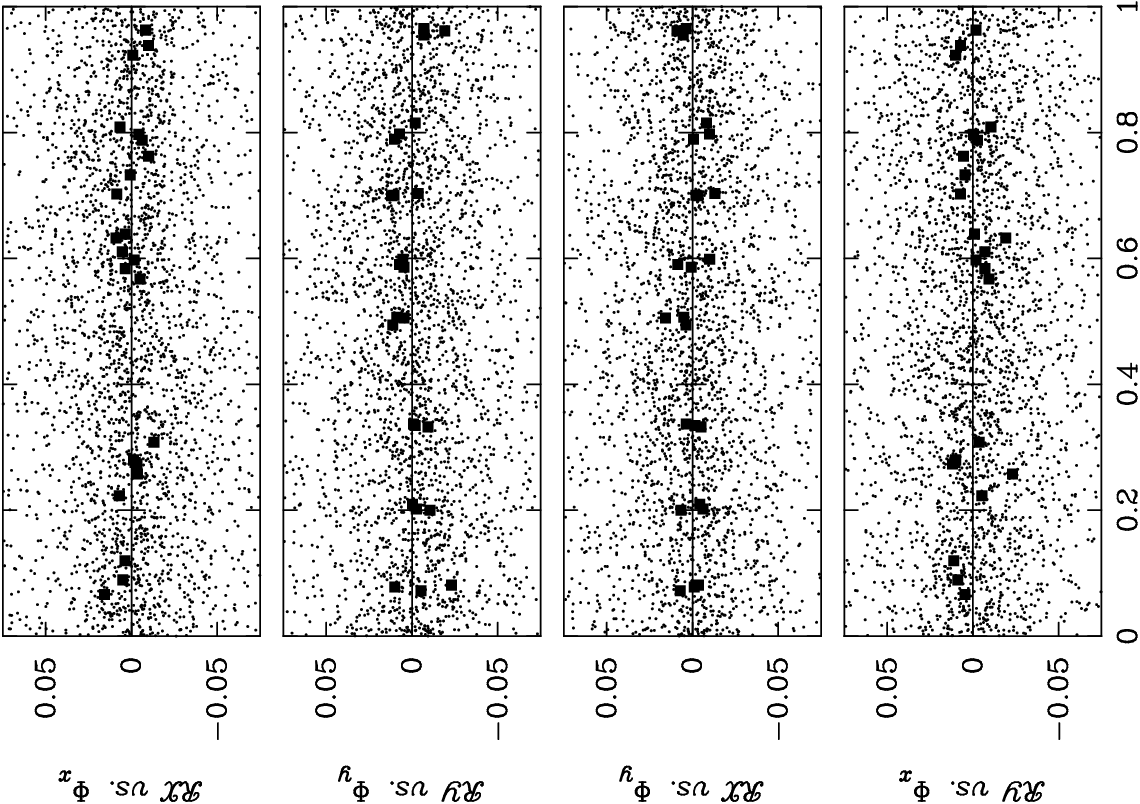}

\vspace*{\baselineskip}
Fig.~11a\hspace{0.1truein}
\includegraphics[angle=-180,scale=0.4]{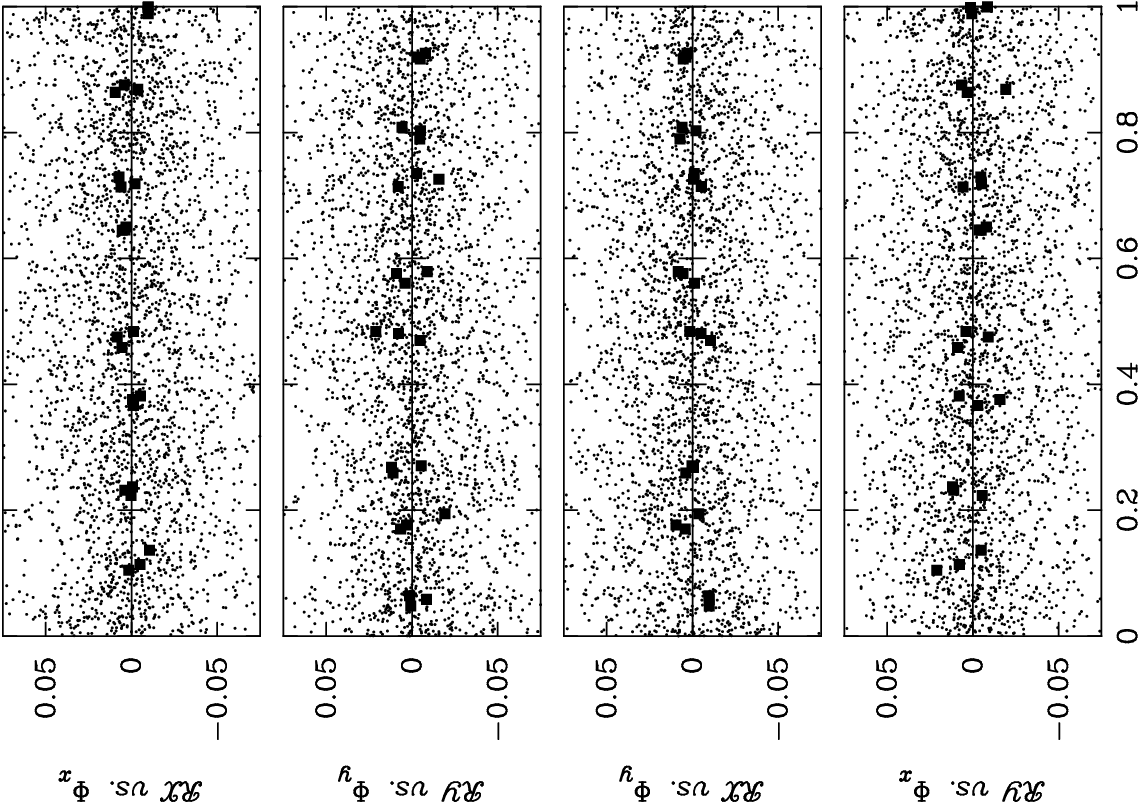}

\vspace*{\baselineskip}
Fig.~11a\hspace{0.1truein}
\includegraphics[angle=-180,scale=0.4]{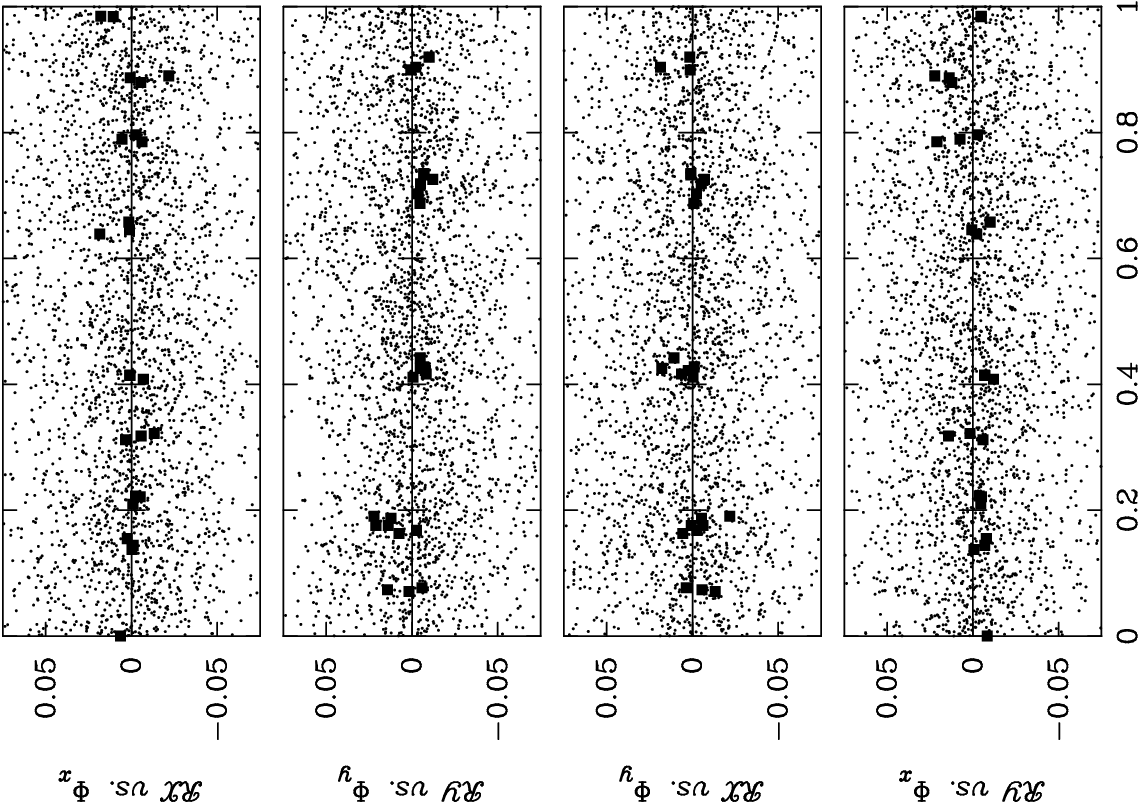}
\caption{Plots for the FOR~J$0238-3443$ field of the position
residuals, ${\cal RX}$ and ${\cal RY}$, as a function of the pixel
phase, $\Phi_{x}$ and $\Phi_{y}$.  The panels $a$, $b$, and $c$ are for
the epochs 2000, 2001, and 2003, respectively.  The solid squares mark
points for the QSO.  The plots show points only for those objects with
S/N~$> 15$.}
\label{rxry-stis2}
\end{figure}

	Figure \ref{rxry-stis2} plots ${\cal RX}$ and ${\cal RY}$
\textit{versus} $\Phi_{x}$ and $\Phi_{y}$ for the FOR~J$0240-3438$
field.  Panels $a$, $b$, and $c$ are for epochs 2000, 2001, and 2003,
respectively.

	No plot in the three panels shows clear evidence for
systematic errors.  The absence of systematic trends in these plots
is consistent with the absence of trends in Figure~\ref{rf-stis2}.

\section{The Proper Motion of Fornax}
\label{sec:pm}

	At this point in the analysis, there are three lists of
coordinates --- one per epoch --- for each field.  Equations (1) - (4)
in \citet{p06} give the form of the transformation between epochs.
This most-general linear transformation has six free parameters $c_{1}
- c_{6}$.  The transformation also includes an additional term that is
linear with $Y$ (see Equations (7) and (8) in \citet{p05}) if the
fitting procedure corrects for the effects caused by decreasing charge
transfer efficiency.  The slope of the linear correction, $b$, is an
adjustable parameter and this correction is applied only to objects
whose S/N is below a specified limit.

The systematic errors in the object coordinates reflected in the trends
in ${\cal RF}$, ${\cal RX}$, and ${\cal RY}$ with $X$, $Y$, $\Phi_{x}$,
and $\Phi_{y}$ will contribute to the error in the centroid of an
object transformed to the standard coordinate system.  Because we are
unable to eliminate the source of these trends, our procedure
propagates these errors into the uncertainty in the proper motion by
increasing the uncertainties in the coordinates of all objects until
the $\chi^2$ of the scatter around the transformation is equal to one
per degree of freedom (see the discussions in \citet{p05} and
\citet{p06}).

\subsection{Motion of the QSOs in the FOR~J$0240-3434$ field}
\label{sec:pmpc}

	The PC was the imaging detector for this field; therefore,
correcting for the effects caused by decreasing charge transfer
efficiency requires fitting for the parameter $b$.  The coordinates of
objects with a S/N~$< 30$ are corrected using Equation~(8) of
\citet{p05}.  For objects with a larger S/N, the size of the correction
decreases linearly with increasing S/N until a value of 100; beyond
this value there is no correction.  This field contains a lensed QSO
with two images, A and B, and so it provides two measurements of the
proper motion.

	Of the 233, 209, and 231 objects measured at the three epochs,
respectively, 191 are common to all epochs.  The greatest difference in
the pointing of \textit{HST} is between the 1999 and 2001 epochs, where
the difference is about 20.6~pixel in the $X$ direction and about
5.6~pixel in the $Y$.  The choice for the individual $\chi^2$ that
triggers fitting for uniform linear motion is 15.  The multiplicative
constant that produces a $\chi^2$ of one per degree of freedom is
1.304.  The fitted value of $b$ is $-6.5\times10^{-5}$.

\begin{figure}
\centering
\includegraphics[angle=-90,scale=0.7]{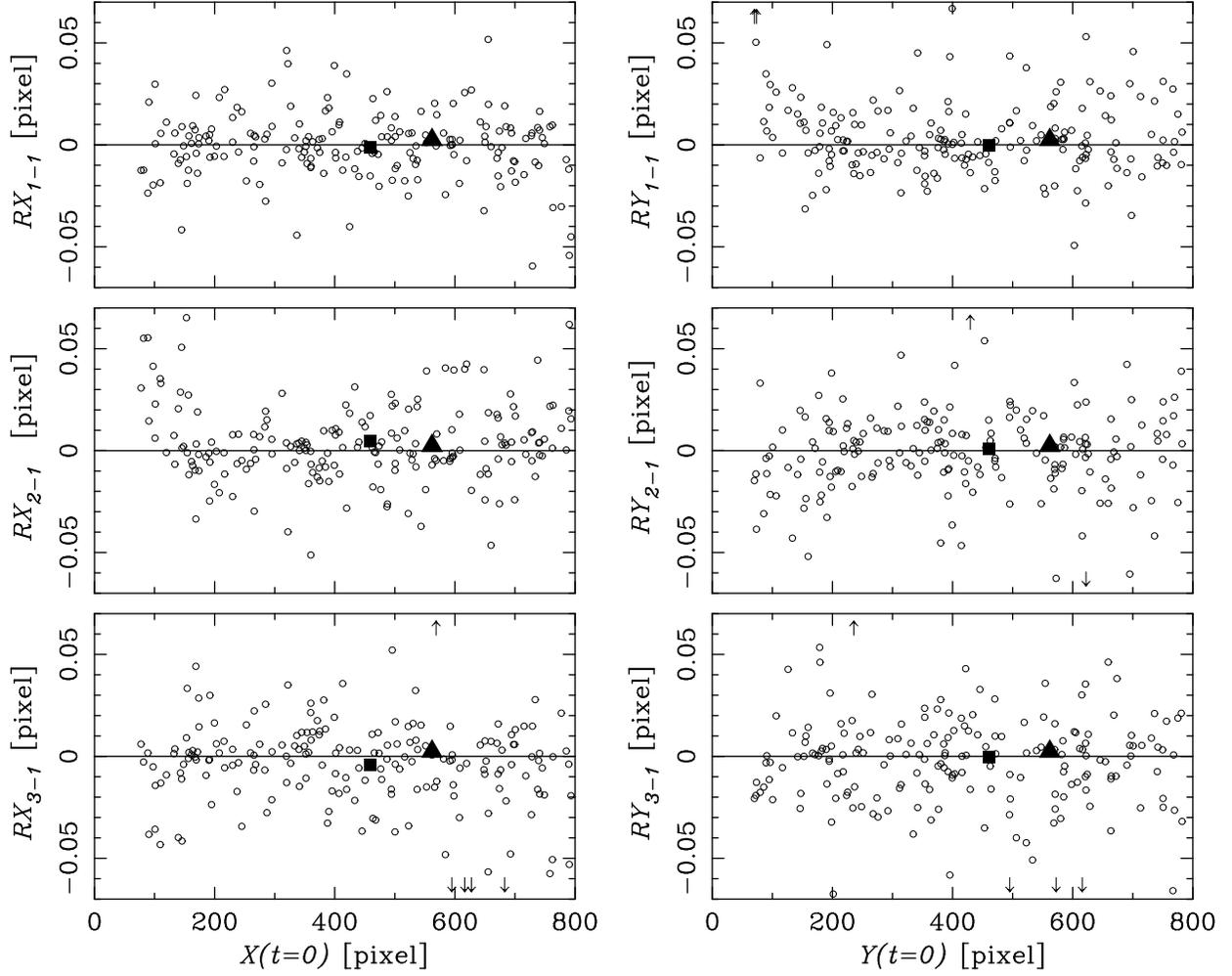}
\caption{Position residuals defined by Equations (10) and (11) in
\citet{p05} for the objects in the FOR~J$0240-3434$ field.  The
epochs increase chronologically from top to bottom.
\textit{Left}: $RX$ vs. $X$.  \textit{Right}: $RY$ vs. $Y$.  The
squares correspond to image A of the QSO and the triangles to image B.
The arrows indicate points beyond the boundaries of the plot.}
\label{fig:pc-rxry}
\end{figure}

\begin{figure}
\centering
\includegraphics[angle=-90,scale=0.7]{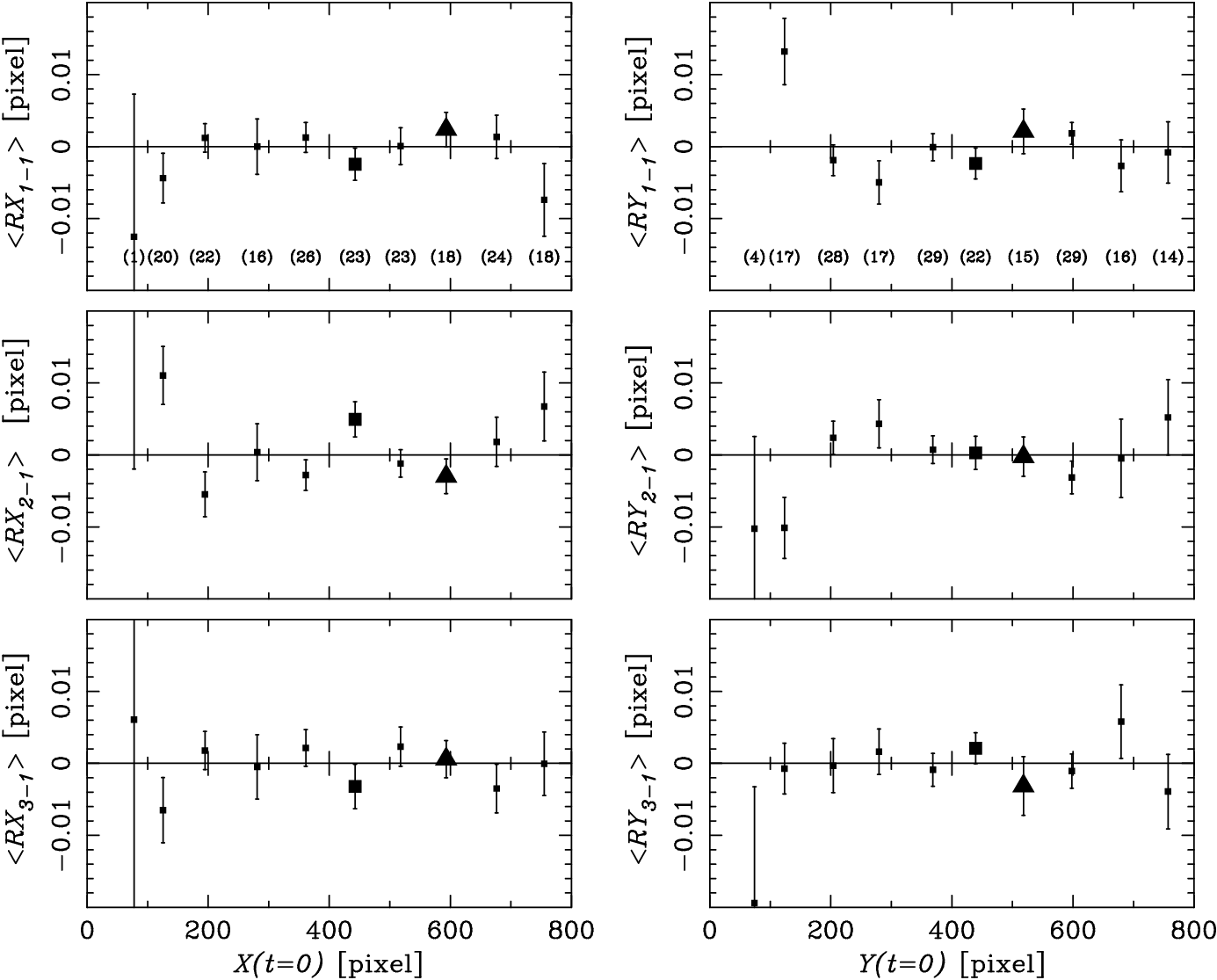}
\caption{Mean position residuals as a function of position for the
FOR~J$0240-3434$ field.  \textit{Left}: $\langle RX \rangle$ vs. $X$.
\textit{Right}: $\langle RY \rangle$ vs. $Y$.  The points are the weighted mean
residuals in 10 equal-length bins in $X$ or $Y$.  The points are
plotted at the mean of the coordinate values in the bin.  The numbers
in parenthesis indicate how many values contributed to the mean.  The
larger squares and triangles mark those points that include
contributions to the mean from images A and B of the QSO,
respectively.}
\label{fig:pc-rxry-a}
\end{figure}

\begin{figure}[p]
\centering
\includegraphics[angle=-90,scale=0.7]{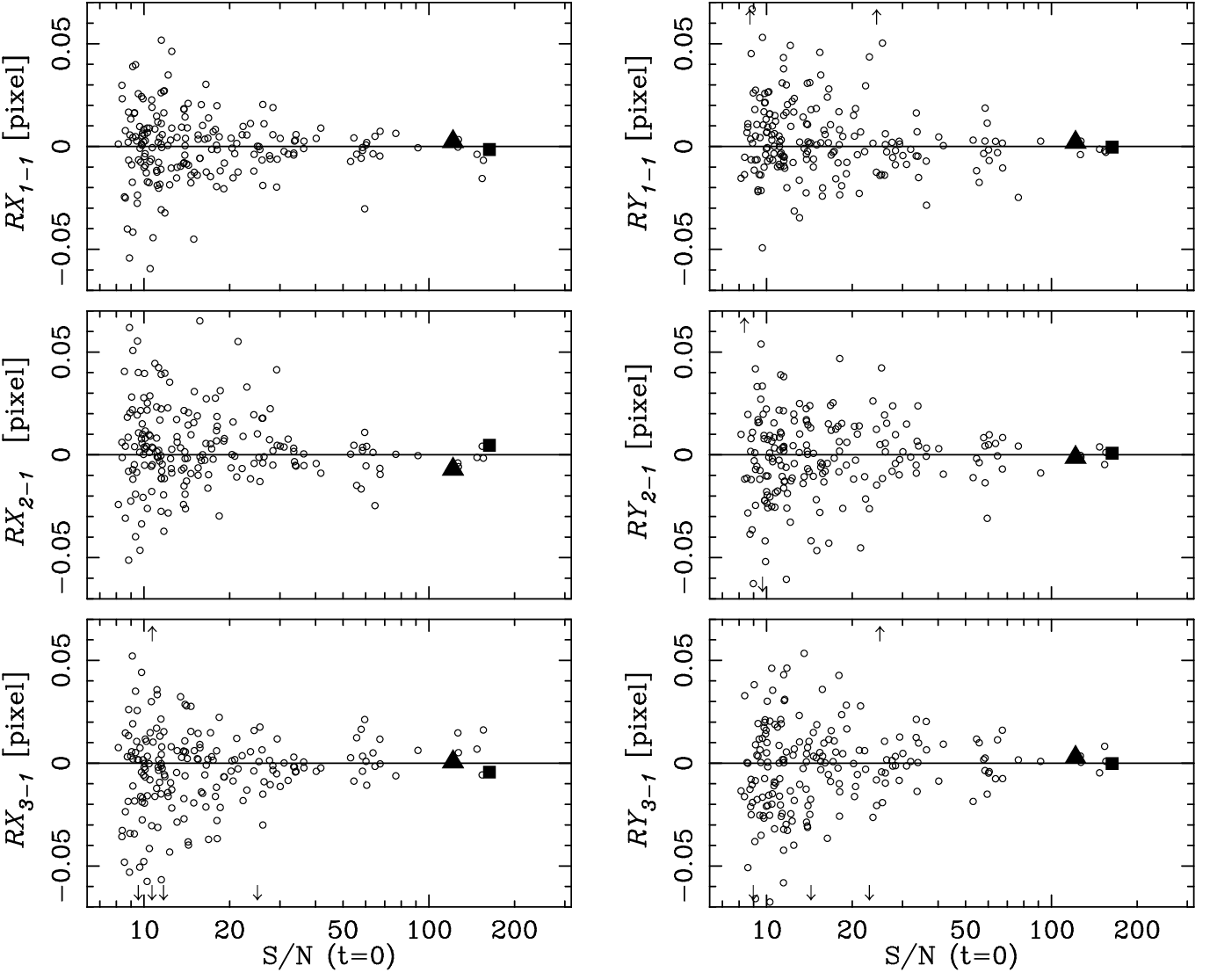}
\caption{Position residuals as a function of S/N for the objects in
the FOR~J$0240-3434$ field.  The epochs increase chronologically from
top to bottom.  \textit{Left}: $RX$ vs. S/N.  \textit{Right}: $RY$
vs. S/N.  The squares correspond to image A of the QSO and the
triangles to image B.  The arrows indicate points beyond the
boundaries of the plot.}
\label{fig:pc-rxry-sn}
\end{figure}

	Figure~\ref{fig:pc-rxry} plots the position residuals, $RX$ and
$RY$, defined by Equations (10) and (11) in \citet{p05} as a function
of position in the standard coordinate system.  A residual is the
difference between the average centroid in the standard coordinate
system and the centroid measured at an epoch, corrected for any fitted
motion, transformed into the standard coordinate system.  The first
digit of the subscripts on $RX$ and $RY$ in the figure indicates the
epoch of the measurement.  None of the plots shows a trend between $RX$
and $X$ or $RY$ and $Y$.  Although this article does not show them, the
plots of $RX$ versus $Y$ and $RY$ versus $X$ also do not show any
trends.
Figure~\ref{fig:pc-rxry-a} plots weighted mean residuals, $\langle RX
\rangle$ and $\langle RY \rangle$, in 10 equal-length bins in $X$ or
$Y$.  The error bar on a point is the usual uncertainty in the
weighted mean.  The values of $\langle RX \rangle$ and $\langle RY
\rangle$ do not show systematic trends for $X$ and $Y$ greater than
approximately 150~pixel.  However, the two points with $X < 150$~pixel
or $Y < 150$~pixel show a combination of large error bars due to a
small number of points contributing to the mean and a deviation from
$\langle RX\rangle = 0$~pixel or $\langle RY\rangle = 0$~pixel due to
a mismatch between the ePSF and PSF near the edges of the chip.  The
points with the largest $X$ or $Y$ also tend to show a greater than
expected deviation, due to the mismatch.  The position residuals do
not show any systematic trends with respect to S/N, as depicted by
Figure~\ref{fig:pc-rxry-sn}.  The plotted values of the S/N are from
the first-epoch data.  As expected, the scatter in the values of $RX$ and
$RY$ increases with decreasing S/N.

\begin{figure}
\centering
\includegraphics[angle=-90,scale=0.48]{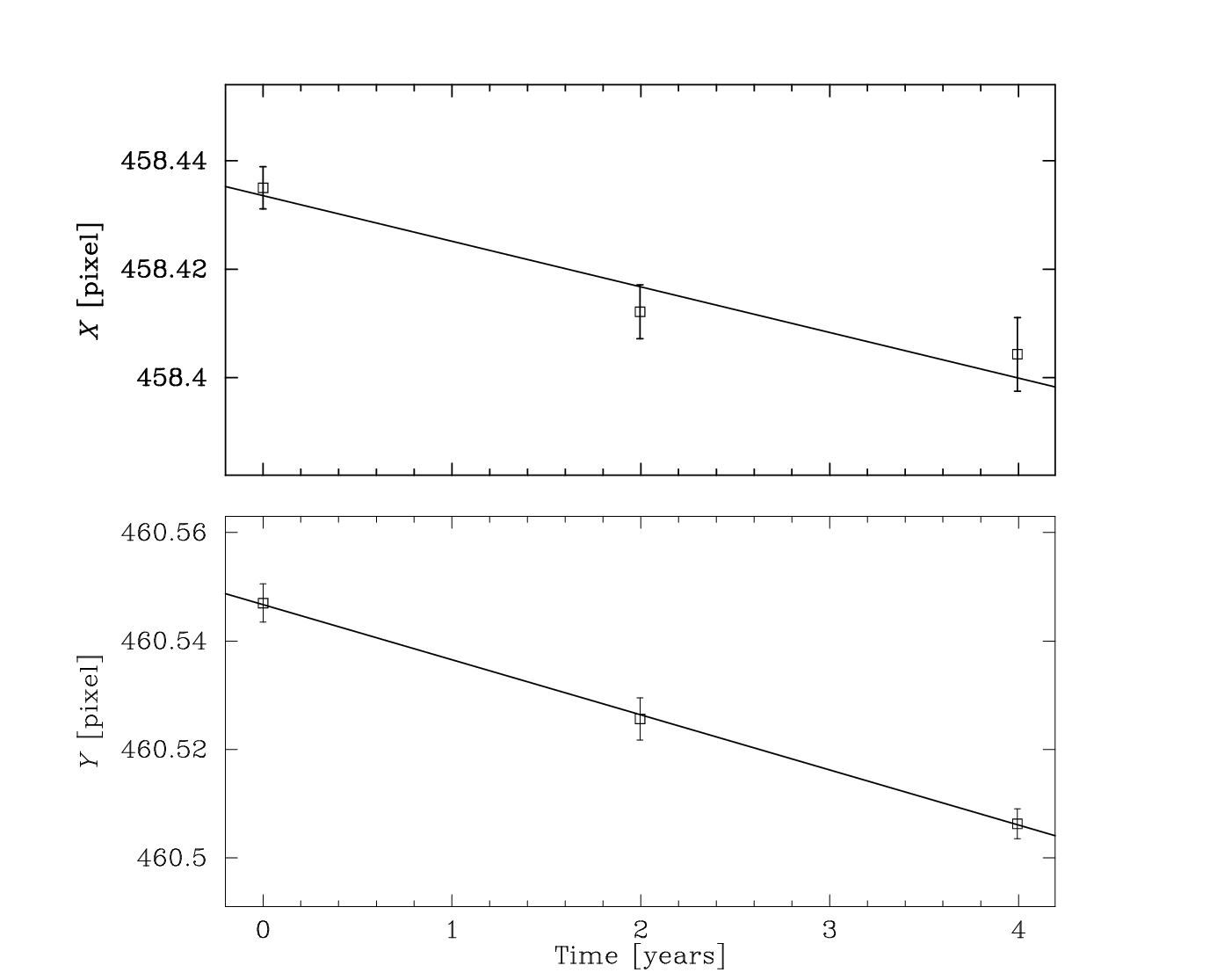}
\caption{Location of image A of the QSO as a function of time for
the FOR~J$0240-3434$ field in the standard coordinate system.}
\label{fig:qso-A}
\end{figure}

\begin{figure}
\centering
\includegraphics[angle=-90,scale=0.48]{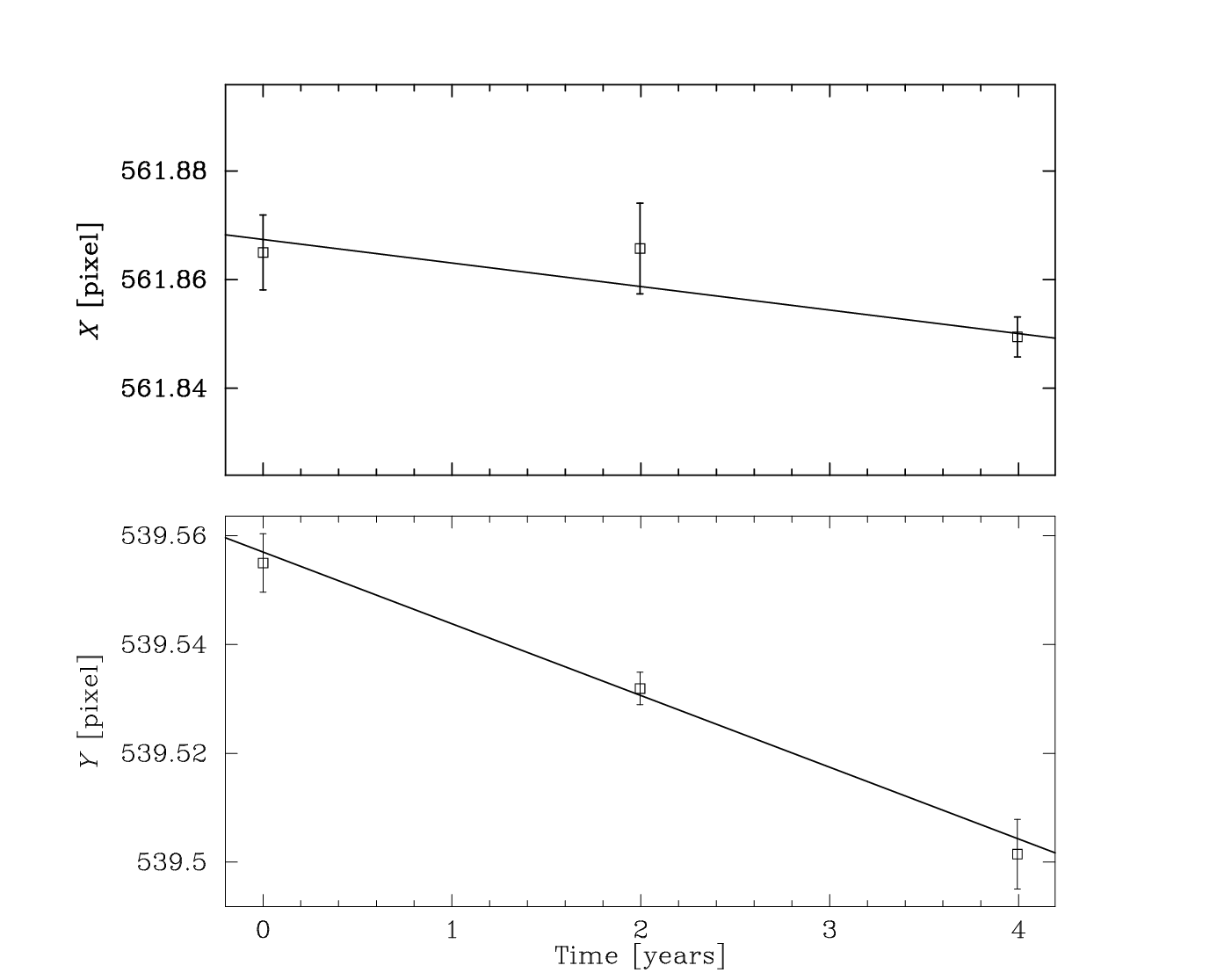}
\caption{Location of the image B of the QSO as a function of time for
the FOR~J$0240-3434$ field in the standard coordinate system.}
\label{fig:qso-B}
\end{figure}

	In the standard coordinate system, in which the stars of Fornax are
at rest, the QSO moves.  Figure~\ref{fig:qso-A} depicts such motion
for image A.  The implied motion from the best-fitting straight lines
is $(\mu_{x},\mu_{y})=(-0.0084\pm 0.0019,-0.0102\pm
0.0014)$~pixel~yr$^{-1}$.  The contribution to the total $\chi^2$ from
the QSO is 1.47 for two degrees of freedom.
Figure~\ref{fig:qso-B} is analogous to Figure~\ref{fig:qso-A} for
image B.  Here, the implied motion is $(\mu_{x},\mu_{y})=(-0.0043\pm
0.0020,-0.0132\pm 0.0024)$~pixel~yr$^{-1}$ and the contribution to the
total $\chi^2$ is 1.37 for two degrees of freedom.  For ease of comparison,
the scales of Figures~\ref{fig:qso-A} and \ref{fig:qso-B} are identical.

\subsection{Motion of the QSO in the FOR~J$0240-3438$ field}
\label{sec:stis1}

	The number of objects with a measured centroid is 442, 407, and
431 in epochs 2000, 2001, and 2002, respectively.  Among these, 310
objects are common to the three epochs.  The greatest difference in the
pointing of \textit{HST} is between the 2000 and 2002 epochs, where the
difference is about 2.5~pixel in the $X$ direction and about 5.3~pixel
in the $Y$.  Because the images used in the analysis were corrected for the
effects caused by the decreasing charge transfer efficiency, the
transformation to the standard coordinate system did not include a term
that would account for such effects.  The choice for the
individual $\chi^2$ that triggers fitting for uniform linear motion is
15.  The multiplicative constant that produces a $\chi^2$ of one per
degree of freedom is 1.100.

\begin{figure}[p]
\centering
\includegraphics[angle=-90,scale=0.7]{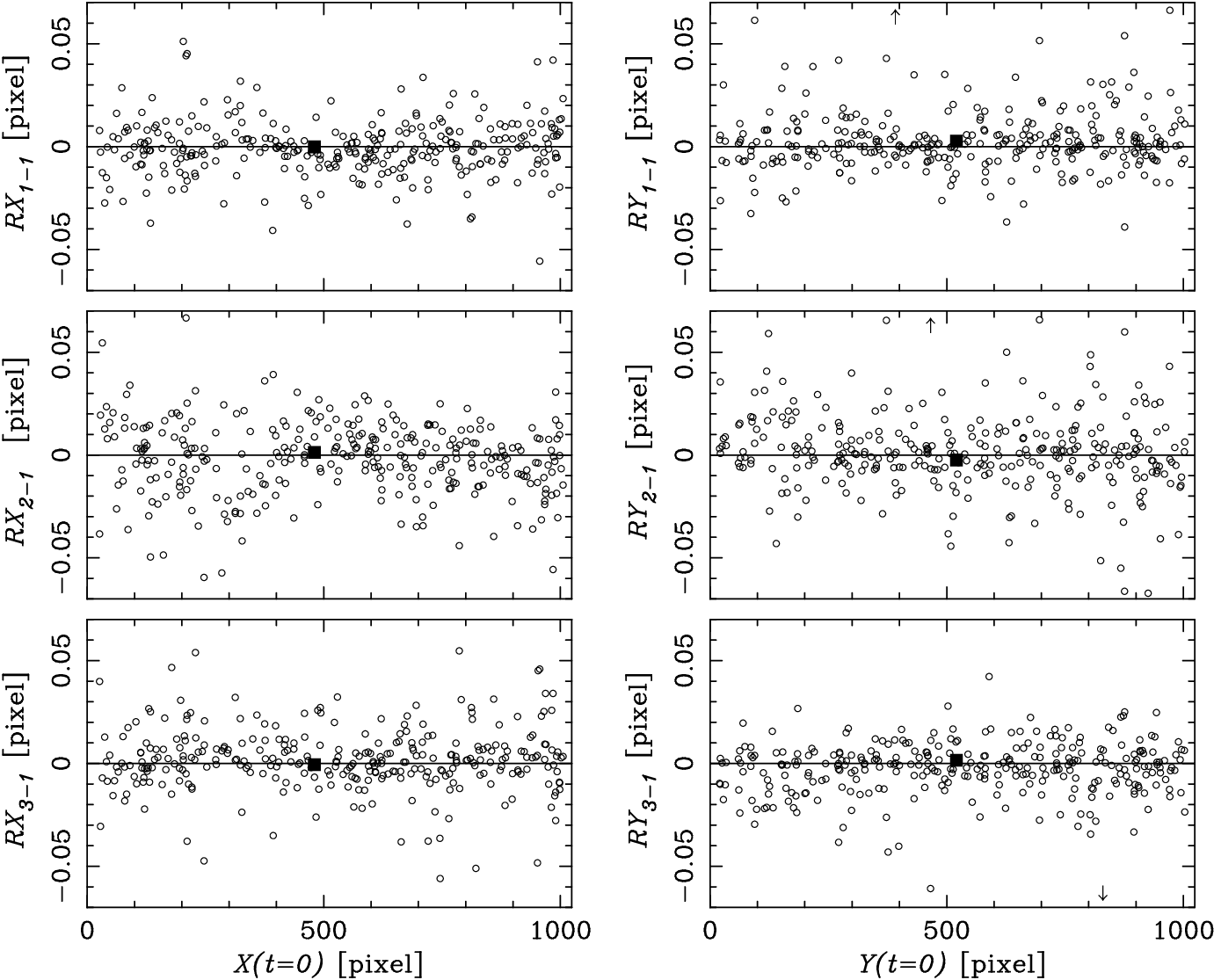}
\caption{Position residuals for the objects in the FOR~J$0240-3438$
field.  The epochs increase chronologically from top to bottom.
\textit{Left}: $RX$ vs. $X$.  \textit{Right}: $RY$ vs. $Y$.  The
squares correspond to the QSO.  The arrows indicate points beyond the
boundaries of the plot.}
\label{fig:stis1-rxry}
\end{figure}

\begin{figure}[p]
\centering
\includegraphics[angle=-90,scale=0.7]{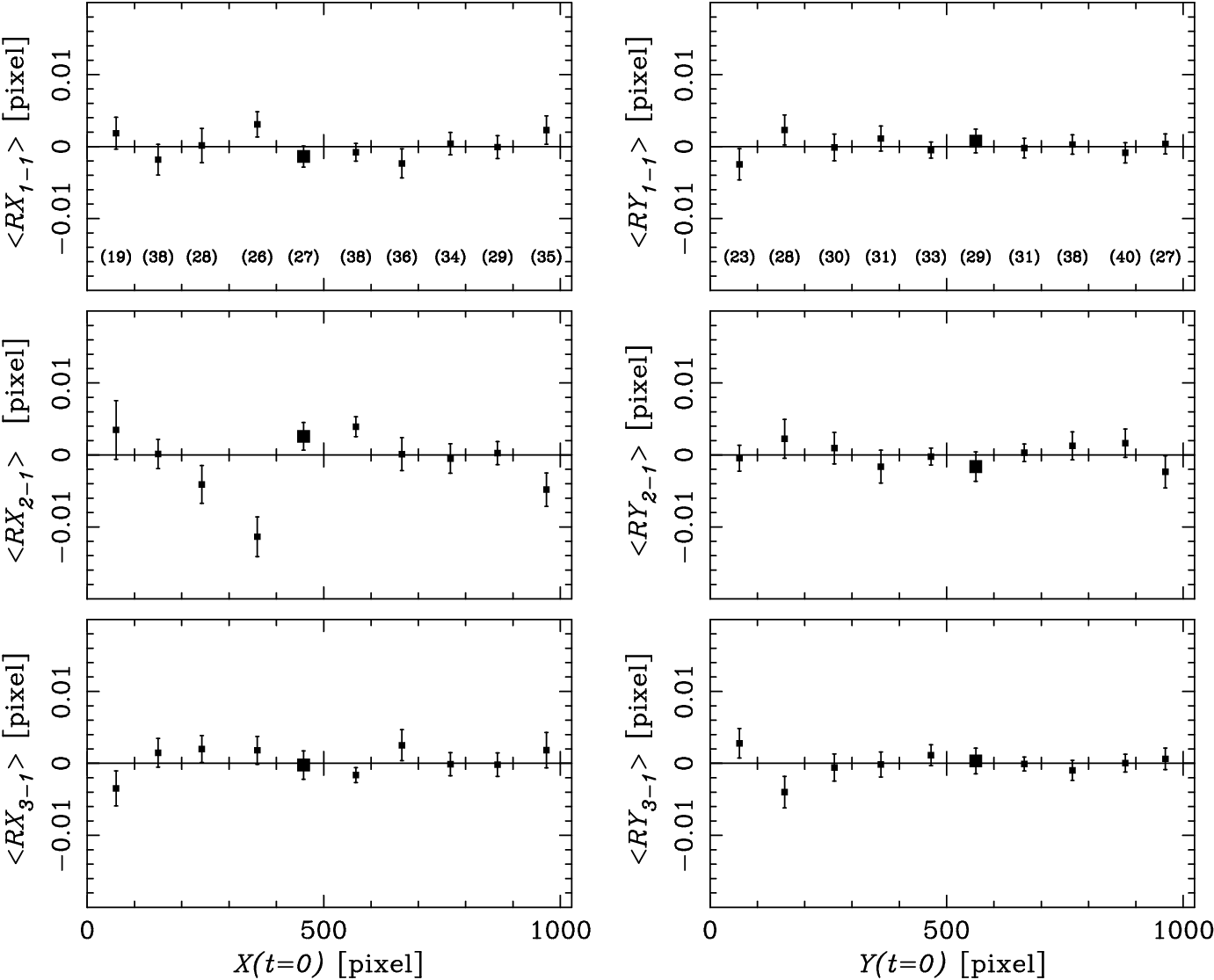}
\caption{Mean position residuals as a function of position for the
FOR~J$0240-3438$ field.  \textit{Left}: $\langle RX \rangle$ vs. $X$.
\textit{Right}: $\langle RY \rangle$ vs. $Y$.  The points are the
weighted mean residuals in 10 equal-length bins in $X$ or $Y$.  The
points are plotted at the mean of the coordinate values in the bin.
The numbers in parenthesis indicate how many values contributed to the
mean.  The larger squares mark those points that include the
contribution to the mean from the QSO.}
\label{fig:stis1-rxry-a}
\end{figure}

\begin{figure}[p]
\centering
\includegraphics[angle=-90,scale=0.7]{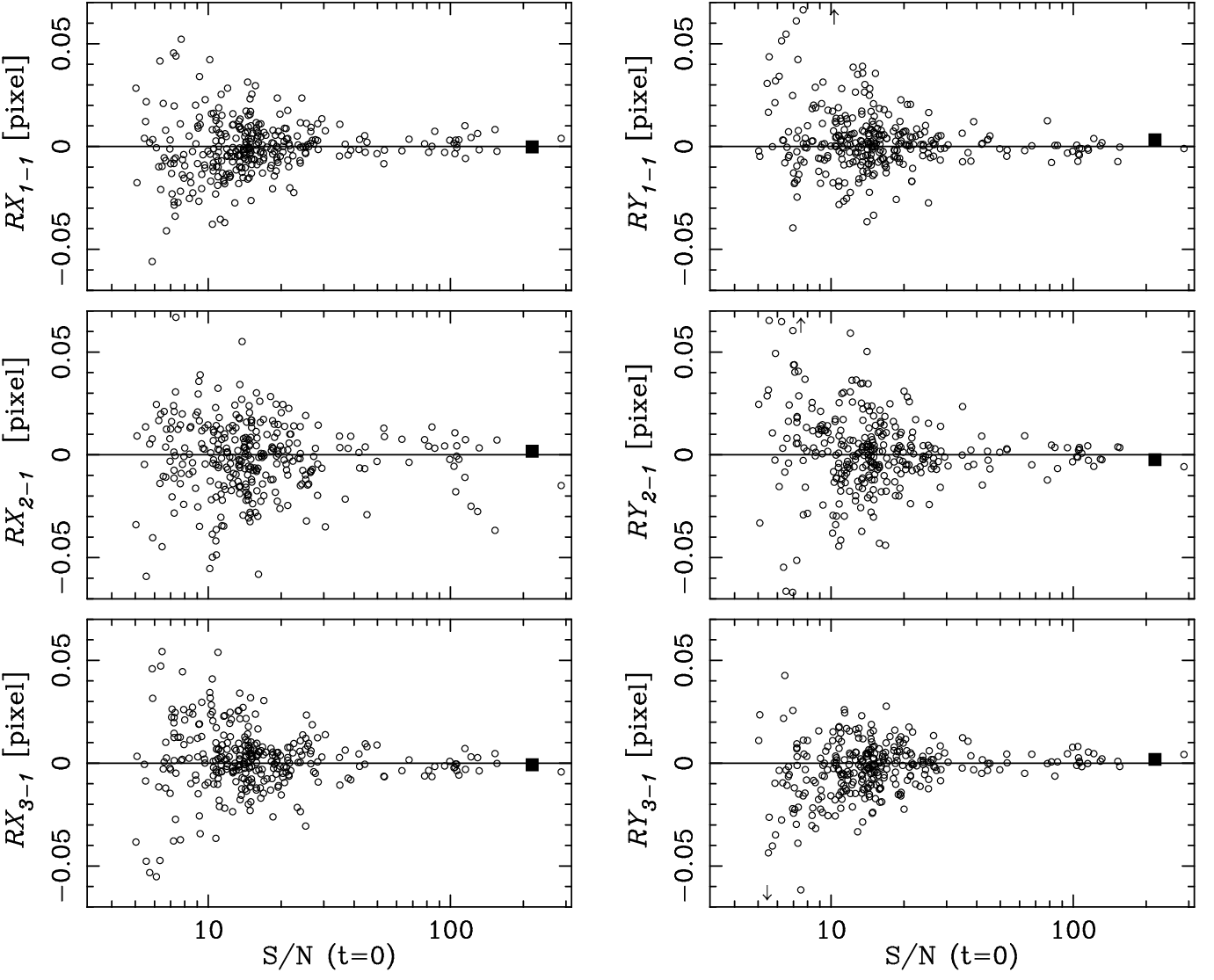}
\caption{Position residuals as a function of S/N for the objects in
the FOR~J$0240-3438$ field.  The epochs increase chronologically from
top to bottom.  \textit{Left}: $RX$ vs. S/N.  \textit{Right}: $RY$
vs. S/N.  The squares correspond to the QSO. The arrows indicate
points beyond the boundaries of the plot.}
\label{fig:stis1-rxry-sn}
\end{figure}

	Figure~\ref{fig:stis1-rxry} plots $RX$ and $RY$ as a function
of position in the standard coordinate system.  Although the plots do
not show any egregious trends, some subtle structures can be
discerned.  The points in the plot of $RX$ \textit{versus} $X$ for the
2001 epoch (middle row) have a greater scatter than those in the
corresponding plots for the two other epochs and show a ``flare up''
around $X \approx 300$~pixel.  The points in the plot $RY$
\textit{versus} $Y$ also have a greater scatter, which, in contrast,
varies little with $Y$.  These observations are more apparent in Figure
\ref{fig:stis1-rxry-a}.  Only the plot of $\langle RX \rangle$
\textit{versus} $X$ for the middle epoch shows a statistically
significant deviation of $\langle RX \rangle$ from 0.  Here, $\langle
RX \rangle$ for $X$ between 300-400~pixel deviates more than 4~$\sigma
$ from 0.  Figure~\ref{fig:stis1-rxry-sn} shows no systematic trends
between the position residuals and S/N.
 

\begin{figure}[t]
\centering
\includegraphics[angle=-90,scale=0.5]{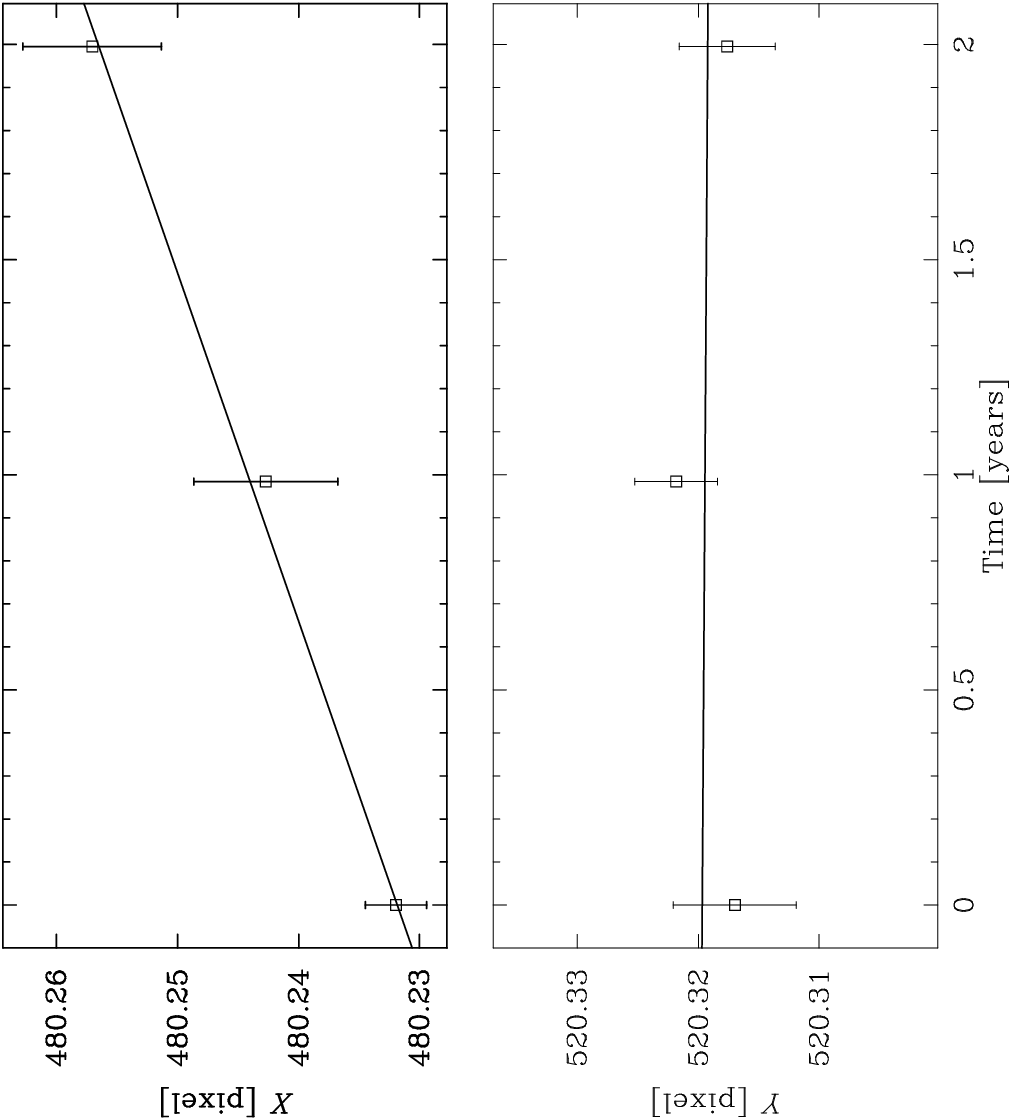}
\caption{Location of the QSO as a function of time for the
FOR~J$0240-3438$ field in the standard coordinate system.  The
vertical axis in each plot has the same scale.}
\label{fig:stis1-qso}
\end{figure}

	Figure \ref{fig:stis1-qso} depicts the motion of the QSO in the
standard coordinate system.  The implied motion from the best-fitting
straight lines is $(\mu_{x},\mu_{y})=(0.0124\pm 0.0030,-0.0002\pm
0.0032)$~pixel~yr$^{-1}$.  The contribution to the total $\chi^2$ from
the QSO is 0.98 for two degrees of freedom.

\subsection{Motion of the QSO in the FOR~J$0238-3443$ field}
\label{sec:stis2}

	The number of objects with a measured centroid is 365, 380,
and 347 in epochs 2000, 2001, and 2003, respectively, and 256 are
common to all three epochs.  The greatest difference in the pointing
of \textit{HST} is between the 2001 and 2003 epochs, where the
difference is about 2.6~pixel in the $X$ direction and about
15.6~pixel in the $Y$.  As for the previous field, the images were
corrected for the effects caused by the decreasing charge transfer
efficiency.  The choice for the individual $\chi^2$ that triggers
fitting for uniform linear motion is 15.  The multiplicative constant
that produces a $\chi^2$ of one per degree of freedom is 1.125.

\begin{figure}[p]
\centering
\includegraphics[angle=-90,scale=0.7]{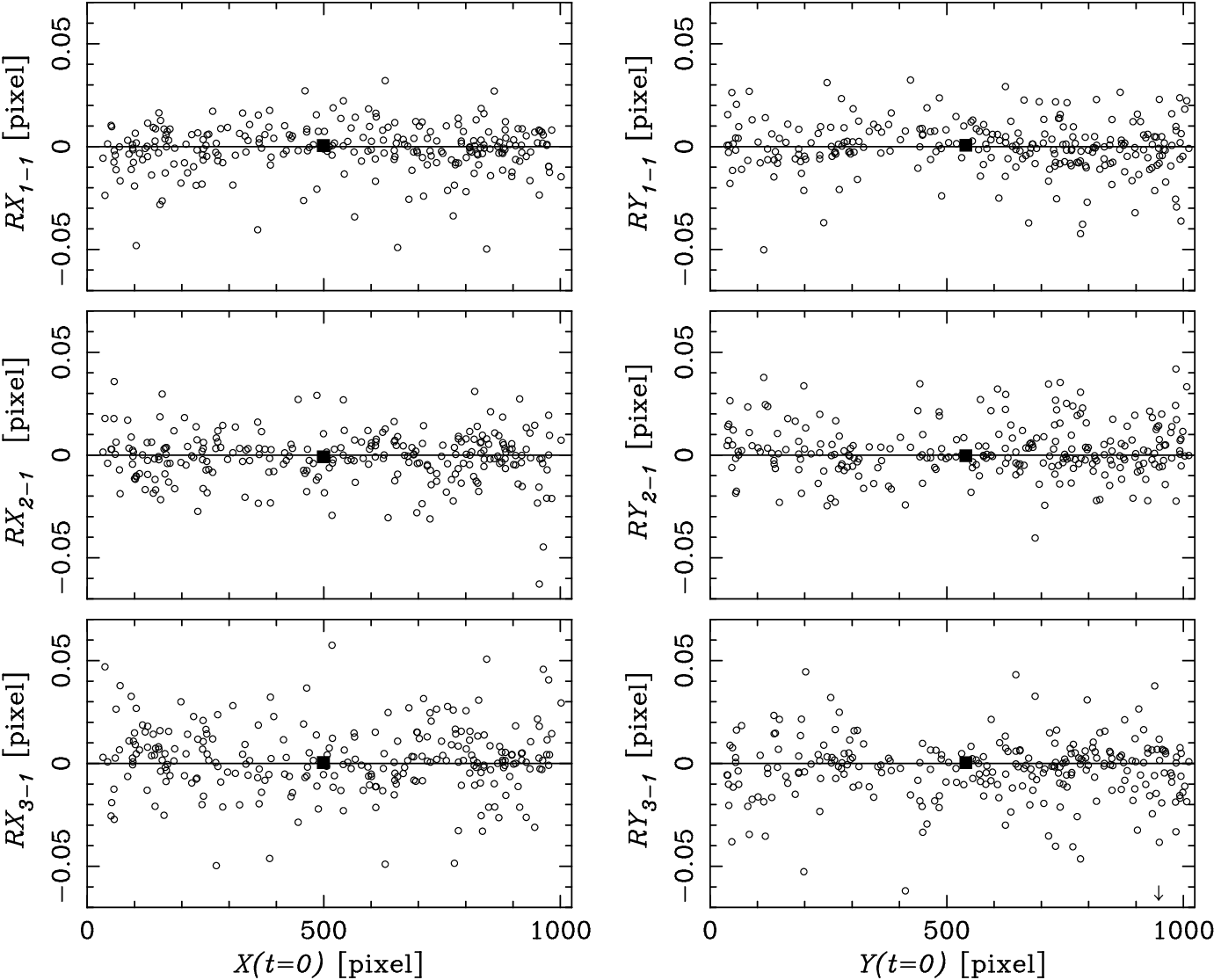}
\caption{Position residuals for the objects in the FOR~J$0238-3443$
field.  The epochs increase chronologically from top to bottom.
\textit{Left}: $RX$ vs. $X$.  \textit{Right}: $RY$ vs. $Y$.  The
squares correspond to the QSO.  The arrows indicate points beyond the
boundaries of the plot.}
\label{fig:stis2-rxry}
\end{figure}

\begin{figure}[p]
\centering
\includegraphics[angle=-90,scale=0.7]{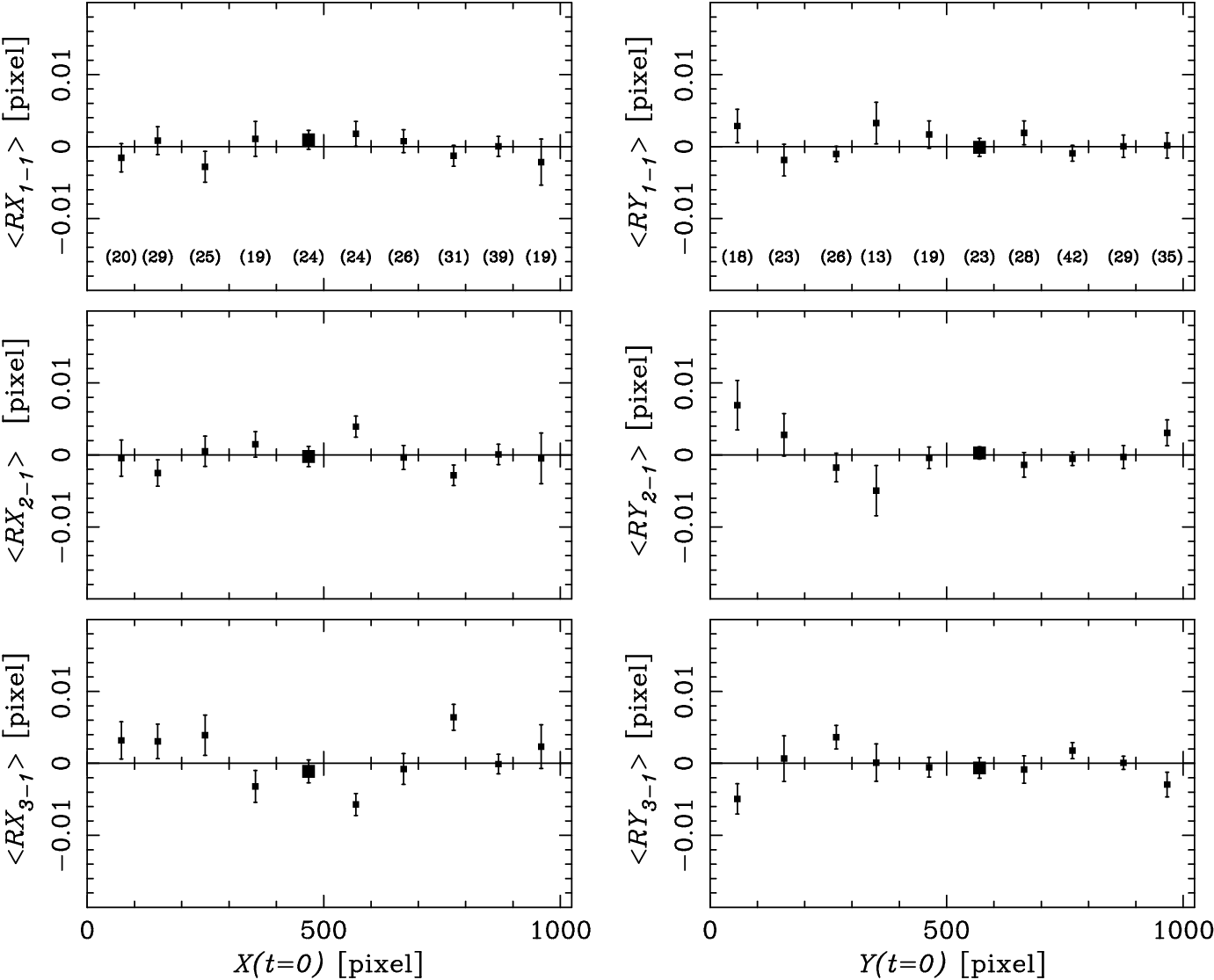}
\caption{Mean position residuals as a function of position for the
FOR~J$0238-3443$ field.  \textit{Left}: $\langle RX \rangle$ vs. $X$.
\textit{Right}: $\langle RY \rangle$ vs. $Y$.  The points are the
weighted mean residuals in 10 equal-length bins in $X$ or $Y$.  The
points are plotted at the mean of the coordinate values in the bin.
The numbers in parenthesis indicate how many values contributed to the
mean.  The larger squares mark those points that include the
contribution to the mean from the QSO. }
\label{fig:stis2-rxry-a}
\end{figure}

\begin{figure}[p]
\centering
\includegraphics[angle=-90,scale=0.7]{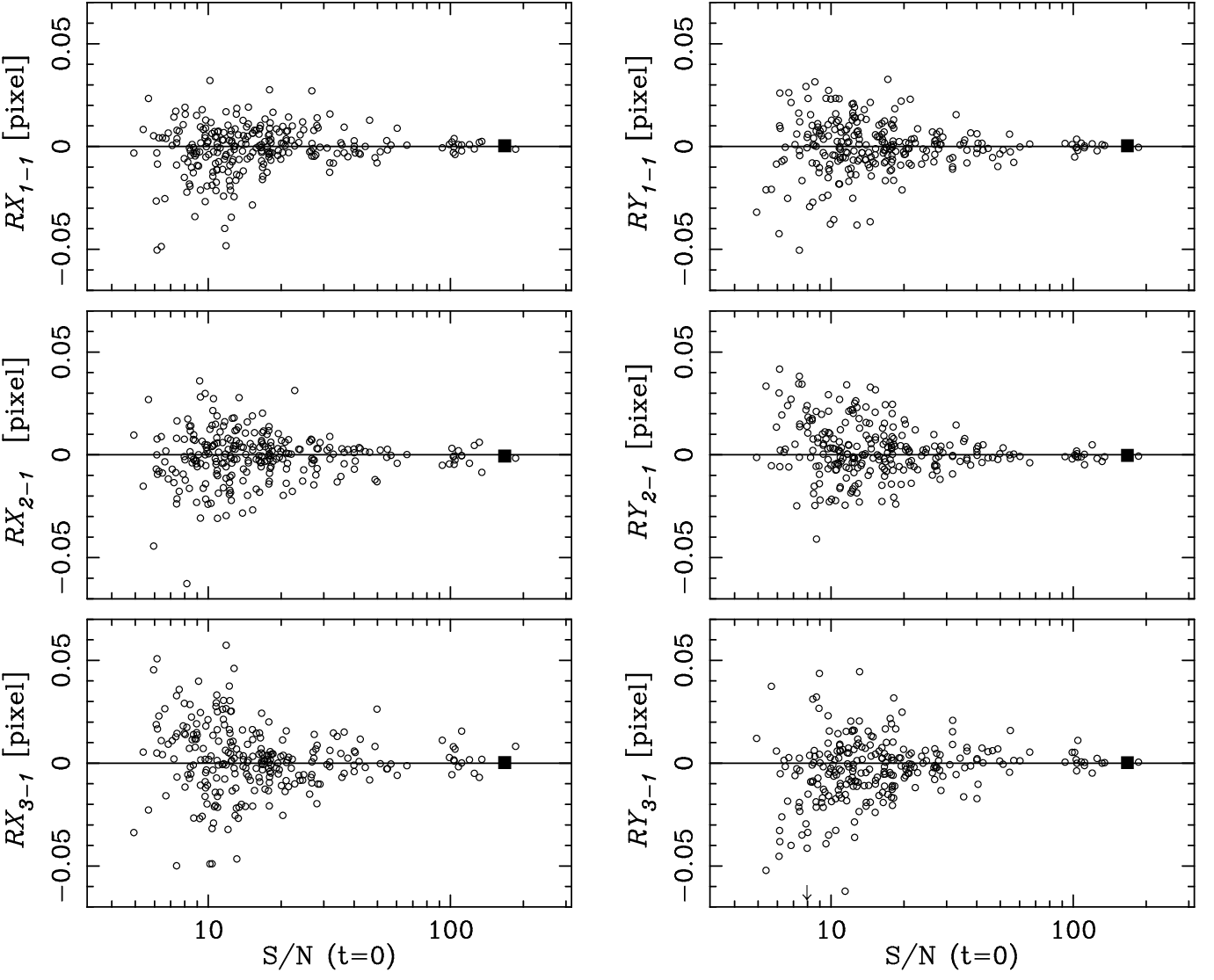}
\caption{Position residuals as a function of S/N for the objects in
the FOR~J$0238-3443$ field.  The epochs increase chronologically from
top to bottom.  \textit{Left}: $RX$ vs. S/N.  \textit{Right}: $RY$
vs. S/N.  The squares correspond to the QSO. The arrows indicate
points beyond the boundaries of the plot.}
\label{fig:stis2-rxry-sn}
\end{figure}

	Figures \ref{fig:stis2-rxry} and \ref{fig:stis2-rxry-a} for the
FOR~J$0238-3443$ field are analogous to Figures \ref{fig:stis1-rxry}
and \ref{fig:stis1-rxry-a}.  The first of these two figures does not
show any alarming trends between the residuals and position.  The
scatter of the points shown in the bottom row of plots --- especially
in $\langle RX \rangle$ \textit{versus} $X$ --- is larger than that in
the other plots.  The larger error bars and somewhat larger scatter
around zero seen in the plot of average residuals confirm this
observation.  Figure~\ref{fig:stis2-rxry-sn}, analogous to
\ref{fig:stis1-rxry-sn}, also does not show any systematic trends
between the position residuals and S/N.

\begin{figure}[t]
\centering
\includegraphics[angle=-90,scale=0.5]{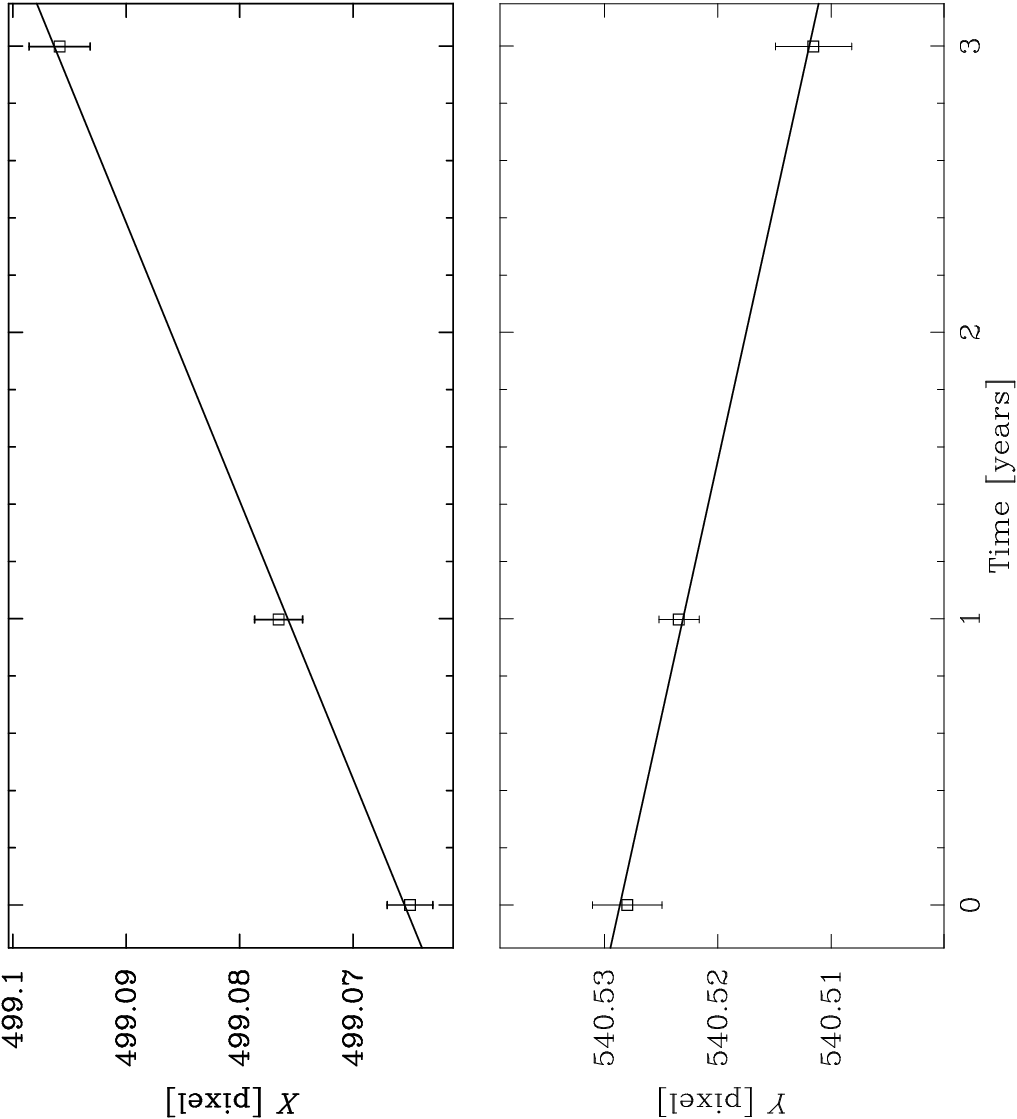}
\caption{Location of the QSO as a function of time for the
FOR~J$0238-3443$ field in the standard coordinate system.  The
vertical axis in each plot has the same scale.}
\label{fig:stis2-qso}
\end{figure}

	Figure \ref{fig:stis2-qso} shows the motion of the QSO in the
standard coordinate system.  The implied motion from the best-fitting
straight lines is $(\mu_{x},\mu_{y})=(0.0103\pm 0.0011,-0.0056\pm
0.0014)$~pixel~yr$^{-1}$.  The contribution to the total $\chi^2$ from
the QSO is 0.33 for two degrees of freedom.

\subsection{Measured Proper Motion}
\label{sec:pm-comp}

	Table~3 gives the measured proper motion for each field in the
equatorial coordinate system and the weighted mean.  Table~4 gives the
proper motion for those objects in the FOR~J$0240-3434$ field for which
it was measured.  Tables~5 and 6 do the same for the FOR~J$0240-3438$
and FOR~J$0238-3443$ fields, respectively.  The first two lines in
Table~4 and the first lines in Tables~5 and 6 correspond to the QSOs.
The subsequent lines list objects in order of decreasing S/N.  After
column (1) listing the ID of an object, columns (2) and (3) give the
$X$- and $Y$-coordinates of the object in the earliest image of the
first epoch.  These images are u50j0201r, o5bl04010, and o5bl03010
for Tables~4, 5, and 6, respectively.  Column (4) gives the
average S/N at the first epoch.  Columns (5) and (6) give the
equatorial components of the measured proper motion.  Each value is the
sum of the measured motion in the standard coordinate system and the
weighted mean proper motion of Fornax given in the bottom line of
Table~3.  In other words, the listed values are with respect to the
QSOs and, thus, the values for the QSOs are listed as 0.  The quoted
uncertainty is the sum in quadrature of the uncertainty in the motion
of an object in the standard coordinate system and the uncertainty of
the weighted mean proper motion of Fornax.  Column (7) gives the
contribution of the object to the total $\chi^{2}$.

	The measured proper motion of Fornax from this article,
$(\mu_{\alpha},\mu_{\delta})=(47.6 \pm 4.6, -36.0 \pm
4.1)$~mas~century$^{-1}$, replaces the preliminary measurement in
\citet{p02}.  \citet{di04} reported another independent measurement,
$(\mu_{\alpha},\mu_{\delta})=(59 \pm 16, -15 \pm
16)$~mas~century$^{-1}$.  For these two independent measurements, the
$\alpha$ components differ by 0.68 times the uncertainty of their
difference, whereas the $\delta$ components differ by 1.27 times.

\begin{figure}[t]
\centering
\includegraphics[angle=-90,scale=0.8]{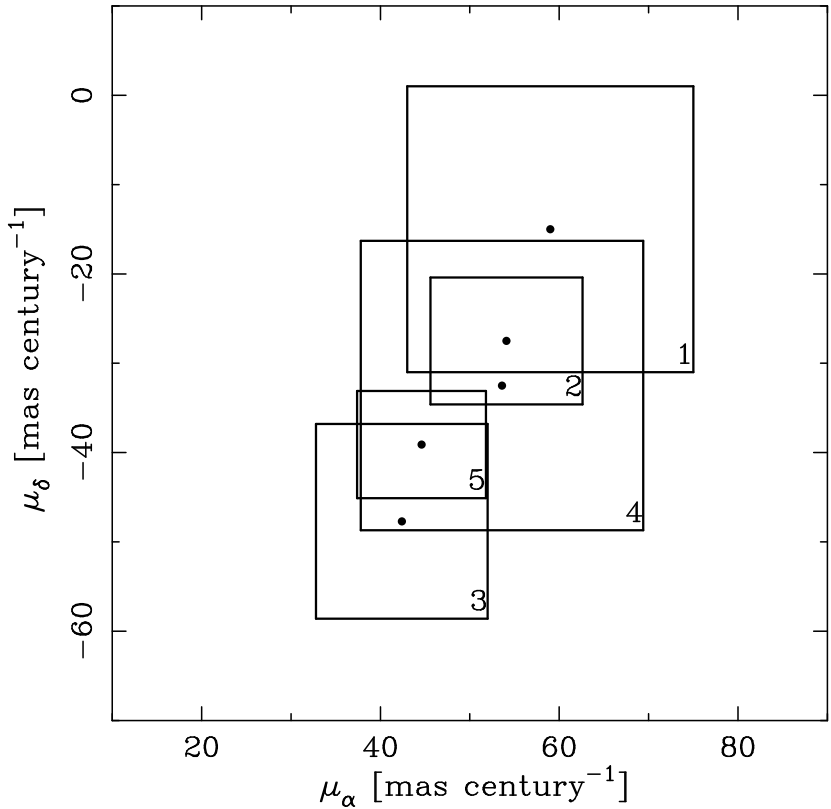}
\caption{Comparison of five independent measurements of the proper
motion of Fornax.  The center of each rectangle, marked with a
circle, is the best estimate of the proper motion, and the sides are
offset by the 1 $\sigma$ uncertainties.  Rectangles 1 - 5 correspond
to the measurements from \citet{di04}, and from this study for the fields
FOR~J$0240-3434$A, FOR~J$0240-3434$B, FOR~J$0240-3438$, and
FOR~J$0238-3443$, respectively.}
\label{fig:pm-comp}
\end{figure}

	Figure~\ref{fig:pm-comp} compares the measurement from
\citet{di04} and the four independent measurements from this article.
A rectangle represents a single measurement: its center, marked with a
circle, is the best estimate of the proper motion and the sides are
offset by the 1-$\sigma$ uncertainties.  Rectangle 1 represents the
measurement from \citet{di04}, while the other four represent the
measurements in this article (see the caption).  The $\alpha$
components of all the measurements differ from their mean by less than
their uncertainties, whereas the $\delta$ components of some
measurements differ by more.  For the measurements from this study,
the largest difference in the $\delta$ components, for 2 and 3, is
1.55 times its uncertainty.  The $\chi^2$ of our measurements around
their weighted mean is 1.3 for $\mu_{\alpha}$ and 2.9 for
$\mu_{\delta}$, each for 3 degrees of freedom.  Including the
\citet{di04} measurement in the weighted mean alters the values by
only 1--2~mas~century$^{-1}$.  The $\chi^2$ of the five measurements
around this mean is 1.9 for $\mu_{\alpha}$ and 4.8 for
$\mu_{\delta}$, each for 4 degrees of freedom.  We conclude that the
agreement among the five independent measurements is reasonable,
implying that the proper motion of Fornax is reliably known.

	The result in \citet{di04} is a weighted mean of two
measurements, one using galaxies and the other QSOs as standards of
rest.  The lensed QSO~J$0240-3434$ is among the standards, and
\citet{di04} give individual values of the proper motion with respect
to image A and B of the QSO.  They are:
$(\mu_{\alpha},\mu_{\delta})=(84 \pm 52, -4 \pm
64)$~mas~century$^{-1}$ for image A and
$(\mu_{\alpha},\mu_{\delta})=(73 \pm 34, -93 \pm
80)$~mas~century$^{-1}$ for image B.  Comparing these to our
measurements in Table~3 shows an agreement within the uncertainties;
however, because of the large uncertainties in the values above they
do not provide a significant check on our results.

\subsubsection{Effect of Changing Charge Transfer Efficiency}
\label{sec:pm-cti}
We have performed a parallel analysis that made no correction for the
effects of charge transfer efficiency for each of the three fields.
The resulting weighted mean $\mu_{\alpha}$ is nearly the same as the
value in Table~3, but the mean $\mu_{\delta}$ is more
negative by 17~mas~century$^{-1}$.  The $\chi^2$ around these means
are 3.0 and 2.9, respectively, both for 3 degrees of freedom.  These
values are larger than those for the results in Table~3, demonstrating
that correcting for the effects of the changing charge transfer
efficiency significantly improves the agreement between the individual
measurements of the proper motion.

\subsection{Galactic Rest-Frame Proper Motion}
\label{sec:grfpm}

	The measured proper motion of the dSph contains contributions
from the motion of the LSR and the peculiar motion of the Sun.  The
magnitude of each contribution depends on the location of the dSph in
the sky.  Removing them produces the Galactic rest-frame proper motion
--- the proper motion measured by a hypothetical observer at the
location of the Sun but at rest with respect to the Galactic center.
Columns (2) and (3) of Table~7 give the equatorial components,
$(\mu_{\alpha}^{\mbox{\tiny{Grf}}},
\mu_{\delta}^{\mbox{\tiny{Grf}}})$, of the Galactic-rest-frame proper
motion.  Their derivation assumes: 220~km~s$^{-1}$ for the circular
velocity of the LSR; 8.5~kpc for the distance of the Sun from the
Galactic center; and $(u_\odot, v_\odot, w_\odot) = (-10.00 \pm 0.36,
5.25 \pm 0.62 , 7.17 \pm 0.38)$~km~s$^{-1}$ \citep{db98} for the
peculiar velocity of the Sun, where the components are positive if
$u_{\odot}$ points radially away from the Galactic center, $v_{\odot}$
is in the direction of rotation of the Galactic disk, and $w_\odot$
points in the direction of the north Galactic pole.  Columns (4) and
(5) give the Galactic rest-frame proper motion in the Galactic
coordinate system,
$(\mu_{l}^{\mbox{\tiny{Grf}}},\mu_{b}^{\mbox{\tiny{Grf}}})$.  Columns
(6) - (8) give the $\Pi$, $\Theta$, and $Z$ components of the space
velocity in a cylindrical coordinate system centered on the dSph.  The
components are positive if $\Pi$ points radially away from the
Galactic axis of rotation, $\Theta$ points in the direction of
rotation of the Galactic disk, and $Z$ points in the direction of the
north Galactic pole.  The derivation of these components assumes
$138$~kpc \citep{sa00} for the heliocentric distance to and $53.3 \pm
0.8$~km~s$^{-1}$ \citep{w06} for the heliocentric radial velocity of
Fornax.  Columns (9) and (10) give the radial and tangential
components of the space velocity for an observer at rest at the Galactic
center.  The component $V_{r}$ is positive if it points radially away
from the Galactic center.  The uncertainties in the listed quantities
derive from Monte Carlo experiments.  The bottom line in Table~7 gives
the weighted mean of each listed quantity.  Note that even though
$\Pi^{2} + \Theta^{2} + Z^{2} = V_{r}^{2} + V_{t}^{2}$ for a
measurement in a given field, this equality may not hold for the mean
values.

The negative $V_{r}$ in Table~7 means that Fornax is moving towards the
Milky Way.  The negative value for $\Theta$ shows that the orbit is
retrograde.

\section{Orbit and Orbital Elements of Fornax}
\label{sec:orbit}

	Knowing the space velocity of a dSph allows a determination of
the orbit of the galaxy for a given form of the Galactic potential.
This work adopts a Galactic potential that has a contribution from a
disk of the form \citep{mn75}
\begin{equation}
\label{diskpot}
\Psi_{\mbox{\small{disk}}}=-\frac{G
M_{\mbox{\small{disk}}}}{\sqrt{R^{2}+(a+\sqrt{Z^{2}+b^{2}})^{2}}},
\end{equation}
from a spheroid of the form \citep{h90}
\begin{equation}
\label{spherpot}
\Psi_{\mbox{\small{spher}}}=-\frac{GM_{\mbox{\small{spher}}}}
{R_{\mbox{\small{GC}}}+c},
\end{equation}
and from a halo of the form
\begin{equation}
\label{logpot}
\Psi_{\mbox{\small{halo}}}=v^{2}_{\mbox{\small{halo}}}\ln
(R^{2}_{\mbox{\small{GC}}}+d^{2}).
\end{equation}
In the above equations, $R_{\mbox{\small GC}}$ is the Galactocentric
distance, $R$ is the projection of $R_{\mbox{\small GC}}$ onto the
plane of the Galactic disk, and $Z$ is the distance from the plane of
the disk.  All other quantities in the equations are adjustable
parameters and their values are the same as those adopted by
\citet{jsh95}: $M_{\mbox{disk}}=1.0\times10^{11}$~M$_{\odot}$,
$M_{\mbox{spher}}=3.4\times10^{10}$~M$_{\odot}$,
$v_{\mbox{halo}}=128$~km~s$^{-1}$, $a=6.5$~kpc, $b=0.26$~kpc,
$c=0.7$~kpc, and $d=12.0$~kpc.

\begin{figure}
\centering
\includegraphics[angle=-90,scale=0.9]{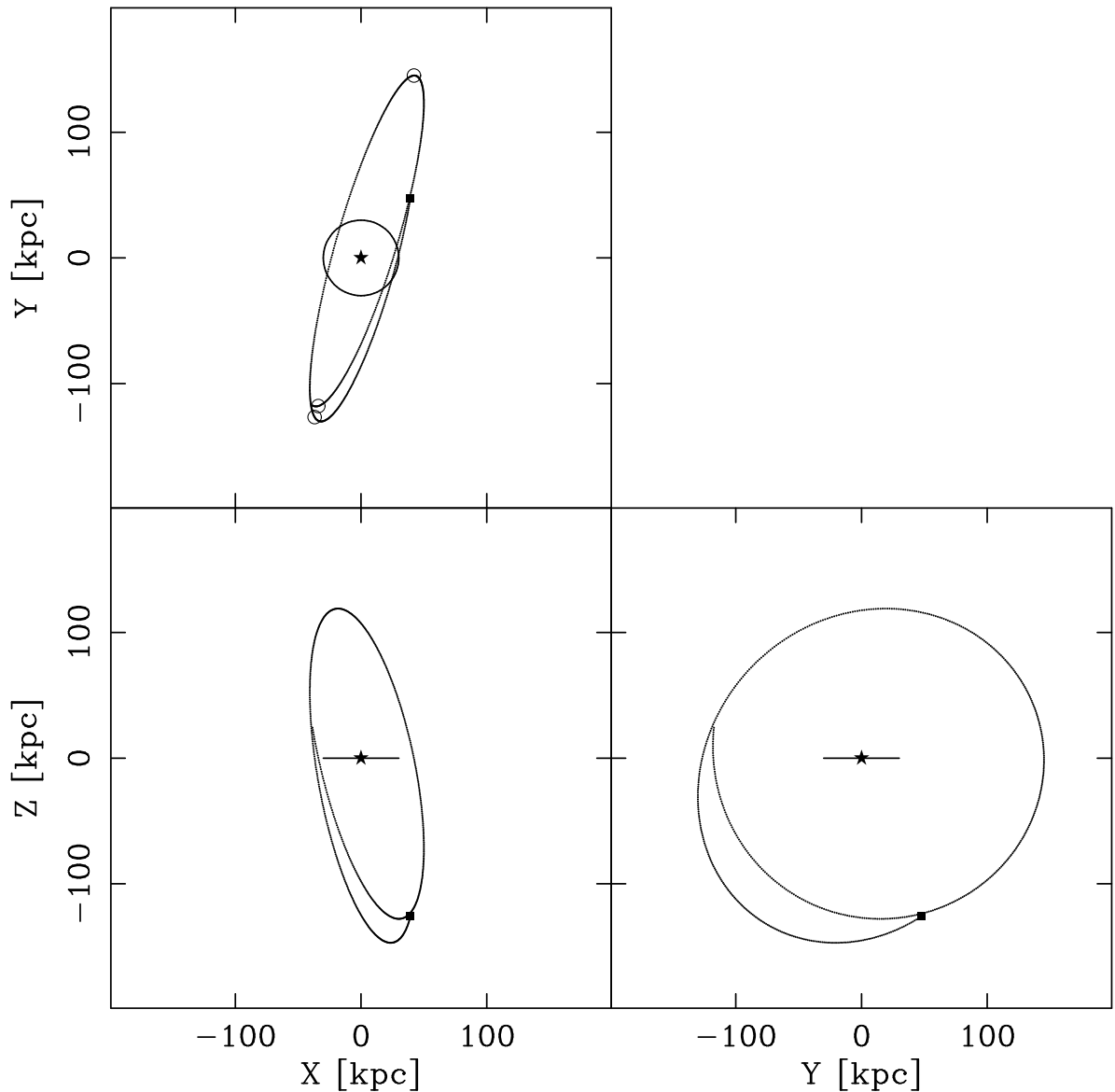}
\caption{Projections of the orbit of Fornax on the $X$-$Y$ plane
(\textit{top left}), the $X$-$Z$ plane (\textit{bottom left}), and the
$Y$-$Z$ plane (\textit{bottom right}).  The origin of the right-handed
coordinate system is at the Galactic center, which is marked with a
star.  The Galactic disk is in the $X$-$Y$ plane and the present
location of the Sun is on the positive $X$ axis.  The squares mark the
current location of Fornax at $(X, Y, Z) = (39, 48, -126)$~kpc.  For
reference, the large circle in the $X$-$Y$ plane has radius of 30
kpc.  The three small circles in the $X$-$Y$ projection mark the
points at which Fornax passes through the plane of the Galactic disk.
The integration starts from the present and extends backward in time
for 6 Gyr.}
\label{fig:orbit}
\end{figure}

	Figure \ref{fig:orbit} shows the projections of the orbit of
Fornax onto the $X$-$Y$ (\textit{top left panel}), $X$-$Z$
(\textit{bottom left panel}), and $Y$-$Z$ (\textit{bottom right
panel}) Cartesian planes.  The orbit results from an integration of
the motion in the Galactic potential given by equations
(\ref{diskpot})--(\ref{logpot}).  The integration extends for 6~Gyr
backward in time and begins at the current location of Fornax with the
negative of the space velocity components given in the bottom line of
columns (6)--(8) of Table~7.  The squares in Figure~\ref{fig:orbit}
mark the current location of the dSph, the stars indicate the center
of the Galaxy, and the three small circles mark the points on the
orbit at which $Z=0$, or, in other words, where the orbit crosses the
plane of the Galactic disk.  The large circle is for reference; it has
a radius of 30~kpc.  In the right-handed coordinate system of
Figure~\ref{fig:orbit}, the current location of the Sun is on the
positive $X$-axis.  The figure shows that Fornax is moving toward the
Milky Way, is closer to apogalacticon than perigalacticon, and that it
has a moderately polar orbit with a small eccentricity.

	Table~8 tabulates the elements of the orbit of Fornax.  The
value of the quantity is in column (4) and its $95\%$ confidence
interval is in column (5).  The latter comes from 1000 Monte Carlo
experiments, where an experiment integrates the orbit using an initial
velocity that is generated by randomly choosing the line-of-sight
velocity and the two components of the measured proper motion from
Gaussian distributions whose mean and standard deviation are the best
estimate of the quantity and its quoted uncertainty, respectively.
The eccentricity of the orbit is defined as
\begin{equation}
\label{eccentricity}
e = \frac{(R_{a} - R_{p})}{(R_{a} + R_{p})}.
\end{equation}
The most likely orbit has about a 1.3:1 ratio of apogalacticon to
perigalacticon, and the 95\% confidence interval for the eccentricity
allows ratios approximately between 1.2:1 and 2.2:1.  The orbit is
retrograde and only $11^{\circ}$ away from being polar.  The orbital
period, 3.2 Gyr, is about 45\% longer than that for Sculptor (2.2 Gyr;
\citet{p06}) and more than a factor of two longer than those for Carina
(1.4~Gyr; \citet{p03}) and Ursa~Minor (1.5~Gyr; \citet{p05}).

\section{Discussion}
\label{sec:disc}
\subsection{Is Fornax a Member of a Stream?}

	\citet{lb82} proposed that most, if not all, dSphs have a
``tidal origin.''  In this picture, the tidal field of the Galaxy
distorts a progenitor galaxy and creates streams of stars and gas which
become gravitationally unbound to the progenitor galaxy, but move on a
similar orbit.  A new dSph forms from this material --- tidal debris
--- when dissipation of energy makes distinct clumps of the debris
gravitationally bound.  Consequently, the galaxies and globular
clusters that formed from the debris move on similar orbits or, in
other words, constitute a ``stream.''  In the case of the ``FLS''
stream \citep{lb82}, Fornax is the surviving progenitor galaxy, whereas
Leo I, Leo II, and Sculptor are the ``condensate'' dSphs.  \citet{lb76}
proposed that another great stream of galaxies is associated with the
Magellanic Stream \citep{m74} and contains the Small and Large
Magellanic Clouds, Draco, and Ursa Minor.  The number of member
galaxies in the streams increased as new satellite galaxies of the
Milky Way were discovered.  \citet{m94} modified the FLS stream by
including in it Sextans and renaming it as the ``FL$^{2}$S$^{2}$''
stream.  The orbital plane of the FL$^{2}$S$^{2}$ stream is only
slightly different from the original plane of the FLS stream.  The idea
of streams is verifiable.  Given distances to and radial velocities of
the members of a stream, the expected proper motions of the members can
be calculated.  Thus, in a more quantitative analysis, \citet{lb95}
calculated likelihoods for the existence of several streams containing
dSphs and globular clusters and predicted the proper motions for the
member galaxies and globular clusters.  Fornax could be a member of one
of four possible streams: streams 1a or 1b (together with Pal 14, Pal
15, and Eridanus), 4a (together with Sextans, Sculptor, and Pal 3), and
4b (together with Sextans and Sculptor).  Table~9 contains the
predicted motions for Fornax; columns (2) and (3) give the $\alpha$ and
$\delta$ components of the proper motion, whereas columns (4) and (5)
give this motion in polar coordinates.  For easy comparison, the bottom
line lists our measurement from Table~3.

	The proper motion of Fornax predicted for stream 4a is in the
closest agreement with our measurement, though even here the difference
is more than 3.2 times the measurement uncertainty in one component.
Given that there must be some uncertainty in the predicted motions,
this difference may not be large enough to rule out membership in
stream 4a.  However, \citet{p06} ruled out the membership of Sculptor
in stream 4a, so the reality of the stream hinges on the agreement
between the predicted and actual proper motions of the remaining three
members.  Currently, neither Sextans nor Pal 3 has a measured proper
motion.

\subsection{Is the Orbit of Fornax in the Kroupa-Theis-Boily Plane?}

	\citet{kr05} noted that viewing 11 of the innermost dwarf
galaxies, satellites of the Milky Way, from a point at infinity with
$\ell = 167\fdg 91$ shows that the galaxies lie on nearly the same
plane, whose poles are at $(\ell, b)=(168^{\circ},-16^{\circ})$ and
$(348^{\circ}, +16^{\circ})$.  The probability is less than $0.5\%$
that such an alignment of galaxies occurs by chance if the galaxies have
an isotropic distribution of orbits.  If the planar alignment is to
persist in time, the dwarf galaxies must have orbits contained in this
plane.   Therefore, its persistence can be tested if proper motions of the
galaxies are known.

	Let the pole of an orbit be in the direction of the orbital
angular momentum vector.  Thus, the Galactic coordinates of the pole
in terms of the orbital elements are:
\begin{equation}
\label{eq:poles}
(\ell,b) = (\Omega+90^{\circ},\Phi-90^{\circ}).
\end{equation}
For prograde orbits, $b < 0$, and for retrograde, $b > 0$.  Using the
best estimates for $\Omega$ and $\Phi$ from Table~8 and uncertainties
that are 1-$\sigma$ errors from Monte Carlo simulations, the orbital
pole of Fornax is at $(\ell,b)=(163^{\circ} \pm 8^{\circ},13^{\circ}
\pm 4^{\circ})$.  The poles of the orbit and plane are separated by
$29.4^{\circ} \pm 4.1^{\circ}$.  The separation is larger than the
measurement uncertainties, however, the agreement between the two poles
does not need to be exact because the plane has a finite thickness.
\citet{kr05} find an rms distance from their plane of 26.4~kpc for the
sample of 11 galaxies --- which extends to about twice the distance of
Fornax.  This thickness is consistent with their claim of a ratio of
height to radius of 0.15 for the disk.  The tilt between the orbit of
Fornax and the plane predicts that the time-averaged rms distance of
Fornax from the plane is 47~kpc.  This calculation assumes a circular
orbit with a radius of 135~kpc, which is the average of the peri- and
apogalacticons.  We conclude that the orbit of Fornax is inconsistent
with the plane.  Whether the orbits of the dSphs are consistent with
the predictions of $\Lambda$-CDM models for the Local Group will be
addressed in a later article when measurements become available for
Draco and Sextans.

\subsection{What Happened to the Gas in Fornax?}

	\citet{shs98} and \citet{sa00} detected a population of blue
and young main-sequence stars in Fornax.  The stars are about $10^{8}$
years old and located in the central region of the galaxy.  Quoting
\citet{shs98}, the population ``has a flattened distribution on the sky
and its major axis is offset by roughly $30^{\circ}$ from the symmetry
axis of the galaxy as a whole.''  This detection implies that Fornax
had to have HI $10^{8}$ years ago.  However, a search for HI with the
VLA by \citet{y99} produced no detection at the limit of $4.6 \times
10^{18}$~cm$^{-2}$ at the galaxy center and $7.9 \times
10^{18}$~cm$^{-2}$ at the distance of one core radius from the center.
\citet{b06} searched for HI near the dwarf galaxies in the Local Group
using the Parkes radio telescope.  The survey confirmed the result of
\citet{y99} that there is no HI at the center of Fornax, but did
identify gas offset about $30^{\prime}$ north and west-northwest from
the center.  The strongest emission is between position angles of about
325$^\circ$ and 360$^\circ$ \citep[see Figure 2 of][]{b06}.
The velocity of the gas is also offset --- about 30~km~s$^{-1}$
below that of the stars.  \citet{b06} calculate an approximately 10\%
probability that this gas is a chance superposition of a high-velocity
cloud.  This probability is not low enough to decisively establish a
physical relationship between the gas and the galaxy.  \citet{b06}
also note that subtracting the emission from the Milky Way is
difficult for these observations.  If there is no HI associated with
Fornax, then why do young stars reside in the central region?  If the
HI is associated, why is it offset in both position and velocity?

	Trying to answer such questions, \citet{y99} offers two
possibilities: (1) there is gas in the central region but it is either
ionized or molecular or (2) HI was displaced from the central region in
the past $10^{8}$ years.  There are no published results of searches
for HII or molecular gas, and so, (1) cannot be ignored or excluded
until such results exist.  In the case of (2), there must be a physical
mechanism for displacing the gas.  A supernova-driven wind blowing the
gas from the center is one possible mechanism \citep{mf99}.  Ram
pressure due to the motion of the galaxy through a gaseous halo of the
Milky Way is another mechanism, \textit{e.g.}, \citet{ga01} or
\citet{ma06}.  In this case, the gas would trail the dSph and be
displaced in the direction opposite to the Galactic rest-frame proper
motion vector.  This vector has a position angle of $120^{\circ} \pm
8.5^{\circ}$, \textit{i.e.} pointing in the southeast direction, which
would predict that the gas should be at a position angle of about
300$^\circ$.  However, the strongest HI emission has a position angle
of about 340$^\circ$ and the elongation of the cloud is not aligned
with the proper motion vector.  \citet{di04} propose a variation of the
ram-pressure picture in which Fornax crossed the Magellanic plane ---
the plane containing the orbit of the LMC as defined by the space
velocity from \citet{vdm02} --- about 190 Myr ago and passed through
gas that was removed from the LMC but which still trails this gas-rich
galaxy.  This picture can be tested using our new proper motion for
Fornax and a recently-measured proper motion for the LMC.

	\citet{ka06} measured the proper motion of the LMC using an
observational technique similar to the one employed in this
contribution.  The measurement is based on \textit{HST} observations of
21 QSOs in the direction of the LMC with the High Resolution Camera of
the Advanced Camera for Surveys.  The QSOs were discovered by their
variability in the MACHO survey \citep{ge03}.  The measured proper
motion of the LMC is $(\mu_{\alpha},\mu_{\delta})=(203 \pm 8, 44 \pm
5)$~mas~century$^{-1}$.

	Integrating the two orbits backwards in time in the potential
given by Eqs.~\ref{diskpot}--\ref{logpot} shows that Fornax crosses the
orbital plane of the LMC about 1.4~Gyr ago and it does so at a distance
of only 1.9~kpc from the orbit of the LMC.  However, when Fornax is
closest to the LMC orbit, the LMC will not arrive at that point for
about another 0.8~Gyr.  Numerical experiments that integrate the orbits
using starting velocities with uncertainties determined by the
observations indicate that Fornax crosses the plane of the LMC orbit
between 0.80~Gyr and 2.0~Gyr ago with 95\% confidence.  Fornax
approaches closer than 20~kpc to the orbit of the LMC in about 50\% of
the simulations.  The Monte Carlo simulations show that the two
galaxies do not come closer than 81~kpc at 97.5\% confidence.

For the orbits adopted here, the recent loss of HI from Fornax 190~Myr
ago cannot be attributed to the ram pressure from gas in the LMC orbital
plane.  The last crossing of the orbital plane by
Fornax occurred about 1.4~Gyr ago, which is 1~Gyr before the cessation
of star formation in the dSph.  The plane crossing occurred 0.6~Gyr before
the cessation of star formation with 97.5\% confidence.  Our
conclusion differs from that of \citet{di04} because of differences in
the adopted orbits for Fornax and, particularly, the LMC.

Though an interaction between Fornax and gas from the LMC cannot
explain the recent cessation of star formation in the dSph, the orbits
of the two galaxies are likely to approach closely.  It is still an
open question whether this intersection of orbits has affected the star
formation history or gas content of Fornax.  The strongest argument
against such an effect is that the closest approach of the two orbits
occurs before the arrival of the LMC at that point.  However, the
amount of gas in the leading and trailing parts of the orbit are
uncertain.  The proper motion for the LMC of \citet{ka06} implies a
Galactocentric space velocity of 380~km~s$^{-1}$, which yields an
orbital period --- calculated using the method in
Section~\ref{sec:orbit}  --- of 4.3~Gyr.  This period is almost three
times longer than that assumed in the canonical models for the
formation of the Magellanic Stream \citep{ga99, yn03, con04}.  This
conflict between the measured motion of the LMC and the models for the
formation of the Magellanic Stream means that it is premature to rule
out the removal of gas from Fornax by interaction with the Stream.  An
interaction with HI leading the LMC could have removed at least some
gas from Fornax.  \citet{pu98} identified what appears to be a leading
tidal tail in HI maps.  However, the likelihood of an interaction is
decreased because the leading tail contains only about 25\% as much gas
as the trailing tail and the gas is organized in three disjoint
complexes rather than in a continuous distribution \citep{bru05}.

\subsection{The Effect of the Galactic Tidal Force on the Structure of
Fornax}
\label{sec:tides}

	If Fornax formed as a spherical system and if the Galactic
tidal force has had an important effect on its structure, then the
projected ellipticity would be the result of the stretching of the
dSph in its orbital plane \citep[\textit{e.g.},][]{jsh95,ola95,pp95}.
Thus, a nearly Galactocentric observer would see the Galactic
rest-frame proper motion vector aligned with the projected major axis.
From Table 1, the position angle of the major axis is $41^{\circ} \pm
1^{\circ}$, whereas this angle for the weighted mean Galactic
rest-frame proper motion is $120^{\circ} \pm 8.5^{\circ}$.  Within the
uncertainties, these two angles are almost orthogonal to each other;
in other words, the true major axis is not in the orbital plane.

	Notwithstanding the disagreement in the two position angles
above, the structure of Fornax exhibits several features which could
have arisen as a consequence of tidal interaction with the Milky Way.
Most recently, \citet{co05} show that: (1) the ellipticity of the
isodensity contours increases with increasing semi-major axis, reaching
a maximum at around $35^{\prime}$, and then decreases beyond
this radius; (2) the position angle of ellipses fitted to the stellar
surface density decreases with increasing semi-major axis; (3)
isodensity contours on the eastern side of the galaxy are closer
together than those on the western; and (4) the map of the stellar surface
density shows stars of the dSph beyond the fitted tidal radius along
the minor axis.

	The feature which is most suggestive of a tidal interaction is
(4).  The direction of the Galactic rest-frame proper motion supports
this picture; however, other aspects of the orbit do not:  the large
perigalacticon, long orbital period, small eccentricity, and
approaching motion of the dSph do not favor the tidal origin for the
observed ``extratidal'' stars.  \citet{co05} propose that instead of
the Galactic tidal force, a collision and merger event with another
system is responsible for this observed feature.

\subsection{Is There Dark Matter in Fornax?}

	Assuming that mass follows light and using the measured
velocity dispersion, \citet{ma91} derived a $M/L_{V}$ of $12.3 \pm 4.5$
in solar units for Fornax, noting that $M/L_{V}$s in the range between
5.3 and 26 are also possible given the uncertainties in the fitted
structural parameters.  The subsequent derivations of $M/L$ by
\citet{w03} and \citet{w06} are within this quoted range.  Knowing the
orbit of a dSph permits another approach to constraining its mass.  The
mass and orbit determine the tidal radius --- the radius beyond which a
star becomes unbound from the dSph due to gravitational interaction
with the Milky Way.  Making the assumption that the tidal radius is
equal to the observed limiting radius then provides a value for the
$M/L$.  The value commonly adopted for the limiting radius is the
limiting radius of a \citet{k66} model fitted to the radial surface
brightness profile (unfortunately often also called the tidal radius).
An exponential --- which has no limiting radius --- is also a good fit
to the surface brightness profiles of dSphs, so the identification of
the limiting radius determined from a King model with the tidal radius
may be incorrect.  Nevertheless, we calculate the $M/L$ of Fornax by
equating the limiting and tidal radii.  A constraint on the $M/L$ of a
dSph with fewer assumptions, also given below, is that it must be large
enough for the dSph to have survived the tidal interaction with the
Milky Way given its measured orbit.

	Equation~(\ref{eq:rt}) \citep{k62,ola92} gives the tidal
radius, $r_t$, as a function of the orbital elements of a dSph moving
in a logarithmic potential for the Milky Way.
\begin{equation}
r_t = \left(\frac{(1-e)^2}{[(1+e)^2/2e]\ln[(1+e)/(1-e)] +1} \,
\frac{M}{M_G}\right)^{1/3} a.
\label{eq:rt}
\end {equation}
In the equation, $e$ is eccentricity of the orbit, $a$ is the
semi-major axis ($a\equiv (R_{a}+R_{p})/2$), $M$ is the mass of the
dSph, and $M_G$ is the mass of the Milky Way within $a$.  Equating
$r_t$ with the observed limiting radius derived by fitting a
\citet{k66} model, $r_k$, gives a value for $M/L$ for a given orbit.
Estimates of the limiting radius for Fornax range from $64^{\prime}$
\citep{co05} to $98^{\prime}$ \citep{w03}.  If $r_k = 64^{\prime}$,
then 10.4\% of the orbits have $M/L_{V} > 5.3$ in 1000 Monte Carlo
simulations similar to those that determined the uncertainties in the
orbital elements; 0.2\% of the orbits have $M/L_{V} > 26$.  If $r_k =
98^{\prime}$, then 100\% of the orbits have $M/L_{V} > 5.3$ and 5.3\%
have $M/L_{V} > 26$.  These results show that there is only a 5.3\%
chance that the global $M/L_{V}$ of Fornax is greater than 26 if the
larger of the limiting radii is identified with the tidal radius.  In
other words, the orbit of Fornax is consistent with the broad range of
$M/L$ values derived from virial equilibrium models and the observed
limiting radius of the dSph and, thus, these calculations do not provide
a strong constraint on the amount of dark matter in Fornax.

	The average measured $M/L_V$ for Galactic globular clusters is
2.3 \citep{pm93}.  Could Fornax have a $M/L_V$ equal to this value and
survive the effect of the Galactic tidal force until the present on our
derived orbit?  Numerical simulations of tidal interactions by
\citet{ola95} and \citet{pp95} show that the ratio of the limiting
radius derived by fitting a theoretical King model \citep{k66}, $r_k$,
to the tidal radius defined by Equation~(\ref{eq:rt}) is a useful
indicator of the importance of the Galactic tidal force on the
structure of a dSph.  These simulations show that: if $r_{k}/r_t
\lesssim 1.0$, the Galactic tidal force has little effect on the
structure of the dSph; at $r_k/r_t \approx 2.0$, the effect of the
force increases rapidly with increasing $r_k/r_t$; and, for $r_k/r_t
\approx 3.0$, the dSph disintegrates in a few orbits.  Assuming that
$M/L_V = 2.3$ and $r_k = 64^{\prime}$, $r_k/r_t > 2.0$ for 0.3\% of
the orbits generated in Monte Carlo simulations.  If $r_k =
98^{\prime}$, the fraction is 10.8\%.  Thus, it is very likely that
Fornax could have survived for a Hubble time on its current orbit
while containing only luminous matter.  We conclude that our orbit of
Fornax and the observed limiting radius do not constrain the amount
of dark matter in the dSph.

\subsection{A Lower Limit for the Mass of the Milky Way}
\label{sec:mmw}

	The space velocity of Fornax imposes a lower limit on the mass
of the Milky Way, which is given by:
\begin{equation}
\label{eq:mmw}
M = \frac{R(V_{r}^{2} + V_{t}^{2})}{2G},
\end{equation}
where $R$ is the Galactocentric distance, and $V_{r}$ and $V_{t}$ are
the radial and tangential components of the space velocity,
respectively, as measured by an observer at the center of the Galaxy.
Equation \ref{eq:mmw} assumes a spherically symmetric mass
distribution and 0 for the total energy of the satellite galaxy.
Setting $R=140$~kpc and using the values from Table~7 for $V_{r}$ and
$V_{t}$, $M=(6.4 \pm 1.9) \times 10^{11} M_{\odot}$.  This estimate
does not place a strong constraint on the amount of mass within
140~kpc of the center of the Milky Way since the value is similar to
the $5.4^{+0.1}_{-0.4}\times 10^{11} M_{\odot}$ within $R=50$~kpc
found by
\citet{sa03}.

\section{Summary}
\label{sec:summary}

	This contribution continues a series of articles, each
reporting the proper motion of a dSph; here, we present a measurement
for Fornax which replaces the value based on only part of the data in
\citet{p02}.  The data come from imaging with the \textit{HST} and STIS
or PC2.  Using the measured proper motion, we integrate the motion of
the dSph to obtain its orbit and then discuss the proposed membership
of Fornax in streams of satellites, the possible alignment of orbits of
the Galactic satellites in a single plane, the association between HI
and the dSph, the proposed interaction between the LMC and Fornax, the
tidal interaction of the dSph with the Milky Way, the dark matter
content of Fornax, and a lower limit on the mass of the Milky Way.  The
major findings are itemized below.

	1. The weighted mean of the four independent measurements of
the proper motion is $(\mu_{\alpha},\mu_{\delta}) = (47.6\pm
4.6,-36.0\pm 4.1)$~mas~century$^{-1}$ in the equatorial coordinate
system for an observer at and moving with the Sun.

	2. The Galactic rest-frame proper motion is that measured by
an observer at the location of the Sun and at rest with respect to the
Galactic center.  In this system, the proper motion is
$(\mu_{\alpha}^{\mbox{\tiny{Grf}}}, \mu_{\delta}^{\mbox{\tiny{Grf}}})
= (24.4\pm 4.6,-14.3\pm 4.1)$~~mas~century$^{-1}$.

	3. The weighted means of the radial and tangential components
of the space velocity with respect to a stationary observer at the
Galactic center are $V_{r} = -31.8 \pm 1.7$~km~s$^{-1}$ and $V_{t} = 196
\pm 29$~km~s$^{-1}$, respectively.  The negative sign of $V_{r}$ means
that Fornax is approaching the Milky Way.

	4. The orbit resulting from the integration of the best
estimate of the motion in a realistic potential of the Milky Way shows
that Fornax is approaching a perigalacticon of 118~kpc on an
approximately polar orbit with an eccentricity of 0.13.  The
apogalacticon of the orbit is 152~kpc and the orbital period is
3.2~Gyr.

	5. Fornax is not a member of stream 1a, 1b, or 4b of
\citet{lb95}, though it could be a member of stream 4a.  Since
\citet{p05} ruled out the membership of the Sculptor dSph in stream 4a,
the reality of this stream depends on whether the proper motions of the
other proposed members, which still need to be measured, agree with the
predictions.

	6. The orbit of Fornax is not in the Kroupa-Theis-Boily plane,
which contains eleven of the Galactic satellite galaxies.

	7. Our findings do not support the hypothesis of
\citet{di04} that Fornax lost its HI about 200~Myr ago as it passed
through the stream of gas trailing the LMC in its orbit.  Our proper
motion for Fornax and that of \citet{ka06} for the LMC show that
Fornax approaches within about 1.9~kpc of the orbit of the LMC, but
1.4~Gyr ago and 0.8~Gyr before the LMC arrives at that point.  An
interaction 1.4~Gyr ago does not explain why star formation in Fornax
continued until 200~Myr ago, though uncertainties in modeling the
Magellanic Stream keep this mechanism for the loss of HI from Fornax
from being definitively ruled out.

\acknowledgments

CP and SP acknowledge the financial support of the Space Telescope
Science Institute through the grants HST-GO-07341.03-A and
HST-GO-08286.03-A and of the National Science Foundation through
grant AST-0098650.  EWO acknowledges support from the Space Telescope
Science Institute through the grants HST-GO-07341.01-A and
HST-GO-08286.01-A and from the National Science Foundation through the
grants AST-0205790 and AST-0507511.  MM acknowledges support from the
Space Telescope Science Institute through the grants HST-GO-07341.02-A
and HST-GO-08286.02-A and from the National Science Foundation through
the grant AST-0098661.  DM is supported by FONDAP Center for
Astrophysics 15010003 and by a Fellowship from the John Simon Guggenheim
Foundation.

\clearpage

\setcounter{table}{0}
\newdimen\digitwidth\setbox0=\hbox{\rm 0}\digitwidth=\wd0
\catcode`@=\active\def@{\kern\digitwidth}

\begin{deluxetable}{lcc}
\tablecolumns{3}
\tablewidth{6.0truein} 
\tablecaption{Fornax at a Glance}
\tablehead{
\colhead{Quantity} &
\colhead{Value}    &
\colhead{Reference} \\
\colhead{(1)}&
\colhead{(2)}&
\colhead{(3)}}
\startdata
Right Ascension, $\alpha$ (J2000.0) &$\phantom{-}02:39:53.1$&van den Bergh (1999
) \\
Declination, $\delta$ (J2000.0) &$-34:30:16.0$&van den Bergh (1999) \\
Galactic longitude, $\ell$& $237.245^{\circ}$& \\
Galactic latitude, $b$    & $-65.663^{\circ}$& \\
Heliocentric distance & $138\pm 8$ kpc      & Saviane et al. (2000)
\\
Luminosity, $L_{V}$ &$(1.4\pm0.4)\times10^{7}~L_{\odot}$&Irwin \&
Hatzidimitriou (1995) \\ 
Ellipticity, $e$ &$0.30\pm0.01$& $^{\prime\prime}$ \\
Position angle & $41^{\circ}\pm 1^{\circ}$ & $^{\prime\prime}$ \\
Core radius  & $13.7^{\prime} \pm 1.2^{\prime}$ & $^{\prime\prime}$ \\
Tidal radius & $71.1^{\prime} \pm 4.0^{\prime}$ & $^{\prime\prime}$ \\
Heliocentric radial velocity&$53.3 \pm 0.8$~km~s$^{-1}$ & Walker et
al. (2006) \\
\enddata
\end{deluxetable}

\begin{deluxetable}{lclcccc}
\tablecolumns{7}
\tablewidth{0pt}
\tablecaption{Fornax: Information about Fields and Images}
\tablehead{&
\colhead{R.A.} &
\colhead{Decl.}&
\colhead{Date} &
               &
               &
\colhead{T$_{exp}$\tablenotemark{a}}\\
\colhead{Field}        &
\colhead{(J2000.0)}    &
\colhead{(J2000.0)}    &
\colhead{$yyyy-mm-dd$} &
\colhead{Detector}     &
\colhead{Filter}       &
\colhead{(s)}          \\
\colhead{(1)}&
\colhead{(2)}&
\colhead{(3)}&
\colhead{(4)}&
\colhead{(5)}&
\colhead{(6)}&
\colhead{(7)}}
\startdata
FOR@J$0240-3434$&02@40@07.70&$-34$@34@20.01&$1999-03-10$&PC2&F606W&$18\times 160
$\\
&&&$2001-03-08$&&&$16\times160$\\
&&&$2003-03-08$&&&$16\times160$\\
\noalign{\vspace{3pt}}
FOR@J$0240-3438$&02@40@38.70&$-34$@38@58.00&$2000-01-31$&STIS&50CCD&$24\times192
$\\
&&&$2001-01-25$&&&$24\times192$\\
&&&$2002-01-29$&&&$24\times192$\\
\noalign{\vspace{3pt}}
FOR@J$0238-3443$&02@38@43.80&$-34$@43@53.00&$2000-03-08$&STIS&50CCD&$24\times193
$\\
&&&$2001-03-08$&&&$24\times192$\\
&&&$2003-03-09$&&&$24\times192$\\
\enddata
\tablenotetext{a}{Number of images times the mean exposure time.  The
actual exposure times of individual images may differ by a few percent
from the mean due to constraints imposed by the orbit of HST.}
\end{deluxetable}

\begin{deluxetable}{lrr}
\tablecolumns{3}
\tablewidth{0pt}
\tablecaption{Measured Proper Motion of Fornax}
\tablehead{
&\colhead{$\mu_{\alpha}$}&\colhead{$\mu_{\delta}$}  \\ 
\colhead{Field}&\multicolumn{2}{c}{(mas century$^{-1}$)}\\
\colhead{(1)}&\colhead{(2)}&\colhead{(3)}}
\startdata
FOR@J$0240-3434A$&$54.1\pm8.5$&$-27.5\pm7.1$\\
\noalign{\vspace{3pt}}
FOR@J$0240-3434B$&$42.4\pm9.6$&$-47.7\pm10.9$\\
\noalign{\vspace{3pt}}
FOR@J$0240-3438$&$53.6\pm15.8$&$-32.5\pm16.2$\\
\noalign{\vspace{3pt}}
FOR@J$0238-3443$&$44.6\pm7.2$&$-39.1\pm6.0$\\
\noalign{\vspace{1pt}}
\hline
\noalign{\vspace{1pt}}
Weighted mean&$47.6\pm4.6$&$-36.0\pm4.1$\\
\enddata
\end{deluxetable}

\begin{deluxetable}{ccccccc}
\tablecolumns{7}
\tablewidth{0pt} 
\tablecaption{Measured Proper Motions For Objects in
the FOR~J$0240-3434$ Field}
\tablehead{ &X&Y& &$\mu_{\alpha}$&$\mu_{\delta}$ &  \\
\colhead{ID}&\colhead{(pixels)}&\colhead{(pixels)}&\colhead{$S/N$}&\colhead{(mas
@century$^{-1}$)}&
\colhead{(mas@century$^{-1}$)} & \colhead{$\chi^2$} \\
\colhead{(1)}&\colhead{(2)}&\colhead{(3)}&\colhead{(4)}&\colhead{(5)}&
\colhead{(6)} & \colhead{(7)}
}
\startdata
  1& 458& 461& 163& $    0 \pm  10 $& $    0 \pm   8 $&  1.47 \\
  2& 562& 540& 122& $    0 \pm  11 $& $    0 \pm  11 $&  1.37 \\
  3& 162& 463&  60& $    6 \pm   9 $& $  -28 \pm  12 $&  1.44 \\
\enddata
\end{deluxetable}

\begin{deluxetable}{ccccccc}
\tablecolumns{7}
\tablewidth{0pt} 
\tablecaption{Measured Proper Motions For Objects in
the FOR~J$0240-34348$ Field}
\tablehead{ &X&Y& &$\mu_{\alpha}$&$\mu_{\delta}$ &  \\
\colhead{ID}&\colhead{(pixels)}&\colhead{(pixels)}&\colhead{$S/N$}&\colhead{(mas
@century$^{-1}$)}&
\colhead{(mas@century$^{-1}$)} & \colhead{$\chi^2$} \\
\colhead{(1)}&\colhead{(2)}&\colhead{(3)}&\colhead{(4)}&\colhead{(5)}&
\colhead{(6)} & \colhead{(7)}
}
\startdata
  1& 480& 520& 218& $    0 \pm  16 $& $    0 \pm  16 $&  0.98 \\
  2& 781& 960&  24& $  116 \pm  34 $& $   24 \pm  27 $&  0.81 \\
  3& 126& 358&  15& $  442 \pm  54 $& $ -317 \pm  47 $&  0.31 \\
\enddata
\end{deluxetable}

\begin{deluxetable}{ccccccc}
\tablecolumns{7}
\tablewidth{0pt} 
\tablecaption{Measured Proper Motions For Objects in
the FOR~J$0238-3443$ Field}
\tablehead{ &X&Y& &$\mu_{\alpha}$&$\mu_{\delta}$ &  \\
\colhead{ID}&\colhead{(pixels)}&\colhead{(pixels)}&\colhead{$S/N$}&\colhead{(mas
@century$^{-1}$)}&
\colhead{(mas@century$^{-1}$)} & \colhead{$\chi^2$} \\
\colhead{(1)}&\colhead{(2)}&\colhead{(3)}&\colhead{(4)}&\colhead{(5)}&
\colhead{(6)} & \colhead{(7)}
}
\startdata
  1& 499& 541& 167& $    0 \pm   8 $& $    0 \pm   7 $&  0.33 \\
  2& 739& 874& 135& $   48 \pm  10 $& $  -80 \pm  12 $&  2.65 \\
  3& 905& 278& 111& $    4 \pm   8 $& $  -29 \pm   9 $&  0.48 \\
  4& 719& 833& 104& $   28 \pm   9 $& $  -63 \pm  12 $&  1.24 \\
  5& 955& 996&  27& $   52 \pm  21 $& $ -143 \pm  22 $&  4.24 \\
  6& 708& 995&  10& $  113 \pm  61 $& $ -226 \pm  40 $&  3.15 \\
\enddata
\end{deluxetable}

\clearpage

\begin{deluxetable}{lrrrrrrrrr}
\tablecolumns{10}
\rotate
\tablewidth{0pt} 
\tablecaption{Galactic-Rest-Frame Proper Motion and Space Velocity of Fornax}
\tablehead{&\colhead{$\mu_{\alpha}^{\mbox{\tiny{Grf}}}$}&
\colhead{$\mu_{\delta}^{\mbox{\tiny{Grf}}}$}&
\colhead{$\mu_{l}^{\mbox{\tiny{Grf}}}$}&\colhead{$\mu_{b}^{\mbox{\tiny{Grf}}}$}&
\colhead{$\Pi$}&\colhead{$\Theta$}&\colhead{$Z$}&\colhead{$V_{r}$}&
\colhead{$V_{t}$}\\ 
\colhead{Field}&\multicolumn{2}{c}{(mas cent$^{-1}$)}
&\multicolumn{2}{c}{(mas cent$^{-1}$)}&\colhead{(km s$^{-1}$)}&\colhead{(km s$^{
-1}$)}&\colhead{(km s$^{-1}$)}
&\colhead{(km s$^{-1}$)}&\colhead{(km s$^{-1}$)}\\
\colhead{(1)}&\colhead{(2)}&\colhead{(3)}&\colhead{(4)}
&\colhead{(5)}&\colhead{(6)}&\colhead{(7)}&\colhead{(8)}&\colhead{(9)}
&\colhead{(10)}}
\startdata
FOR@J$0240-3434A$&$30.9\pm8.5$&$-5.8\pm7.1$&$0.2\pm7.2$&$31.4\pm8.5$&$172\pm51$&
$-21\pm47$&$116\pm23$&$-27.7\pm3.1$&$206\pm54$\\
\noalign{\vspace{5pt}}
FOR@J$0240-3434B$&$19.2\pm9.6$&$-26.0\pm10.8$&$22.1\pm10.8$&$23.5\pm9.6$&$109\pm
57$&$-158\pm71$&$94\pm26$&$-36.6\pm4.1$&$211\pm66$\\
\noalign{\vspace{5pt}}
FOR@J$0240-3438$&$30.4\pm15.8$&$-10.8\pm16.2$&$5.2\pm16.2$&$31.8\pm15.8$&$170\pm
94$&$-54\pm110$&$118\pm43$&$-29.8\pm6.2$&$212\pm96$\\
\noalign{\vspace{5pt}}
FOR@J$0238-3443$&$21.4\pm7.2$&$-17.3\pm6.0$&$12.9\pm6.1$&$24.3\pm7.2$&$121\pm43$
&$-100\pm39$&$96\pm19$&$-33.1\pm2.7$&$180\pm44$\\
\noalign{\vspace{5pt}}
\hline
\noalign{\vspace{5pt}}
Weighted mean&$24.4\pm4.6$&$-14.3\pm4.1$&$9.6\pm4.1$&$26.8\pm4.6$&$137\pm27$&$-8
0\pm27$&$103\pm12$&$-31.8\pm1.7$&$196\pm29$\\
\enddata
\end{deluxetable}

\clearpage
\hoffset=0.0truein

\begin{deluxetable}{lcccc}
\tablecolumns{5}
\tablewidth{0pt} 
\tablecaption{Orbital elements of Fornax}
\tablehead{Quantity&Symbol&Unit&Value&95\% Conf. Interv.\\
\colhead{(1)}&\colhead{(2)}&\colhead{(3)}&\colhead{(4)}&\colhead{(5)}}
\startdata
Perigalacticon&$R_{p}$&kpc&$118$&$(66,137)$ \\
\noalign{\vspace{1pt}}
Apogalacticon&$R_{a}$&kpc&$152$&$(144,242)$\\
\noalign{\vspace{1pt}}
Eccentricity &$e$&&$0.13$&$(0.11,0.38)$\\
\noalign{\vspace{1pt}}
Period&$T$&Gyr&$3.2$&$(2.5,4.6)$\\
\noalign{\vspace{1pt}}
Inclination&$\Phi$&deg&101&$(94,107)$ \\
\noalign{\vspace{1pt}}
Longitude&$\Omega$&deg&73&$(58,90)$ \\
\enddata
\end{deluxetable}

\begin{deluxetable}{lcccc}
\tablecolumns{5}
\tablewidth{0pt} 
\tablecaption{Predicted Proper Motion of Fornax}
\tablehead{
&
\colhead{$\mu_{\alpha}$}&
\colhead{$\mu_{\delta}$}&
\colhead{$\vert {\boldmath \mu}\vert $}&
\colhead{PA}\\
\colhead{Stream No.}&
\colhead{(mas century$^{-1}$)}&
\colhead{(mas century$^{-1}$)}&
\colhead{(mas century$^{-1}$)}&
\colhead{(deg)}\\ 
\colhead{(1)}&\colhead{(2)}&\colhead{(3)}&\colhead{(4)}&\colhead{(5)}}
\startdata
1a  & 30  & --15 &34  & 117 \\
\noalign{\vspace{1pt}}
1b  & 15  & --27  &31  &151  \\
\noalign{\vspace{1pt}}
4a  & 38 & --23 & 44 & 121 \\
\noalign{\vspace{1pt}}
4b  & 6.3 & --19  & 20 & 162 \\
\noalign{\vspace{1pt}}

Our Result&$47.6 \pm 4.6$&$-36.0\pm 4.1$&$60.0\pm4.4$& $127\pm4.1$\\
\enddata
\end{deluxetable}

\end{document}